\documentclass[12pt,preprint]{aastex}
\usepackage{apjfonts}
\usepackage{amsmath}
\usepackage{amssymb}
\usepackage{lscape}
\usepackage{color}

\newcommand{\etal}{et~al.~}

\def \Se#1{Section~\ref{sec:#1}}
\def \Ses#1#2{Section~\ref{sec:#1}-\ref{sec:#2}}

\def \Fig#1{Figure~\ref{fig:#1}}
\def \Figs#1#2{Figures~\ref{fig:#1}-\ref{fig:#2}}

\def \Tbl#1{Table~\ref{tbl:#1}}
\def \Tbls#1#2{Tables~\ref{tbl:#1}-\ref{tbl:#2}}

\def \eg{{e.g.~}}
\def \ie{{i.e.~}}

\def \spose#1{\hbox to 0pt{#1\hss}}
\def \ltsima{$\; \buildrel < \over \sim \;$}                    
\def \lsim{\lower.5ex\hbox{\ltsima}}  
\def \gtsima{$\; \buildrel > \over \sim \;$}                    
\def \gsim{\lower.5ex\hbox{\gtsima}} 
\def \ltsim{\mathrel{\spose{\lower 3pt\hbox{$\sim$}}
     \raise 2.0pt\hbox{$<$}}}
\def \gtsim{\mathrel{\spose{\lower 3pt\hbox{$\sim$}}
     \raise 2.0pt\hbox{$>$}}}

\def\pmb#1{\setbox0=\hbox{#1}%
\kern-.025em\copy0\kern-\wd0 \kern.05em\copy0\kern-\wd0 
\kern-.025em\raise.0433em\box0} 

\slugcomment{Accepted to ApJS.}

\shorttitle{Constraining stellar population models}
\shortauthors{Roediger \etal 2013}

\begin{document}


\title{CONSTRAINING STELLAR POPULATION MODELS $-$ \\ I. AGE, METALLICITY AND ABUNDANCE PATTERN \\ COMPILATION FOR GALACTIC GLOBULAR CLUSTERS}

\author{Joel C. Roediger and St\'ephane Courteau}
\affil{Department of Physics, Engineering Physics \& Astronomy, Queen's 
University, Kingston, Ontario, Canada}

\author{Genevieve Graves}
\affil{Department of Astrophysical Sciences, Peyton Hall, Princeton, NJ 08540, USA}

\and

\author{Ricardo P. Schiavon}
\affil{Astrophysics Research Institute, IC2, Liverpool Science Park, Liverpool John Moores University, 146 Brownlow Hill, Liverpool, L3 5RF, UK}

\email{jroediger,courteau@astro.queensu.ca,graves@astro.princeton.edu,\\R.P.Schiavon@ljmu.ac.uk}


\begin{abstract}
We present an extenstive literature compilation of age, metallicity, and 
chemical abundance pattern information for the 41 Galactic globular clusters 
(GGCs) studied by \cite{Sc05}.  Our compilation constitutes a notable 
improvement over previous similar work, particularly in terms of chemical 
abundances.  Its primary purpose is to enable detailed evaluations of and 
refinements to stellar population synthesis models designed to recover the 
above information for unresolved stellar systems based on their integrated 
spectra.  However, since the Schiavon sample spans a wide range of the known 
GGC parameter space, our compilation may also benefit investigations related to 
a variety of astrophysical endeavours, such as the early formation of the Milky 
Way, the chemical evolution of GGCs, and stellar evolution and 
nucleosynthesis.  For instance, we confirm with our compiled data that the GGC 
system has a bimodal metallicity distribution and is uniformly enhanced in the 
$\alpha$-elements.  When paired with the ages of our clusters, we find evidence 
that supports a scenario whereby the Milky Way obtained its globular clusters 
through two channels, in situ formation and accretion of satellite galaxies.  
The distributions of C, N, O, and Na abundances and the dispersions thereof per 
cluster corroborate the known fact that all GGCs studied so far with respect to 
multiple stellar populations have been found to harbour them.  Finally, using 
data on individual stars, we also confirm that the atmospheres of stars become 
progressively polluted by CN(O)-processed material after they leave the main 
sequence and uncover evidence which suggests the $\alpha$-elements Mg and Ca 
may originate from more than one nucleosynthetic production site.

\par

We estimate that our compilation incorporates all relevant analyses from the 
literature up to the mid-2012.  As an aid to investigators in the fields named 
above, we provide detailed electronic tables online of the data upon which our work is based\footnote{\url{www.astro.queensu.ca/people/Stephane\_Courteau/roediger2013/index.html}}.
\end{abstract}

\keywords{Astronomical data bases: miscellaneous,
          Galaxy: abundances,
          Galaxy: globular clusters: general,
          Galaxy: stellar content}


\section{Introduction}\label{sec:Intro}

The stellar content of galaxies represents the time integral, up to the epoch 
of observation, of both their star formation and chemical enrichment 
histories.  This implies that an understanding of stellar populations is 
essential in the study of the galaxy formation problem.  The nature of 
galaxies' stellar populations may be pursued through multi-band photometric 
and/or spectroscopic datasets.  The latter tend to be more highly valued for 
their greater number of population tracers (\ie absorption line/band strengths 
or full spectra versus broad-filter fluxes), reduced sensitivity to dust \citep{Ma05} and multiplexed ability to simultaneously constrain stellar ages, 
metallicities, chemical abundance patterns, dynamics and mass distributions 
\citep[\eg see][]{CvD12}.  The present work concerns spectroscopic-based 
stellar population studies.

Stellar population synthesis (SPS) models are the tools which connect 
observations of stellar systems to their physical parameters.  In order to 
apply them with confidence, the accuracies of their predictions must be 
subjected to in-depth evaluations first.  The standard approach to SPS model 
evaluations is to verify that the predictions uniquely reproduce, to within 
some desired tolerance, the benchmark values of various parameters for well-characterized stellar systems obtained by way of independent and trustworthy 
techniques \citep{SB99,Sc02a,Sc02b,Sc04,Th03,Pr04,LW05,Men07,Sc07,GS08,Ko08,Pe09,Vaz10,Th11b}).

The stellar systems best suited for evaluations of SPS models are the many star 
clusters found in the Milky Way and its nearby satellite galaxies.  The reasons 
for this are two-fold.  First, those clusters are defined by comparitively 
simple star formation and chemical enrichment histories, making them the 
closest tangible approximation to the most basic stellar system treated by SPS 
models.  In other words, they best embody the concept of the so-called simple 
stellar population: a collection of stars born from an instantaneous burst of 
star formation and having a uniform chemical composition.  Second, the relative 
proximity of these clusters allows their stellar content to be resolved well 
below their respective main sequence turn-offs.  This makes it possible to 
accurately constrain their ages via isochrone fitting to colour-magnitude 
diagrams (CMDs) and abundance patterns from the synthesis of high-resolution 
spectra of individual members.

Of all the star cluster systems found in the nearby Universe, the Galactic 
globular clusters (GGCs) are most valuable for SPS model evaluations, albeit in 
the regime of old ages and low to solar metallicities, akin to quiescent 
galaxies, spiral bulges, and extragalactic globular cluster systems.  The value 
of GGCs in this respect is tied to the fact that their stellar contents have 
been the most extensively studied to date through CMDs and spectral syntheses.  
Somewhat ironically, it is because of the special attention paid to GGCs that 
we now know of many instances where they systematically deviate from the 
textbook definition of a simple stellar population, largely through 
inhomogeneities in the abundances of several light elements (see \citealt{Gr12a} for a recent and detailed review).  Given the necessity of model evaluations 
though, the emergent complexity of GGCs is insufficient grounds to void their 
status as the premier sample for such purposes.  Instead, modellers should 
adjust the aim of their evaluations to reproducing the {\it mean} abundance 
patterns of those GGCs known to harbour multiple populations, but progress 
along these lines is only in its infancy \citep{Co11}.

Amongst the many public SPS models, that of \citet[hereafter S07]{Sc07} is one 
of three designed to recover the abundances of light elements for an observed 
stellar population, in addition to the usual diagnostics of age and 
metallicity; the other two models are from \cite{Th11a} and \cite{CvD12}.  Of 
these three models, that of S07 stands out as the one whose abundance 
predictions have been the most rigorously tested thus far over a considerable 
metallicity range\footnote{The \cite{CvD12} models, however, are only intended 
for use on stellar populations of approximately solar metallicity.} 
($\Delta$[Fe/H] $\sim$ 1.2 dex).  \cite{GS08} found that the S07 model 
reproduces, to within $\pm$0.1 dex, the known ages, metallicities, and 
abundance ratios of the GGCs 47 Tuc, NGC 6441 and NGC 6528, as well as the 
Galactic open cluster M67.  Despite their success, the sample analysed was 
rather small, totalling five clusters, implying that a more extensive 
evaluation covering a wider range in GGC properties, \eg horizontal branch 
morphology, is still needed to establish the ultimate robustness of this 
model.  For instance, these authors noted that the calcium abundance recovered 
for the metal-poor GGC NGC 6121 was $\sim$0.3 dex lower than that measured by 
\cite{Iv99}.  This discrepancy led them to caution about the use of the S07 
model in the regime [Fe/H] $\lesssim$ -1.0, precisely where the reliability of 
this model has been poorly validated to this point.

In their analysis, \cite{GS08} used the library of high-quality integrated blue 
spectra measured by \citet[hereafter S05]{Sc05} to recover the ages, 
metallicities and abundance patterns for the four GGCs in their sample.  
Indeed, the work of S05 was inspired by the need for in-depth evaluations of 
the accuracies of spectroscopic-based SPS model predictions.  As such, the 
authors selected their targets (41 in all) to be representative of the entire 
GGC system, spanning a wide range of metallicities, horizontal branch 
morphologies, concentrations, Galactocentric coordinates, and Galactocentric 
distances.  To this day, the S05 library remains the only one of its kind.  In 
addition to it, SPS model evaluations require a matching database of 
independent age, metallicity, and abundance pattern estimates for the 41 S05 
GGCs against which the model predictions may be compared.  Existing 
compilations of GGC parameters do not satisfy these needs, however, because 
they either provide metallicity information alone (\citealt{Ha96}, 2010 
edition\footnote{\url{http://www.physics.mcmaster.ca/$\sim$Harris/mwgc.dat}}; 
hereafter Ha10) or overlook the abundances of certain light elements (\eg C, N) 
and cover only a fraction of the full S05 sample \citep[hereafter Pr05]{Pri05}.  These shortcomings have arguably been at the heart of the statistically weak 
tests of the S07 and \cite{Th11a} models to date\footnote{Recall that the \cite{CvD12} model has a limited metallicity range, meaning that it can only be 
tested on metal-rich Galactic star clusters at this time \citep[\eg M67, NGC 6528;][]{Co13b}.}.  Stringent validations of such SPS models are thus stymied 
until a more complete database of independent stellar population information 
for this sample is assembled.

In this paper, we wish to rectify this situation by presenting the most 
extensive combined compilation yet of available GGC ages, metallicities, and 
abundance patterns.  To do so, we draw on existing compilations of GGC 
parameters \citep[hereafter MF09; Ha10]{MF09}, as well as the vast literature 
on the chemical compositions of the S05 clusters.  The application of our data 
set to a statistically robust evaluation of the S07 and other SPS models will 
be presented elsewhere (Roediger et al., in prep.; hereafter Paper II).

The layout of the present paper is as follows.  In \Ses{D&R-CompMeth}{D&R-A&ZSources}, we discuss the GGC sample, methodology and some of the principal data 
sources which underlie our compilation.  Granted that our sample is 
representative of the GGC system as a whole, we use our compilation in \Se{D&R-SPGGC} to draw insights into a variety of topics related to the stellar 
populations of these objects.  Comparisons of our work with other prior 
compilations of GGC stellar population data are presented in \Se{D&R-CPW}.  
Finally, we conclude and contemplate other possible uses of this compilation in 
\Se{Concs}.


\section{Data \& results}\label{sec:D&R}

Spectroscopic-based SPS models are designed to predict the full spectra and/or 
strengths of absorption line/band indices over a wide range of ages and 
metallicities for simple stellar populations of any specified abundance 
pattern.  The ability of the S07 model to fit for chemical abundances is 
realized by inverting its functionality, that is, by perturbing the specified 
abundance pattern until the same age and metallicity are obtained amongst all 
possible index-index pairs under consideration for an observed stellar system.  
Practically speaking, the model steps through the $n$-dimensional parameter 
space spanned by the available data in a hierarchical fashion, beginning with 
indices most sensitive to age and metallicity effects (\eg H$\beta$, Fe5250, 
Fe5335) and ending with those that trace chemical abundances (\eg Mg{\it b}, 
Ca4227).  In this way, the model simultaneously predicts the best-fit age, 
metallicity and light-element abundance pattern ([Mg/Fe], [C/Fe], [N/Fe], 
[Ca/Fe]) for a given system.  While the S07 model can, in principle, fit for 
the ratios [O/Fe], [Na/Fe], [Si/Fe], [Cr/Fe], and [Ti/Fe] as well, their values 
are fixed at this time\footnote{[O/Fe] is fixed at 0.0 or +0.5, depending on 
the adopted isochrone (solar-scaled versus $\alpha$-enhanced), and [Cr/Fe] at 
0.0, while [Na/Fe], [Si/Fe], and [Ti/Fe] track [Mg/Fe].} since they are not 
reliably traced by existing Lick indices.  \cite{GS08} presented an efficient 
algorithm, ``EZ\_Ages'', to carry out the required inversion of the S07 model 
so that it can be applied to the measured indices of any stellar system.  
Further details on either the S07 model or EZ\_Ages are provided in those 
introductory papers, as well as in Paper II.

For a robust evaluation of the S07 and other SPS models (Paper II), we will 
draw upon the S05 library of intermediate-dispersion, high-$S/N$ integrated 
blue spectra for 41 GGCs, as do most other investigators for such purposes.  An 
important aspect of this library is the sample's wide coverage of the known GGC 
parameter space (see \Fig{GGCDistbns} and \Tbl{Sample}), which makes it fairly 
representative of the entire GGC system as well as the stellar populations of 
whole galaxies (\eg early-types) or their sub-components (\eg bulges).  Either 
a suite of absorption-line/band strengths measured from these data or the full 
spectra themselves may be fitted using one of several different SPS models to 
recover the ages, metallicities, and abundance patterns of these GGCs.  The 
performance of any given model is then judged by comparing these fitted 
parameters against the most complete compilation yet of similar but independently-derived information for the S05 GGCs, the latter of which is the primary 
focus of this work.

Although the preceding discussion has largely focussed on the S07 model, it 
must be appreciated that our compilation is perfectly general and can be 
applied to the evaluation of {\it any} spectroscopic-based SPS model \citep[e.g.][]{Th11a,CvD12}.  In fact, such an undertaking would undoubtedly help 
highlight the merits of the particular ingredients and/or fitting techniques 
adopted by different models.  Our compilation may also be useful to any other 
field concerned with GGCs, such as the formation of the Milky Way, or stellar 
evolution and nucleosynthesis.  In the following sub-section, we describe the 
methodology by which our compilation was assembled given the wealth of 
literature data pertaining to our needs.  For reasons that will be made clear, 
this discussion will largely revolve around the abundance patterns of our 
clusters.

\subsection{Compilation methodology}
\label{sec:D&R-CompMeth}
The over-arching principle for our compilation of the available literature on 
the stellar populations of the S05 GGCs is that it be as comprehensive and 
complete of a record as possible.  For each one of our clusters then, we have 
strived to obtain as many estimates as we could for its age, metallicity, and 
abundances of light elements, with the extent of the latter group being 
dictated by those elements currently treated within SPS models (\ie Mg, C, N, 
Ca, O, Na, Si, Cr, Ti)\footnote{We explicitly include the latter five elements 
to enable the most complete evaluations possible of {\it all} SPS models that 
predict abundance patterns.  In future revisions to the present work, we 
envisage adding information on heavier elements which yield other unique 
insights from the perspective of galaxian stellar population analyses \citep[\eg Sr, Ba;][]{Co13a}.}.  While extensive and homogeneous compilations already 
exist with respect to GGC ages and metallicities, and cover large fractions of 
the S05 sample (\eg MF09, Ha10; see \Se{D&R-A&ZSources}), resources of similar 
quality on the individual chemistries of these clusters is more limited and 
heterogeneous.

The premium compilation of GGC chemical compositions until now was presented by 
Pr05, who gathered $\alpha$-element abundances (Mg, Ca, Si, Ti) from high-resolution spectroscopic analyses in the literature for a sample of 45 GGCs.  Their 
results prove less than ideal for the purposes of SPS model evaluations since 
the elements C, N, O, Na, and Cr, which SPS models now cover, were omitted and 
their sample has a rather small overlap with that of S05 (15 clusters only).  
To improve upon the shortcomings of Pr05, we extracted from the literature 
measurements of the relevant chemical abundance ratios for each S05 cluster 
from as many references as possible.  In so doing, we have found that 
specifying a complete abundance pattern for any one S05 GGC often required data 
from at least two references; for example, the abundances of carbon and 
nitrogen are usually studied together but separately from those of the other 
elements listed earlier.  Our desire for completeness therefore naturally 
encouraged us to draw on multiple sources when assembling the abundance 
patterns we are advocating for use here.  In doing so, we have combined the 
results from {\it all} chemical composition studies on each cluster\footnote{We 
have created an electronic data table which lists these results for each 
cluster in our sample.  These data tables may be retrieved online at \url{www.astro.queensu.ca/people/Stephane\_Courteau/roediger2013/index.html}}.

The abundance pattern we adopt as the benchmark for each cluster was built by 
calculating the mean value and root-mean-square (rms) dispersion of the 
available independent measurements for each of the elements listed above.  This 
aspect of our compilation embodies some noteworthy advantages.  First, merging 
results as we do should reduce the {\it statistical} error in the value that we 
recommend for any given abundance ratio, albeit at the price of increasing its 
{\it systematic} error.  We do not consider this a drawback but rather another 
advantage of our approach since systematic errors (\eg sample selection, solar 
abundances, atomic parameters) might be a significant source of discrepancy 
between abundance patterns predicted by SPS models and star-by-star spectral 
syntheses.  Having a metric for the degree of systematic error involved in the 
latter, via the dispersions, will undoubtedly be very helpful for judging the 
reality of model predictions.  The last advantage of our approach is that it 
should also naturally reflect the existence of putative multiple stellar 
populations when present within a given cluster (again, via the dispersions).  
A hallmark of the multiple population phenomenon is that, amongst the members 
of an affected cluster, the abundances of several elements either correlate 
(Al-Si) or anti-correlate (C-N, Na-O, Mg-Al) with one another \citep{Gr12a}.  
Modulo the particulars on sample selection, we then expect to find large 
spreads in the abundance ratios of these elements between the stars from either 
a single study or multiple ones\footnote{As discussed below and in \Se{D&R-SPGGC}, the C, N, and O abundances of individual GGC stars also depend on their 
evolutionary status, which results in an additional source of dispersion 
amongst these parameters}.  Moreover, by combining such scattered measurements 
into a single estimate for a cluster's abundance pattern, we can be assured 
that the corresponding dispersions reflect the presence of multiple populations 
by being comparitively large to those of species which are excluded from the 
above trends (\eg Ca).  Note that any and all claims we make herein as to the 
causes of inflated rms dispersions (re: systematics versus multiple 
populations) are ultimately suggestive and not based on thorough quantitative 
analyses.

Another major principle for our compilation involves concentrating, where 
possible, on studies whose results pertain solely to evolutionary stages from 
the main sequence (MS) through the asymptotic giant branch (AGB).  We impose 
such a restriction because the onset of thermal flashes, third dredge-ups, dust-gas separation (winnowing) and mass loss during the final (post-AGB) stage in 
the evolution of low-mass stars can give rise to surface abundances which 
poorly reflect the chemical composition of the gas from which they formed \citep[e.g.][]{SL09}.  Third dredge-up episodes, in particular, would pollute the 
surfaces of such highly-evolved stars with CNO-processed material from the hydrogen-burning shell (if present), effectively lowering the abundance of carbon 
and raising that of nitrogen there, relative to those of MS stars \citep[e.g.][]{Mo04}.  Stars ascending the sub-giant and red giant branches (SGB and RGB, 
respectively) can also have their surface chemistries of C, N, and O affected 
by mixing episodes (\eg \citealt{Ib64}, \citealt{SM79}; \Se{D&R-SPGGC}), but 
given that most spectroscopic analyses of individual GGC members do not 
penetrate to fainter magnitudes than this phase precludes our rejection of such 
data.  In other words, limiting our mean carbon and nitrogen abundances only to 
measurements obtained from MS stars would significantly thin out our 
compilation.  Instead, we embrace such data and simply caution SPS modellers 
to consider the evolutionary stage down to which their predicted [C/Fe] and 
[N/Fe] values correspond.  In \Se{D&R-SPGGC} we highlight the possibility that 
carbon depletion and nitrogen enhancement as a function of position along the 
SGB/RGB may be crudely quantified.

In addition to highly-evolved stars, we also exclude from our compilation 
(again, where possible) data corresponding to ``exotic'' stages of stellar 
evolution, such as very hot ($T_{eff} >$ 11 500 K) horizontal branch (HB) 
stars.  In this case, the surface abundances of elements are often perturbed by 
effects like radiative levitation and gravitational sedimentation \citep[e.g.][]{Pa06}.  Unlike the case of mixing along the SGB/RGB though, it is unclear that 
empirical corrections for these processes are forthcoming simply because it is 
rare to find individual GGCs with mixtures of exotic plus ``normal'' ($T_{eff} 
<$ 11 500 K) HB stars, let alone homogeneous abundance analyses thereof.  Thus, 
unless data from the MS, SGB, RGB, (cooler) HB, and AGB for a cluster are all 
not available, we deem abundance ratios based on the most advanced and exotic 
stages of stellar evolution unsuitable for our purposes and omit them from our 
compilation.

In light of the above caveats, we wish to provide the exact rationale behind 
our compiled abundance pattern for each S05 cluster.  We do just this, in brief 
and on a per cluster basis, in the Appendix, with attention being paid to the 
following related themes: adopted references, omitted data, systematic errors, 
and evidence of multiple populations from both our data and other methods 
(where applicable).  \Tbls{Refs-1}{Refs-2} also provide the relevant references 
from which our recommended ages, metallicities, and abundance patterns for the 
S05 GGCs were drawn.  In the following section, we specifically address our 
sources and methodology used to arrive at the ages and metallicities of the S05 
clusters.  Much of that information will therefore not be repeated in the 
Appendix.

\subsection{Age and metallicity sources}
\label{sec:D&R-A&ZSources}
Our selection of sources for age and metallicity information on the S05 GGCs 
embraces similar principles as described above regarding their abundance 
patterns.  In terms of their ages, a cursory review of the relevant literature 
reminds us of genuine discrepancies on a per cluster basis.  While isochrone 
fitting of one form or another to CMDs has long been the standard by which 
GGC ages are obtained, the results therefrom appear to be plagued by rather 
large uncertainties.  The origins of these discrepancies are most likely tied 
to the values of various parameters assumed by the scientist (\ie distance, 
reddening, metallicity, etc.) and/or each isochrone set (\ie mixing length, 
helium abundance, etc.).  Since little is to be practically learned by merging 
together the available absolute age determinations for any given cluster 
(unlike the case with their chemical abundances), we prefer our compiled values 
of this parameter to come from a single source.

The majority of the ages adopted in our compilation come from MF09.  These 
authors have performed the most extensive and homogeneous age analysis of GGCs 
to date, totalling 64 clusters in all and using HST/ACS photometry plus many 
flavours of isochrones \citep{Be94,Gi00,Pi04,Do07}.  The ages in this work 
from the Dotter \etal isochrones were cast in terms of both the \cite{ZW84} and 
\cite{CG97} metallicity scales.  By normalizing their results from each 
isochrone set to the corresponding mean absolute age of their 13 lowest-metallicity GGCs, MF09 found that the {\it relative} ages were robust to the particular 
choice of isochrone (see their Table 4) and carried a formal precision between 
2-7\%.  For the 25 S05 GGCs which overlap with the MF09 sample, we adopt their 
relative ages based on the Dotter \etal isochrones and Carretta \& Gratton 
scale.  From their \S 6.1, we assume a normalization factor of 12.80 $\pm$ 0.17 
Gyr to put these ages on an absolute footing.  Note that the uncertainty in the 
normalizing factor does not account for systematics, \eg bolometric 
corrections, but we anticipate this issue will be thoroughly addressed in 
forthcoming work on absolute GGC ages by the same group, as alluded to in MF09.

To bolster the reliability of their {\it relative} ages, MF09 also compared 
them against those of \cite{DAng05}, the formerly largest homogeneous GGC age 
compilation, totalling 55 clusters in all.  In doing so, they found mutual 
consistency between the two datasets to within their own published error bars, 
where De Angeli et al.'s HST and ground-based sub-samples yielded mean 
residuals of -0.04 $\pm$ 0.07 and -0.02 $\pm$ 0.08, respectively.  Furthermore, 
MF09 failed to detect any trends in the residuals as a function of 
metallicity.  We conduct a similar comparison in \Fig{CompAge}, but in terms of 
{\it absolute} ages and with respect to multiple prior age compilations \citep{Ro99,SW02,Do10}, where the relative ages from Rosenberg \etal were transformed 
assuming a zeropoint of 13.2 Gyr (see their \S 4).  To ensure the comparison is 
fair, we limited it to the twelve S05 GGCs common to all four compilations 
examined therein.  The mutual overlap between the samples of MF09 and other age 
compilations in the literature \citep{CK95,Ch96,Ri96,Bu98,SW98,Va00,MW06} is 
actually poorer than this and so we have omitted those results from \Fig{CompAge}.  Instead, we compare in the Appendix (wherever possible) our adopted ages 
against {\it all} other estimates in the literature known to us, on a per 
cluster basis.  Such an exercise provides us with a first-order impression of 
the systematic discrepancies involved between any two individual age 
determinations.  There, we also provide the normalization factors or constants 
we have used to transform those ages originally expressed on relative scales 
into absolute terms.

The most striking feature from \Fig{CompAge} is the presence of several 
clusters in each panel whose ages from prior compilations disagree egregiously 
with those of MF09.  These disagreements are found in different age and 
metallicity regimes\footnote{The age regimes quoted here refer, in a 
qualitative manner, to the values from MF09.}, depending on the source under 
consideration: young ages and low metallicities for \cite{Ro99}, old ages and 
high metallicities for \cite{SW02}, and at both age and metallicity extrema for 
\cite{Do10}.  Of course, the identification of these inconsistencies hinges on 
how representative the published uncertainties are of the total error budget.  
The error bars shown in \Fig{CompAge} are largely statistical in nature and do 
not consider the systematic effects of uncertainties in, amongst many others, 
distance modulus, reddening and stellar evolution model.  With exception to the 
results of Dotter et al., this criticism may be unwarranted or overstated since 
the investigators employed distance- and reddening-independent methods 
(Rosenberg et al.; Salaris \& Weiss) and/or provided {\it relative} ages only 
(Rosenberg et al.; MF09).  Overall then, \Fig{CompAge} leaves us with the 
impression that systematic uncertainties in absolute age dating of GGCs resides 
(at worst) at the 2-3 Gyr level.  Neglecting ages which exceed that of the 
Universe \citep[13.76 $\pm$ 0.11 Gyr;][]{Ko11}, this estimate is corroborated 
by our cluster-by-cluster comparisons of individual age determinations in the 
Appendix.  A more detailed examination than that of this issue lies beyond the 
scope of this paper.

While the existence of significant systematics in absolute age determinations 
might suggest that, for now, it be best to evaluate age predictions from SPS 
models in a relative sense, it is not clear that this can be done in practice.  
One complication is that the S05 sample does not contain the same clusters upon 
which MF09 base their normalization.  Another is that, for the sake of 
completeness of our compilation, we appeal to five other sources of age 
information for 13 of the 16 S05 GGCs which do not overlap with the MF09 
sample.  Based on \Fig{CompAge}, it is clear that transformations of these 
results onto the MF09 scale would at best be crudely defined.  Therefore, until 
a more complete source of relative ages for the S05 GGCs becomes available, SPS 
modellers will have to bear in mind the systematics underpinning the absolute 
ages against which they test their predictions.  To assist in this awareness, 
we explicitly caution the reader in the Appendix about those clusters in our 
sample for which their adopted age {\it does not} come from MF09.  Our 
extensive list of references which provide independent age measurements for the 
S05 GGCs is summarized globally in \Tbl{Refs-1} and for individual clusters in 
\Tbl{Refs-2}.  Note that the specific reference of our adopted age for each 
cluster is shown in boldface in the latter.

Our knowledge of the metallicities of the S05 GGCs has greatly improved with 
the work of \cite{Ca09c}.  Based on high-resolution optical spectra for about 
2000 RGB stars belonging to 19 GGCs (13 of which are in the S05 sample), these 
authors have created the premier database of homogeneous and spectroscopic 
metallicities for GGCs.  Through it, they have defined a new GGC metallicity 
scale spanning almost the full range of values exhibited by these systems in 
this parameter space, from [Fe/H] $\sim$ -2.4 to -0.3 dex.  Since Carretta et 
al.'s sample overlaps with those from previous metallicity scales \citep{ZW84,CG97,Ru97,KI03} by ten clusters or more, they were also able to derive 
transformations between those and their own.  This enabled them to express the 
metallicities of all 133 GGCs from the 2003 version of the \cite{Ha96} 
catalogue, which encompasses our whole sample, in terms of their own scale.

\begin{figure*}
 \begin{center}
  \includegraphics[width=0.9\textwidth]{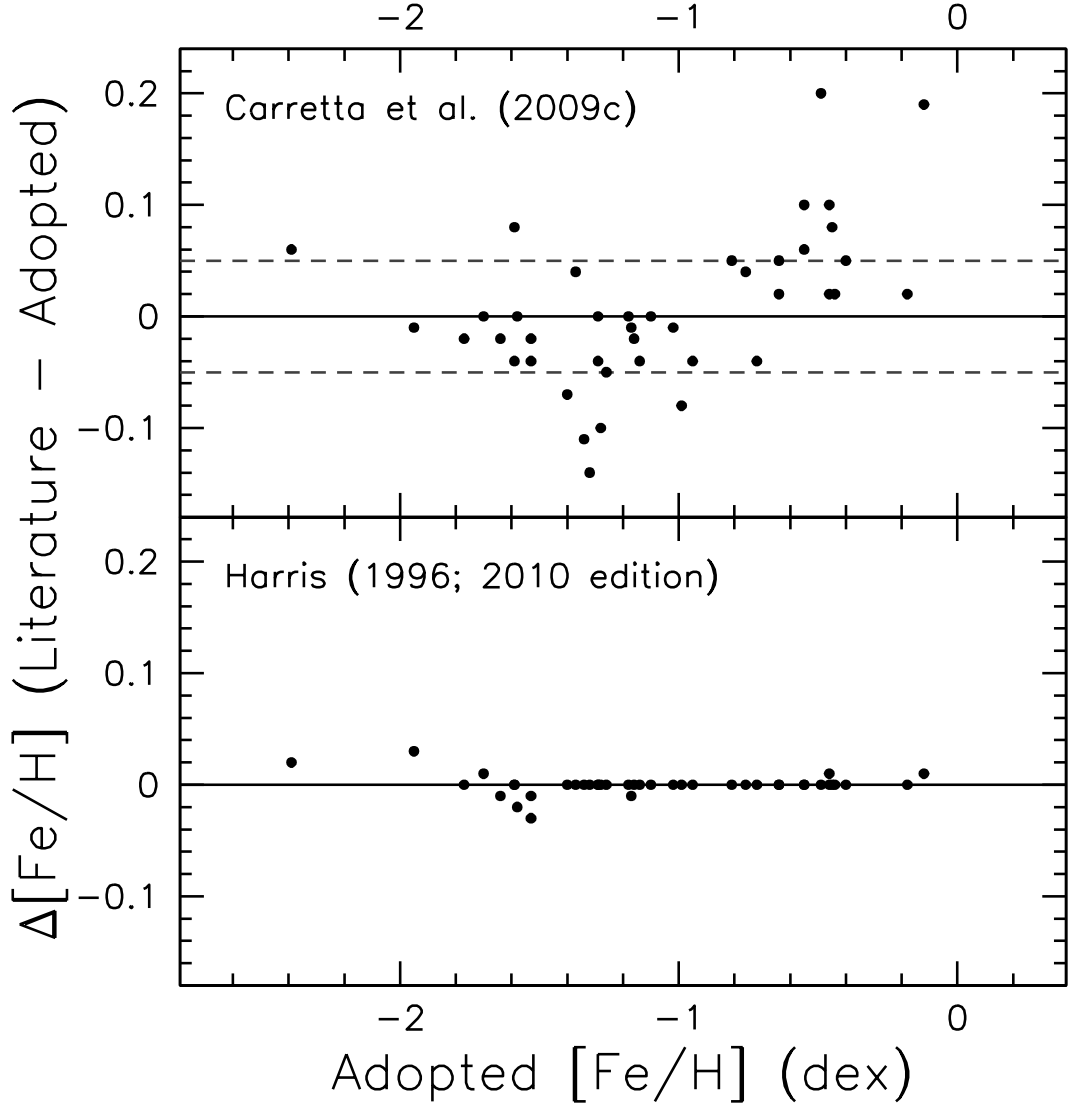}
  \caption{Differences between our adopted metallicities for the S05 GGCs 
against other independent estimates thereof in the literature, either from the 
Ha10 catalogue ({\it lower panel}) or \cite{Ca09c} ({\it upper panel}).  The 
differences are plotted against our adopted values in both panels, while 
the dashed lines in the upper panel outline the 68$^{\text{th}}$-percentile of 
the distribution about the locus of equality (solid line).  The same lines 
would essentially overlap with equality in the lower panel and are thus omitted 
therefrom.  The median error per point is 0.12 and 0.14 dex in the lower and 
upper panels, respectively.}
  \label{fig:CompFeH}
 \end{center}
\end{figure*}

Ha10 improves upon the work of \cite{Ca09c} by merging the latter's results 
with those of \cite{AZ88}, after first transforming them to the Carretta \etal 
scale, as well as metallicities for individual clusters from other studies.  
Given its complete coverage of the S05 sample and the sheer popularity of this 
database, we would ideally adopt the metallicities from Ha10 for our 
compilation.  However, since Ha10 do not provide uncertainties on their values, 
we instead calculate the mean metallicities of our clusters using the same 
references and weighting scheme as Ha10.  The bottom panel of \Fig{CompFeH} 
shows the differences between the Ha10 metallicities and our replicas thereof.  
Considering the complete S05 sample, the agreement between the two datasets is 
superb, with 68\% of the data points exhibiting differences of $\pm$0.01 dex or 
less.  We also find that our adopted metallicities compare favorably with those 
from Carretta \etal (\Fig{CompFeH}, {\it top}), albeit with 68\% of the data 
points showing differences up to $\pm$0.05 dex.  This agreement is not all that 
surprising as the Ha10 metallicities, and thus our replicas as well, are 
weighted to the data of Carretta \etal by a factor of three more than those 
from any other source.

\subsection{The stellar populations of Galactic globular clusters}
\label{sec:D&R-SPGGC}
The results of our literature compilation on the stellar population properties 
of the S05 GGCs are presented in \Tbls{GGCSPs-1}{GGCSPs-2} and \Figs{AgeDistbn}{XFeDistbn}.  In \Tbl{GGCSPs-1}, we list the recommended age (column 2), mean 
metallicity (column 3), and mean Mg, C, N, and Ca abundances (columns 4-7) for 
each cluster, while \Tbl{GGCSPs-2} gives their mean O, Na, Si, Cr, and Ti 
abundances (columns 2-6).  Entries therein which we consider suspect with 
respect to our compilation principles (elucidated above) appear in boldface; 
the reader is referred to the Appendix for the rationale behind each of these 
flags.  Recall that, because of practical limitations which bar the computation 
of relative ages from both literature data and SPS models for {\it all} of the 
S05 GGCs, we have cast all of our compiled ages into absolute terms.  In light 
of the apparent systematics which afflict absolute age estimates, we look 
forward to future work from expert groups which properly address this issue.
Until then, SPS modellers will simply have to be mindful of these uncertainties 
in our compiled data, which we optimistically gauge to be $\sim$2-3 Gyr, when 
evaluating age predictions.

\begin{figure*}
 \begin{center}
  \includegraphics[width=0.9\textwidth]{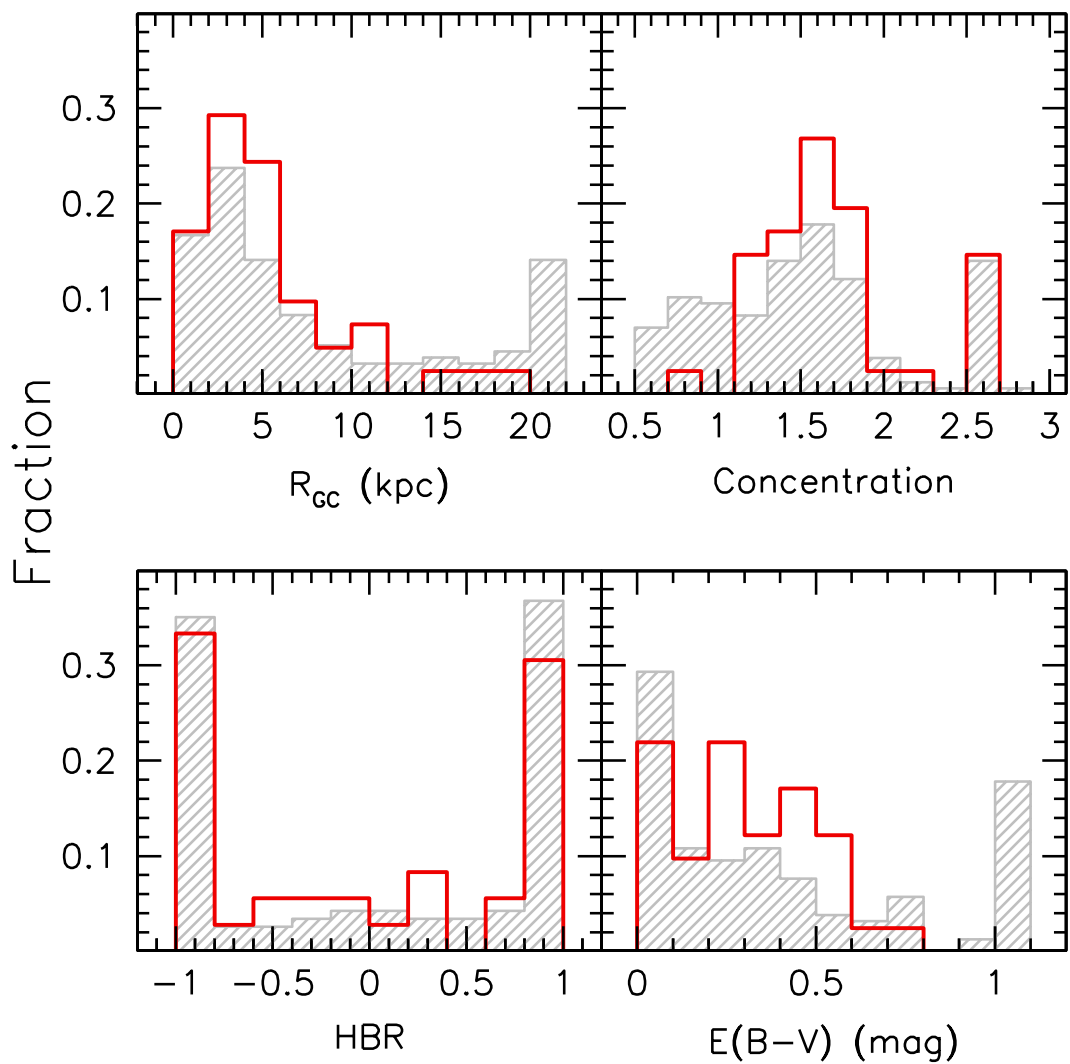}
  \caption{Percent distributions of Galactocentric radii [R$_\text{{GC}}$; 
{\it upper-left}], concentrations [{\it upper-right}], horizontal branch ratios 
[HBR; {\it lower-left}] and reddenings [$E$($B-V$); {\it lower-right}] for the 
Galactic globular cluster (GGC) samples from the full 2010 edition of the 
\citet[hereafter Ha10]{Ha96} catalogue (grey) and the \citet[S05]{Sc05} 
spectral library (red).  Note that the rightmost bins in the upper-left and 
lower-right panels actually span the (unbounded) ranges of $\ge$22 kpc and 
$\ge$0.5 mag, respectively.  All distributions were created using data from the 
Ha10 catalogue.}
  \label{fig:GGCDistbns}
 \end{center}
\end{figure*}

While the original intent of the present compilation was for evaluations of SPS 
models, we foresee its broad applicability to a variety of astrophysical 
endeavours since the S05 sample is representative of the whole GGC system (\Fig{GGCDistbns}; \Tbl{Sample}).  To demonstrate this point, we use our compilation 
in the following sub-sections to garner some brief insight into the early 
formation and chemical evolution of the Milky Way, atmospheric mixing during 
stellar evolution and the sites of explosive stellar nucleosynthesis.

\subsubsection{Ages \& metallicities}
\label{sec:D&R-A&M}
\begin{figure*}
 \begin{center}
  \includegraphics[width=0.9\textwidth]{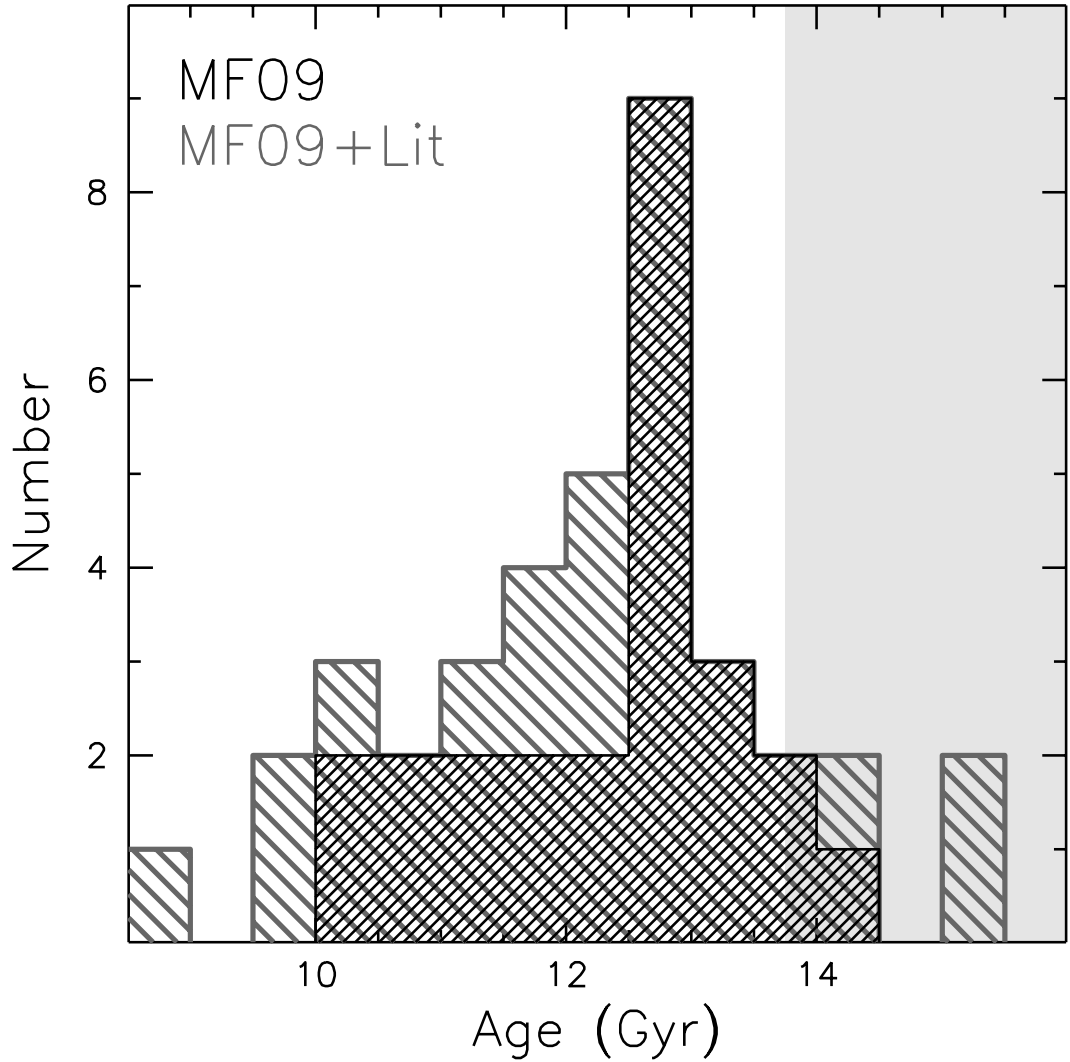}
  \caption{Adopted age distribution for the S05 GGC sample based on either the 
MF09 compilation alone (black), or MF09 plus additional sources of ages from 
the literature (for those clusters not studied by MF09; dark gray).  The shaded 
region corresponds to ages in excess of that of the Universe \citep{Ko11}.}
  \label{fig:AgeDistbn}
 \end{center}
\end{figure*}

In \Figs{AgeDistbn}{XFeDistbn} we show the distributions of all the stellar 
population diagnostics presented in \Tbls{GGCSPs-1}{GGCSPs-2} for the whole S05 
sample.  Referring to \Fig{AgeDistbn} and the MF09 results shown therewithin 
(black histogram), it is seen that the S05 sample has an age distribution which 
is both strongly-peaked (between 12.5 and 13 Gyr) and skewed to very old ages.  
Those clusters whose ages come from other sources in the literature are 
represented by the gray histogram.  These additional estimates tend to broaden 
the overall distribution to both younger and older ages, as well as the strong 
peak described by the MF09 results.  The fact that over half of these other 
age determinations are found to have extreme values in relation to those from 
MF09 bolsters our position on treating them with caution.  Note that the shaded 
region in \Fig{AgeDistbn} demarcates ages which exceed that of the Universe 
\citep{Ko11} and while some of our clusters are found there, the statistical 
errors on their values alone overlap with the allowed (unshaded) region.

From the MF09 ages in \Fig{AgeDistbn}, it is tempting to conclude that the S05 
GGCs originated from a two-component star formation history.  This history 
could be described as consisting of either a sharp burst superimposed upon a 
comparitively steady background (lasting $\gtrsim$4 Gyr) or a vigourous early 
episode which quickly peaked and was then regulated down to a more sustainable 
level.  By broadening the overall distribution to more extreme ages, the 
additional literature data shown in \Fig{AgeDistbn} seems to agree better with 
the first of these two scenarios.  Modulo systematics, these data would temper 
this scenario though by spreading the burst component over a longer timescale 
($\sim$1.0-1.5 Gyr).

\begin{figure*}
 \begin{center}
  \includegraphics[width=0.9\textwidth]{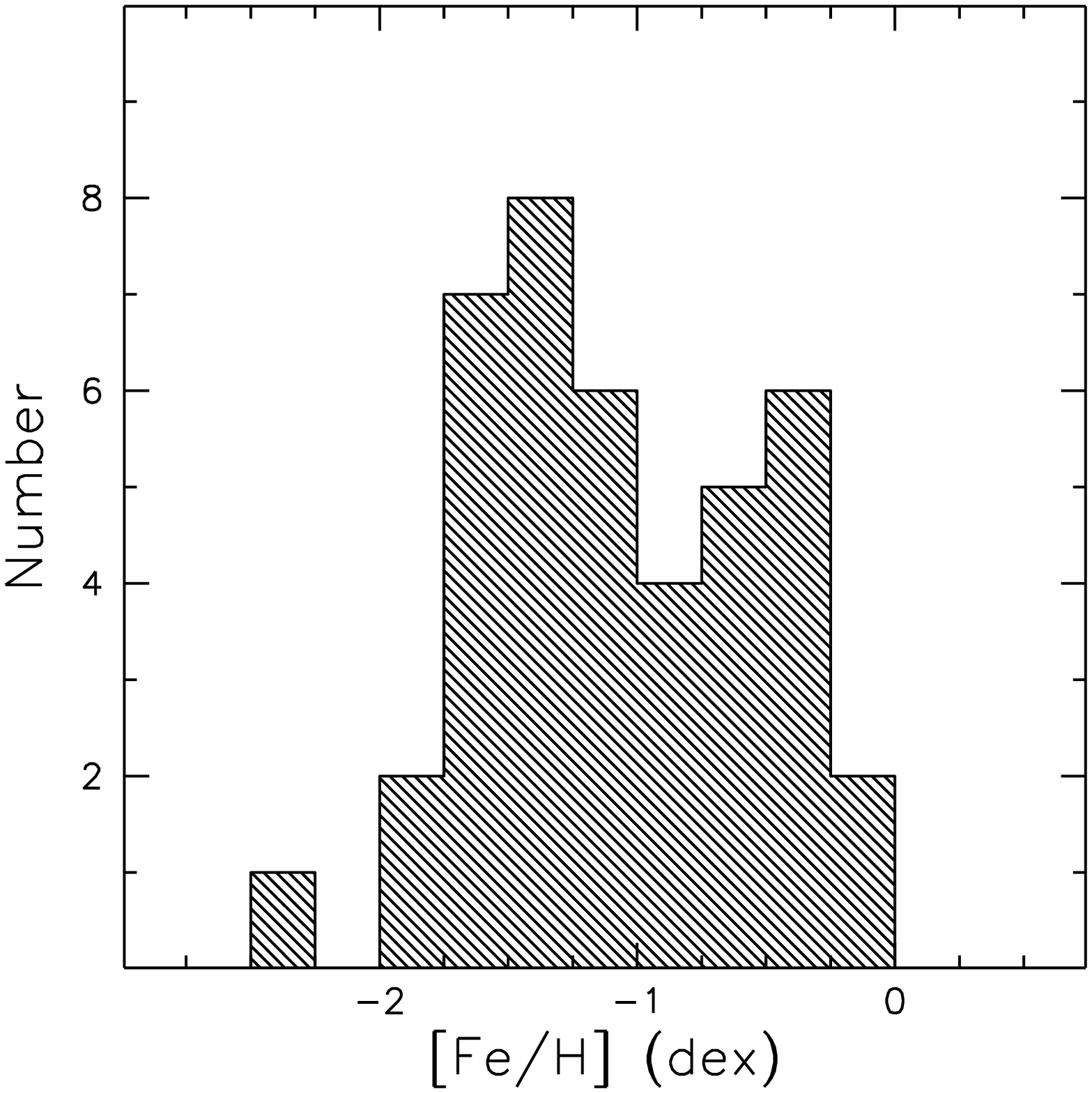}
  \caption{As in \Fig{AgeDistbn}, but for our adopted [Fe/H] values.}
  \label{fig:FeHDistbn}
 \end{center}
\end{figure*}

\begin{figure*}
 \begin{center}
  \includegraphics[width=0.9\textwidth]{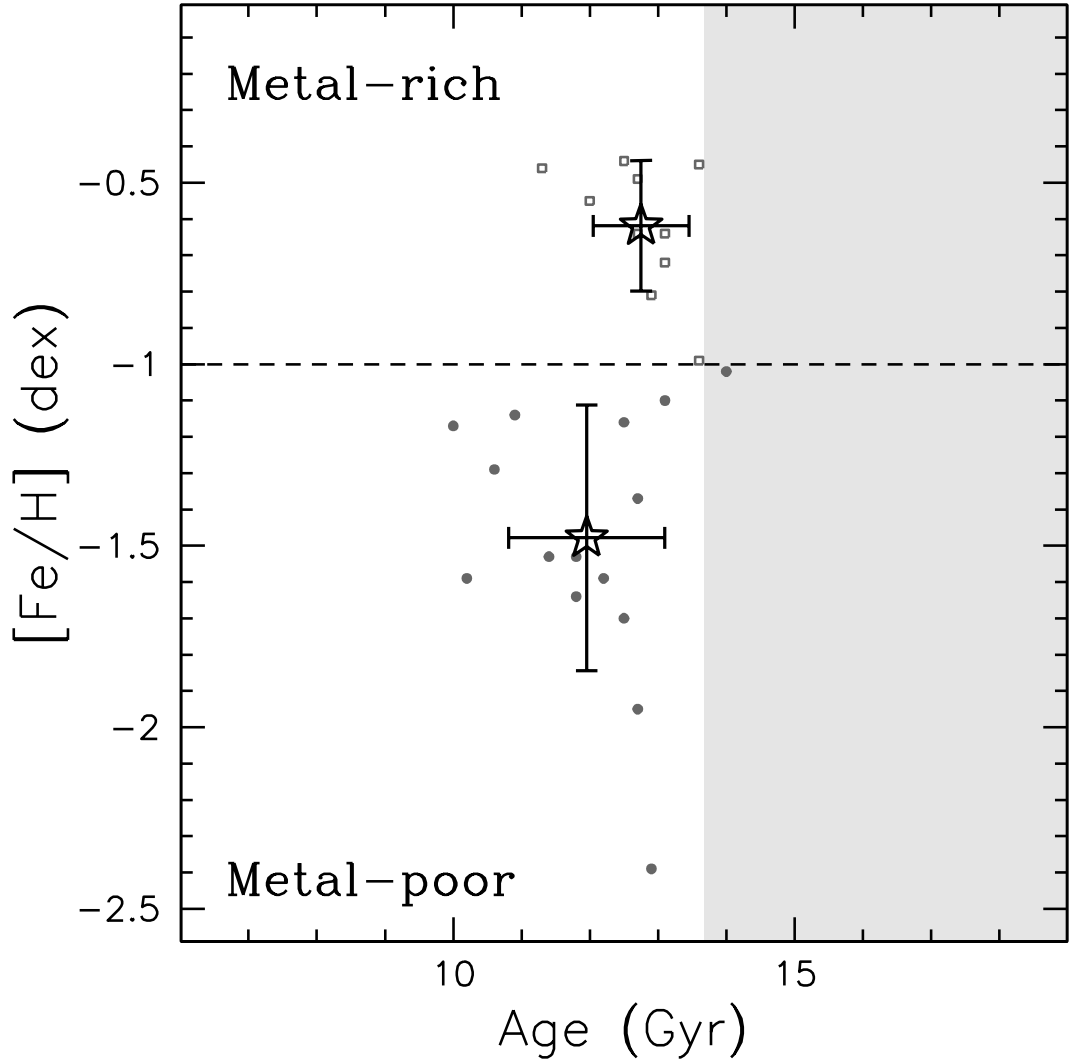}
  \caption{Metallicity versus age for S05 GGCs included in the MF09 sample.  
The mean ages and metallicities (and corresponding rms uncertainties) of the 
metal-poor (circles) and metal-rich (squares) GGCs in our sample are denoted by 
the stars and error bars, while the dashed line at [Fe/H] = -1.0 dex roughly 
marks the transition between these two groups, as seen in \Fig{FeHDistbn}.  The 
shaded region corresponds to ages in excess of that of the Universe 
\citep{Ko11}.}
  \label{fig:FeHvsAge}
 \end{center}
\end{figure*}

The metallicity distribution for the S05 GGCs (\Fig{FeHDistbn}) also appears to 
support the idea that these clusters arose from at least two distinct channels 
given its clear bimodal shape (in agreement with \citealt{Zi85}), with a 
peak-to-peak separation of about $\sim$1.0 dex.  When we examine the available 
MF09 ages of the clusters comprising each metallicity sub-group however (\Fig{FeHvsAge}), there does not appear to be a strong correlation between the two 
parameters.  The metal-poor and metal-rich GGCs of our sample, separated at 
[Fe/H] = -1.0 dex, which overlap with MF09 have a mean age and rms dispersion 
of 12.0 $\pm$ 1.1 Gyr and 12.8 $\pm$ 0.7 Gyr, respectively; the distinction of 
these two groups by age only worsens when we consider all of our adopted 
values.

The situation seen in \Fig{FeHvsAge} can be attributed to the presence of 
several old GGCs ($>$12 Gyr) in our metal-poor sub-sample, whereas only one of 
our metal-rich GGCs has an age of $<$12 Gyr.  Moreover, our metal-rich sub-sample harbours a high incidence of very old clusters in that three (four, if the 
metal-rich/-poor boundary lay slightly lower; \eg [Fe/H] = -1.05) of our six 
oldest objects are contained therewithin.  Since our sample is representative 
of the entire GGC system, these findings bear some implications with regards to 
the formation of the Milky Way, in particular its halo.  For instance, the 
parameter spread in \Fig{FeHvsAge} would be hard to explain within a scenario 
in which {\it all} of the S05 GGCs formed in situ since one would expect the 
metal-poor GGCs to be {\it older} than the metal-rich ones, not younger.  
Instead, this spread is consistent with a picture in which the GGC system arose 
from its members either forming in situ or being accreted from satellite 
galaxies.  Although the possible correlation of these two channels with 
metallicity remains unclear, we interpret the older, metal-rich and younger, 
metal-poor clusters as the descendants of the former and latter, respectively.  
Similar conclusions have been reached by other analyses of the GGC age-metallicity relation using much larger samples \citep[\eg MF09,][]{FB10,Do11}.  Further 
insight into this topic might be achieved by searching for correlations between 
the above parameters and those from the Ha10 catalogue, particularly velocity 
information, but such a task is beyond the scope of this work.  \cite{SW02} and 
MF09 already investigated the relationship between age and galactocentric 
radius for GGCs, and found none.

\subsubsection{Chemical abundance distributions \& atmospheric mixing}
\label{sec:D&R-CAD&AM}
\begin{figure*}
 \begin{center}
  \begin{tabular}{c c}
   \includegraphics[width=0.5\textwidth]{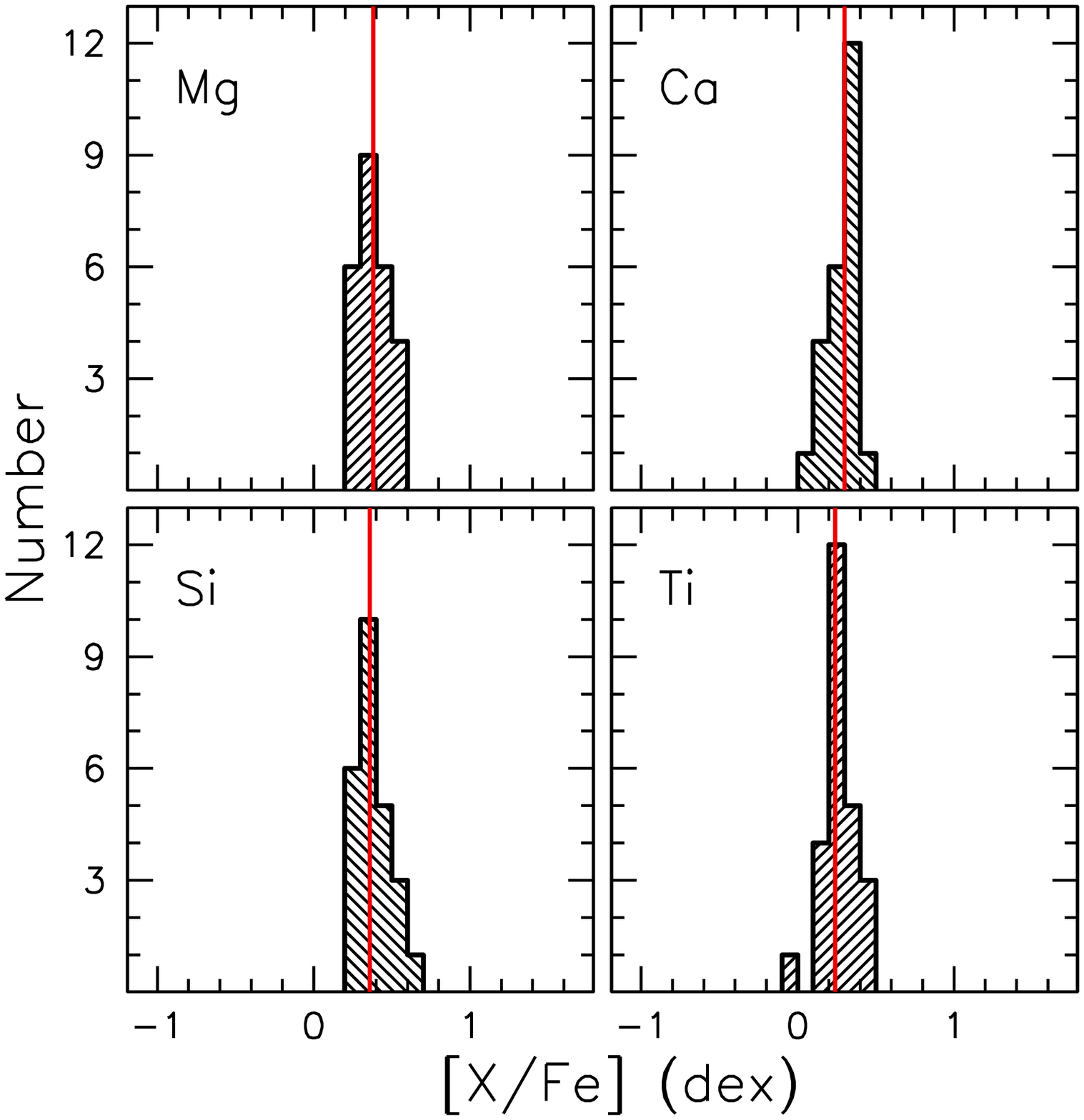} &
   \includegraphics[width=0.5\textwidth]{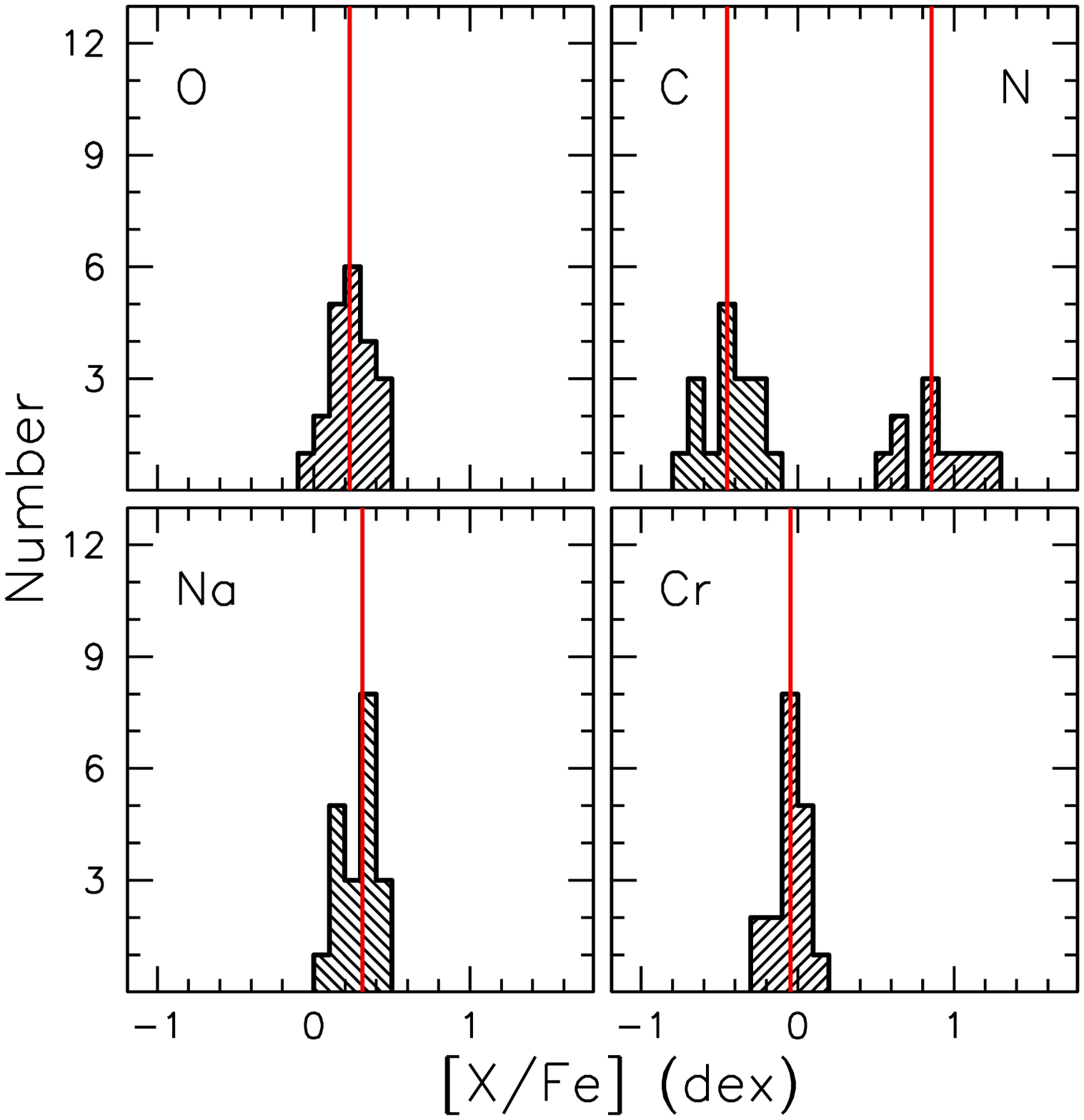}
  \end{tabular}
  \caption{As in \Fig{AgeDistbn}, but for our adopted abundance patterns.  The 
specific abundance ratio plotted in each panel is indicated in the top-left or 
top-right corner while the median value of each distribution is marked with a 
red line.  Note that less certain entries from Tables~\ref{tbl:GGCSPs-1} and 
\ref{tbl:GGCSPs-2} have been omitted from these distributions.}
  \label{fig:XFeDistbn}
 \end{center}
\end{figure*}

Further critical insight into the stellar populations that comprise GGCs can be 
gleaned from the distributions of their mean chemical abundance ratios, as 
shown in \Fig{XFeDistbn}.  These, in turn, afford us with further information 
on the formation of the Milky Way, as well as certain aspects related to 
stellar evolution.  From \Fig{XFeDistbn}, we first note that the distributions 
of the mean abundances of the $\alpha$-elements (Mg, Ca, Si, Ti) amongst these 
systems all show sharp peaks towards super-solar values.  The respective median 
values for the [Mg/Fe], [Ca/Fe], [Si/Fe] and [Ti/Fe] distributions are +0.38, 
+0.30, +0.36 and +0.24 dex, which implies that the GGC system, on the whole, 
formed over rapid timescales.  On the other hand, slightly broader 
distributions are found for mean abundances of carbon, nitrogen, and oxygen 
amongst our GGCs, in that we obtain rms dispersions of $\gtrsim$0.15 dex for 
them compared to $\lesssim$0.1 dex for the $\alpha$-elements.  Apart from 
possible undersampling effects\footnote{We possess $\alpha$-element abundances 
for 25 of our GGCs, while C, N, and O abundances are known for half of our 
sample, at best.}, we interpret the broader distributions of these elements as 
reflecting the combined and well-known phenomena of atmospheric mixing and 
multiple stellar populations within these clusters.  We concentrate on the 
former for the remainder of this sub-section and take up the latter in the next.

\begin{figure*}
 \begin{center}
  \begin{tabular}{c c}
   \includegraphics[width=0.5\textwidth]{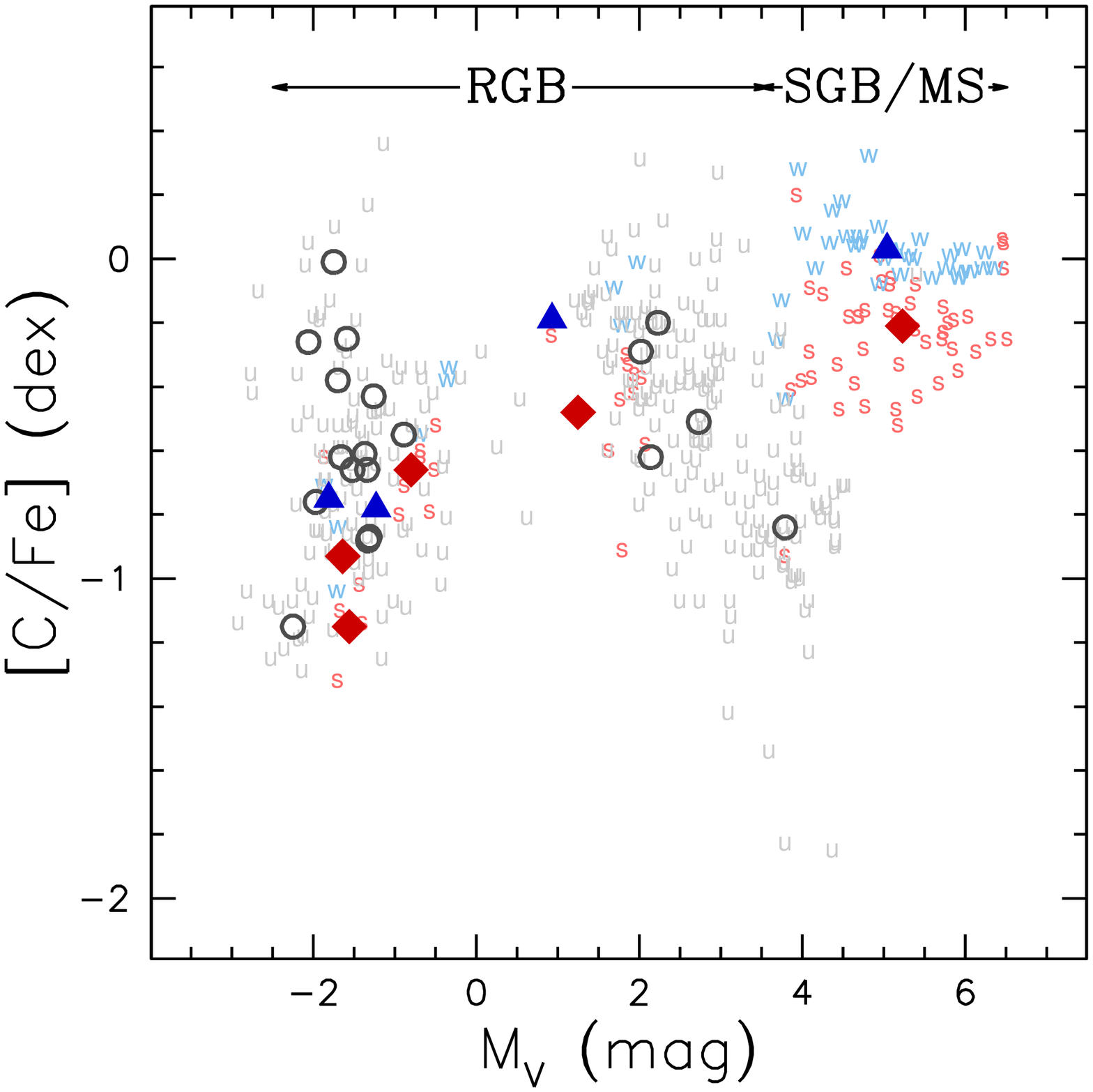} & 
   \includegraphics[width=0.5\textwidth]{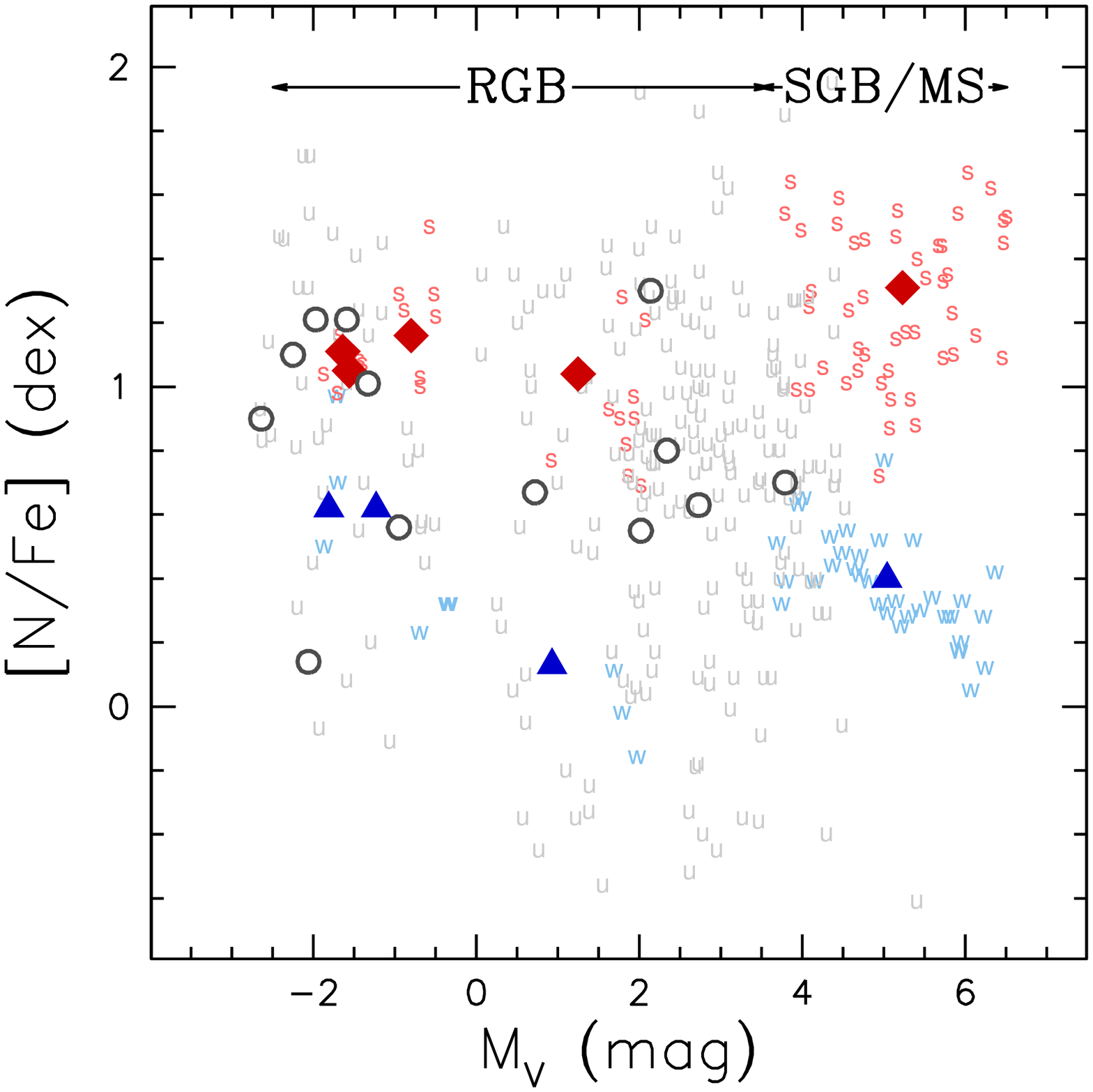}
  \end{tabular}
  \caption{Carbon ({\it left}) and nitrogen ({\it right}) abundance versus 
absolute visual magnitude for individual stars belonging to the S05 GGCs 
(coloured letters).  The mean values of the data from the individual studies 
are represented with the coloured points.  The point types (and colours) for 
both the individual stars and whole samples have been chosen to indicate the 
corresponding CN-strength: s/diamond (pink/red) if strong, w/triangle 
(light/dark blue) if weak and u/circle (light/dark grey) if unknown.  The 
horizontal axis has been delineated into zones which roughly correspond to the 
red giant branch (RGB) and sub-giant branch/main sequence (SGB/MS) phases of 
stellar evolution present within GGCs.}
  \label{fig:CNvsMV}
 \end{center}
\end{figure*}

To bolster the above argument, we show with coloured points in \Fig{CNvsMV} the 
mean abundances and luminosities of stars from each individual study of the 
carbon and nitrogen abundances in the S05 GGCs.  Unfortunately, not all studies 
we cite in this regard could be represented in these plots since luminosity 
information is not available to us in many cases.  This may be responsible for 
the apparent gap in the data in the range -0.5 $< M_V <$ +0.5 mag.  The point 
types in \Fig{CNvsMV} reflect whether the assorted samples consist of CN-weak 
(triangles) or CN-strong (diamonds) stars; circles are used when CN strengths 
are unknown to us.  Moreover, looking from bright to faint luminosities, the 
evolutionary status of the sample changes from predominantly RGB stars (leftmost-to-middle) to SGB/MS stars (rightmost).

\begin{figure*}
 \begin{center}
  \includegraphics[width=0.9\textwidth]{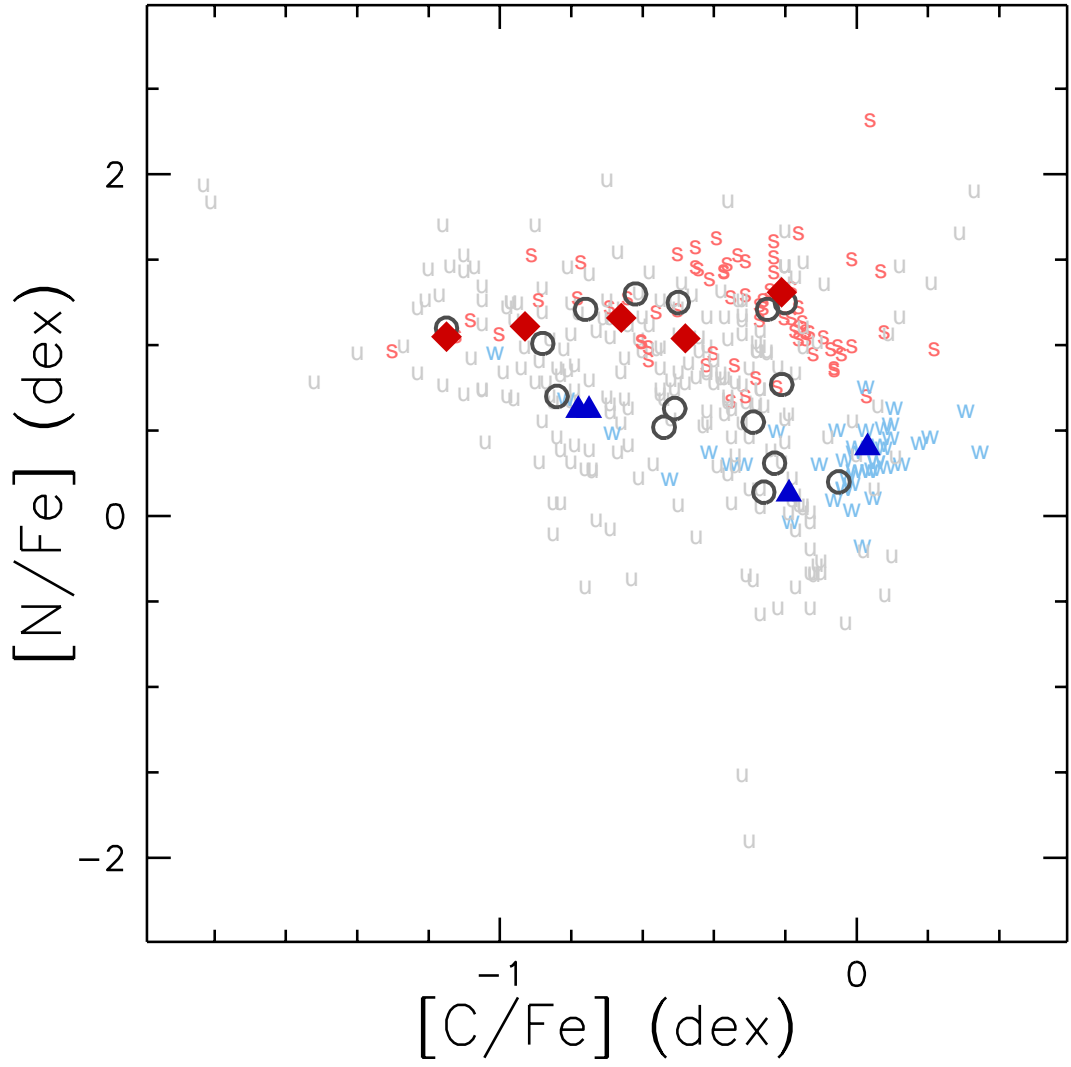}
  \caption{As in \Fig{CNvsMV} but in terms of nitrogen abundance versus carbon 
abundance.}
  \label{fig:NvsC}
 \end{center}
\end{figure*}

Concentrating only on the mean values for CN-weak and CN-strong stars, an 
evolutionary trend is apparent from \Fig{CNvsMV} whereby [C/Fe] tends to 
monotonically decrease as a given star leaves the MS and ascends the RGB.  Over 
the same evolutionary path, [N/Fe] for the star will decrease to a minimum at 
about $M_V \sim$ +1 mag and rise thereafter.  These same trends are also 
conveyed by the coloured points in \Fig{NvsC}, where we circumvent the need for 
luminosity information and thus benefit from better statistics.  The 
significant scatter amongst the data points representing both individual stars 
and mean values for unclassified samples may be due to mundane issues like 
systematics or more nuanced ones like mixtures of CN-weak and CN-strong stars 
within any one study.  An example of each case might be the solar-like [C/Fe] 
value of upper-RGB stars in NGC 6121 at $M_V \sim$ -2 mag \citep{SmV05} for the 
former and the [N/Fe] value of lower-RGB stars in NGC 7078 at $M_V \sim$ +2 mag 
\citep{Co05} for the latter.  The mixture interpretation is supported by 
analyses of {\it single} clusters that have found large spreads in [C/Fe] and 
[N/Fe] of stars at any common evolutionary phase from the MS through the tip of 
the RGB (\eg NGC 0104, 6205, 6254, 6397, 6752, 7006; \citealt{Ca05}, 
\citealt{SmG05}, and references therein).

The evolutionary trends seen in \Figs{CNvsMV}{NvsC} are commonly associated 
with a combination of the first dredge-up \citep[e.g.][]{Ib64} followed by deep 
atmospheric mixing \citep[e.g.][]{SM79} that occur in the atmospheres of low-mass stars after they complete core-hydrogen burning.  The first dredge-up is 
defined by the mixing of partially-processed material from the stellar interior 
into the outer atmosphere as the convective envelope grows in size during the 
star's SGB phase.  It is responsible for the gentle decline observed in both 
[C/Fe] and [N/Fe] from the MS to about the midpoint of the RGB ($M_V \sim$ +1 
mag).  Once the star reaches the RGB bump, deep mixing is thought to set in and 
create the rapid rate of depletion and (now) enhancement of atmospheric carbon 
and nitrogen, respectively.  Deep mixing accomplishes this by bringing CN(O)-processed material from the hydrogen-burning shell into the outer atmosphere once 
the shell overcomes the molecular weight barrier left by the first dredge-up 
and expands into the outer convective envelope.  Both the luminosity of the 
bump (\ie onset of mixing) and the depletion rate of carbon are known to 
decrease as a function of metallicity \citep{FP90,Mart08}.  These two 
dependencies may help explain some of the scatter seen in \Fig{CNvsMV} at high 
luminosities ($M_V \gtrsim$ 0 mag).

In light of the fact that evolved stars undergo episodes of atmospheric mixing, 
it seems conceivable that the distributions of [C/Fe] and [N/Fe] in our 
compilation would be somewhat broader than those of unaffected species, as seen 
in \Fig{XFeDistbn}.  The reason for this is that spectroscopic studies of 
resolved GGC members have historically measured [C/Fe] and [N/Fe] from a 
variety of evolutionary stages.  Were measurements of these ratios for MS stars 
to become available for our entire sample, we would anticipate a tightening of 
the corresponding distributions.  Until that time comes, the reality of these 
mixing episodes should compel SPS modellers to carefully consider the 
luminosity biases of published spectroscopic studies of individual GGC members 
when evaluating the accuracies of their [C/Fe] and [N/Fe] predictions\footnote{The depth of the data being modelled must also be considered.  For instance, the 
luminosities of individual GGC members down to which the S05 spectra are 
sensitive has yet to be firmly established (but see Barber et al., in prep.).}.

\subsubsection{Multiple populations}
\label{sec:D&R-MP}
In addition to mixing phenomena, \Fig{CNvsMV} also shows that some intrinsic 
degree of broadening in the [C/Fe] and (especially) [N/Fe] distributions for 
the S05 GGCs is to be expected on account of the multiple populations found 
within many of them.  From those plots, it is seen that the dichotomy in CN 
band strength (weak/strong) extends down to the MS, the CN band is much more 
sensitive to [N/Fe] than [C/Fe], and CN-weak stars are characterized by lower 
[C/Fe] and higher [N/Fe] values than CN-strong stars.  The union of these facts 
then creates the potential for the mean values of [C/Fe] and [N/Fe] from any 
given study to be biased either low or high depending on how accurately the 
sample embodies the true CN distribution of the associated cluster.  In fact, 
we might be able to infer the as yet unknown CN strengths of certain samples 
based on the relative proximity of circlular points to the triangular or 
diamond points in \Fig{CNvsMV}.

Another hallmark of the multiple population phenomenon is the anti-correlation 
of [O/Fe] and [Na/Fe] ratios exhibited by stars from all major evolutionary 
stages within affected clusters \citep{Gr12a}.  The existence of this anti-correlation would thus tend to bias estimates of a given cluster's mean abundances 
of these species if not accounted for during sample selection.  We suspect that 
this effect may be at least a contributing factor to the relatively broader 
distribution for [O/Fe] seen in \Fig{XFeDistbn}, compared to those of abundance 
ratios that are not known to vary from star-to-star.  This suspicion could be 
tested by investigating whether the breadth of our [O/Fe] and [Na/Fe] 
distributions is {\it simultaneously} consistent with the observed ranges in 
these abundance ratios for individual stars from the large, homogeneous Na-O 
anti-correlation study of \cite{Ca09a,Ca09b}.  This is beyond the scope of the 
present work however.

It is worth mentioning at this juncture that the exact origins of the multiple 
populations observed in GGCs remains unknown.  The perpetuation of chemical 
abundance variations down to unevolved, MS stars makes strongly certain that 
the existence of a second (and sometimes third; \citealt{Ca09a}) generation of 
stars is tied to an external agent.  However, at least three candidates could 
be responsible for the pattern of enhanced nitrogen, sodium, aluminum, and 
(possibly) helium abundances plus depleted carbon, oxygen, and magnesium 
abundances that typifies the younger generations: (i) massive AGB stars \citep{Ve01}, (ii) massive rotating MS stars \citep{De07} and (iii) as in (i) but for 
intermediate masses \citep{VDA08}.  One way to help distinguish between these 
candidates is to study whether the sum of the CNO elements varies between the 
populations in each affected cluster.  Simply put, massive stars are expected 
to alter the individual abundances of these elements but leave their sum 
unchanged, while intermediate-mass AGB stars, by way of the third dredge-up, 
will not.  Evidence thus far of variations in the CNO sum within individual 
clusters has been contentious \citep{Iv99,Ca05,CM05,Ca08,Mi08,Ve09,Yo09,Vi10}.  
With some work, our extensive compilation of carbon, nitrogen, and oxygen 
abundances for individual GGC stars may be helpful in shedding further light on 
this issue, but will not be investigated further here.

\subsubsection{Source(s) of $\alpha$-elements}
\label{sec:D&R-SAE}
While \Figs{CNvsMV}{NvsC} point to the existence of atmospheric mixing episodes 
within the evolved stars of the S05 GGCs, \Fig{XFeDistbn} shows that the mean 
abundances of $\alpha$-elements amongst these clusters remain more or less 
homogeneous.  This homogeneity is accounted for within the context of mixing by 
the fact that $\alpha$-elements are exempt from the CN(O) cycle, such that 
their abundances likely reflect the chemistry of the gas from which these GGCs 
were born.  Moreover, that the $\alpha$-element abundance ratios for the S05 
GGCs are all greater than the solar value by factors of 1.7-2.4 implies that 
these systems must have formed quite rapidly, on timescales less than that of 
Type Ia supernovae ($\sim$1 Gyr).

\begin{figure*}
 \begin{center}
  \begin{tabular}{c c}
   \includegraphics[width=0.5\textwidth]{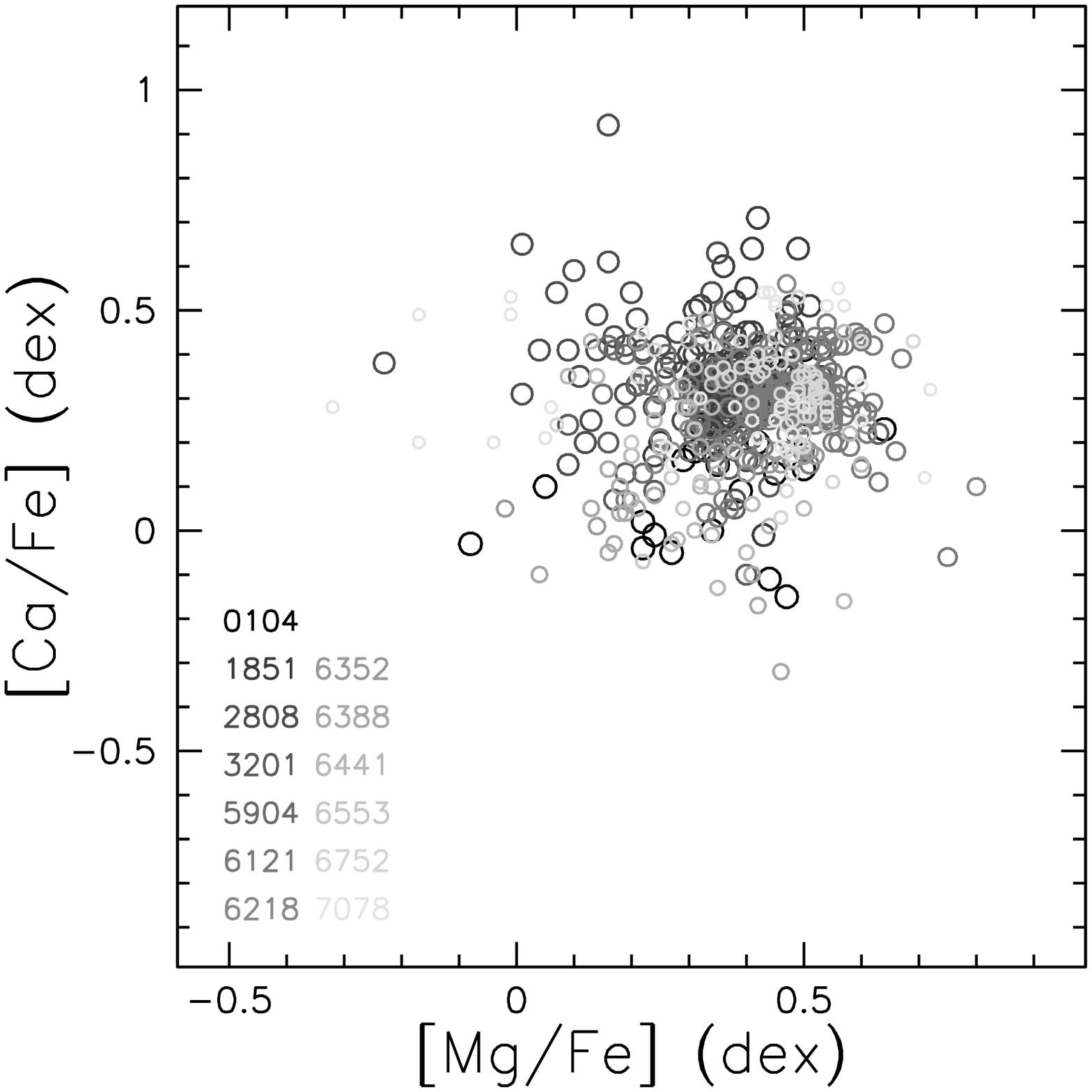} & 
   \includegraphics[width=0.5\textwidth]{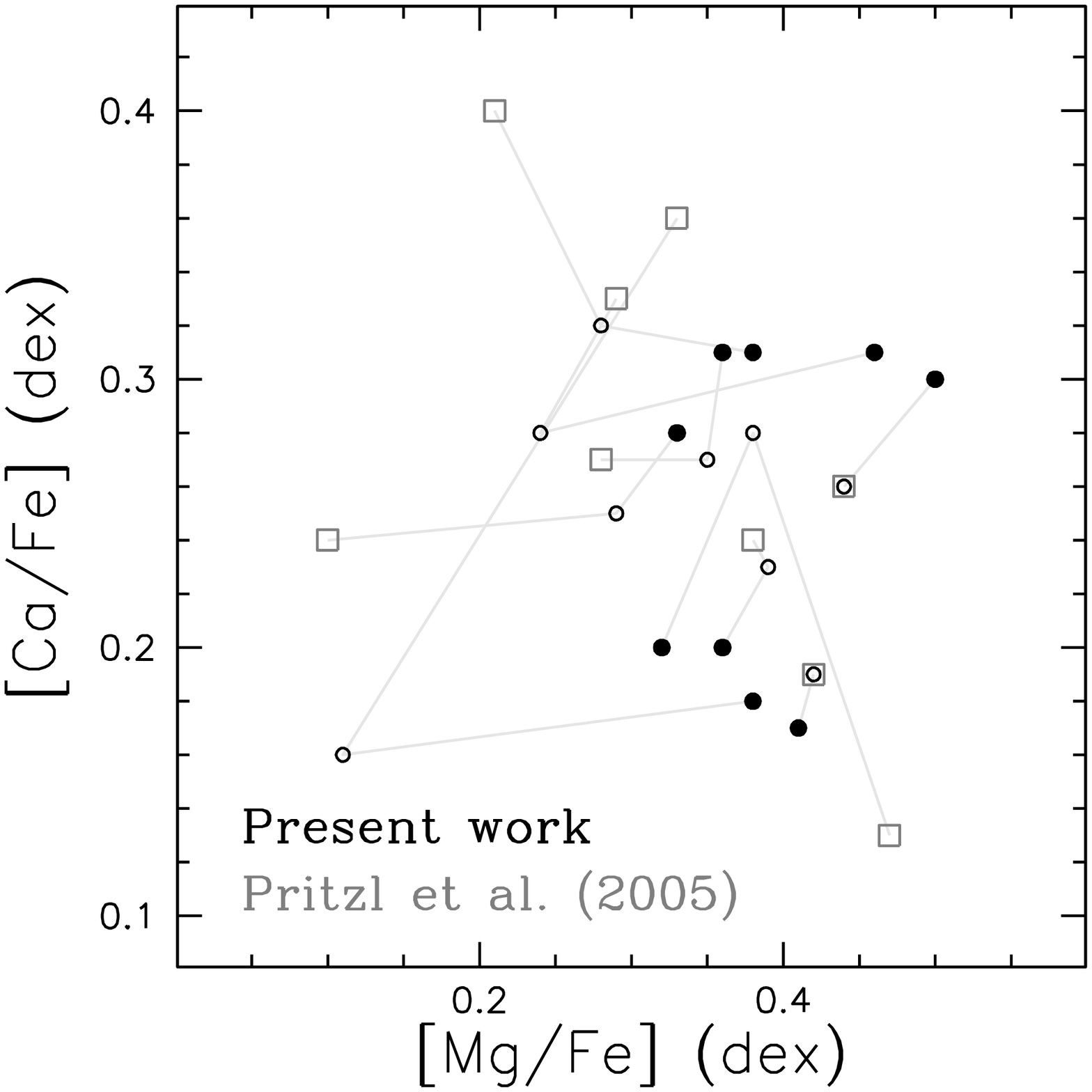}
  \end{tabular}
  \caption{Calcium versus magnesium abundance for individual stars ({\it left}) 
or whole clusters ({\it right}) belonging to the S05 sample.  In the left-hand 
panel, data for individual clusters are represented by a unique combination of 
grayscale shading and point size, with the shading key expressed in the 
bottom-left corner (numbers refer to NGC IDs of the clusters).  The mean 
abundances of these species are plotted in the right-hand panel for the nine 
S05 clusters shown at left that overlap with the sample of \cite{Pri05}.  
There, we show the values obtained by these authors (open squares), in the 
present work (filled circles) and from the same data and method used by Pritzl 
et al., except without making corrections for choice of log{\it gf} values nor 
solar abundance scale (open circles).  To help demonstrate the effects that 
different approaches taken to compilation work such as ours, the different 
point types are connected by a line for each cluster.}
  \label{fig:CavsMg}
 \end{center}
\end{figure*}

An interesting corollary on the chemical enrichment of GGCs, based on their 
$\alpha$-element abundances, is presented in \Fig{CavsMg}.  The left-hand panel 
shows [Ca/Fe] versus [Mg/Fe] for individual stars from our compilation and 
belonging to the 13 S05 clusters having the most measurements in this regard.  
These data are clearly uncorrelated and scatter about their centroid at 
([Mg/Fe],[Ca/Fe]) $\sim$ (+0.4,+0.3) with rms dispersions (0.13-0.14 dex) 
comparable to the median errors in the individual stellar abundances (0.12-0.14 
dex).  These evidences are enough to suggest that the production sites of these 
two chemical species are not one and the same.

In the right-hand panel of \Fig{CavsMg}, we show with open squares the mean 
magnesium and calcium abundances from Pr05 for the nine clusters in the left-hand panel that fall within their sample.  Excluding the single outlier at 
([Mg/Fe],[Ca/Fe]) $\sim$ (+0.10,+0.24), we find the rather surprising result 
that these abundances are {\it anti}-correlated for this sub-sample and data.  
Not surprisingly, the corresponding values from our compilation (filled 
circles) exhibit no such correlation.  This discrepant behaviour between our 
results and those of Pr05 could arise from the different approaches taken with 
respect to the following issues: (i) scope of input data, (ii) averaging 
method, and (iii) systematics.  The latter refers to Pr05's attempt to 
homogenize all of their input data to the same log{\it gf} and solar abundance 
scale, something we neglect to do.  To appreciate the potential role of 
systematics, we also show in this panel, with open circles, the mean abundances 
we derive based on the same references and averaging method used by Pr05, but 
without log{\it gf} and solar abundance corrections.  For each cluster, we 
connect with a line the abundances from the three distinct methods.  Comparing 
the positions of open squares and circles, it is clear that the choice of 
atomic constants and solar abundance pattern often plays a significant role in 
setting the values of [X/Fe] for any element X, by as much as $\pm$0.2 dex.  On 
the other hand, the offsets between open and filled circles may be regarded as 
the effect of our using more references per cluster and straight averages, as 
well as neglecting systematic corrections.  Our methodology clearly affects our 
adopted abundance patterns as well, but this should not reflect any fundamental 
flaws in our results.

The prospect of anti-correlated magnesium and calcium abundances amongst the 
GGC population supports the inference that the production sites of these two 
species are not one and the same or, even more intriguing, that the yields of 
Type II supernovae fluctuate on an element-by-element level.  Note that such a 
trend is also found amongst the points in the right-hand panel of \Fig{CavsMg} 
which represent the results of our attempt to mimic the approach of Pr05 (open 
circles), again modulo a single outlier.  When the full sample of either 
compilation is considered, however, this anti-correlation changes to a weak 
positive correlation.  The fact that neither sample is complete though implies 
that further investigation of the ratios [Mg/Fe] and [Ca/Fe] amongst individual 
GGC stars on a larger, more homogeneous basis may be warranted.

The suggestion that the abundances of magnesium and carbon do not track one 
another within any given stellar population is not new.  For instance, several 
studies of the central stellar populations of early-type galaxies have 
concluded that [Mg/Fe] increases modestly with velocity dispersion amongst 
these systems, but that [Ca/Fe] is uniform, at about the solar value \citep{Va97,Wo98,Tr98,HW99,Sa02,Th03,Ce03,Ce04,Sm09,Wo11}.  \cite{Pro05} casted doubt on 
the authenticity of these results by pointing out that the Ca4227 index, a 
popular tracer of calcium abundance, has its blue pseudocontinuum contaminated 
by a CN band.  \cite{Grav07} and \cite{Co13b} have shown, however, that the 
[Ca/Fe] ratio remains uniform amongst red sequence galaxies even when the 
abundances of carbon and nitrogen are properly accounted for {\it a priori}.

In the Milky Way, \cite{Fu07} found that while [Ca/Fe] decreases with 
increasing [Fe/H] for RGB stars in the bulge, their [Mg/Fe] ratios remain more 
or less uniform at $\sim$ +0.3 dex.  Earlier claims to this same effect were 
made by \cite{MWR94}, \cite{Zo04}, and \cite{AB06}.  These discrepant trends 
may only apply to the Galactic bulge though, since the calcium abundances of 
both metal-poor field and thick disk stars behave in a fashion consistent with 
that exhibited by the other $\alpha$ elements \citep{Wh89,Ed93,Re03}.

The sum of the above discrepancies results in a confusing picture, to say the 
least, of how stellar systems enrich themselves in the $\alpha$-elements.  One 
possible solution is that the abundance pattern of a system, and thus the 
source(s) of its chemical evolution, depends sensitively on the intensity of 
the star formation from which it was created.  For instance, the spheroidal 
systems which seem to exhibit genuine differences in the behaviours of their Mg 
and Ca abundances (\ie early-type galaxies, Milky Way bulge, and GGCs) likely 
formed most of their stellar mass over rapid timescales.  In other systems, the 
star formation history could very well have been more protracted, leading to 
potentially different sources of chemical enrichment.  A most useful test of 
this proposed solution would be to see if and how chemodynamical simulations 
could reproduce the precise pattern of $\alpha$-element abundances we find in 
our GGCs.

\subsection{Comparison with previous work}
\label{sec:D&R-CPW}
Having presented our compilation and advertised some of the immediate science 
that can be gleaned from it, it would also be prudent to assess the robustness 
of our results by comparing them to those from previous similar work.  
The comparison of our adopted ages and metallicities against other sources of 
such information has already been performed elsewhere (\Se{D&R-A&ZSources}; 
Appendix) so that we need only focus here on the chemical abundance patterns of 
our clusters.

\begin{figure*}
 \begin{center}
  \includegraphics[width=0.9\textwidth]{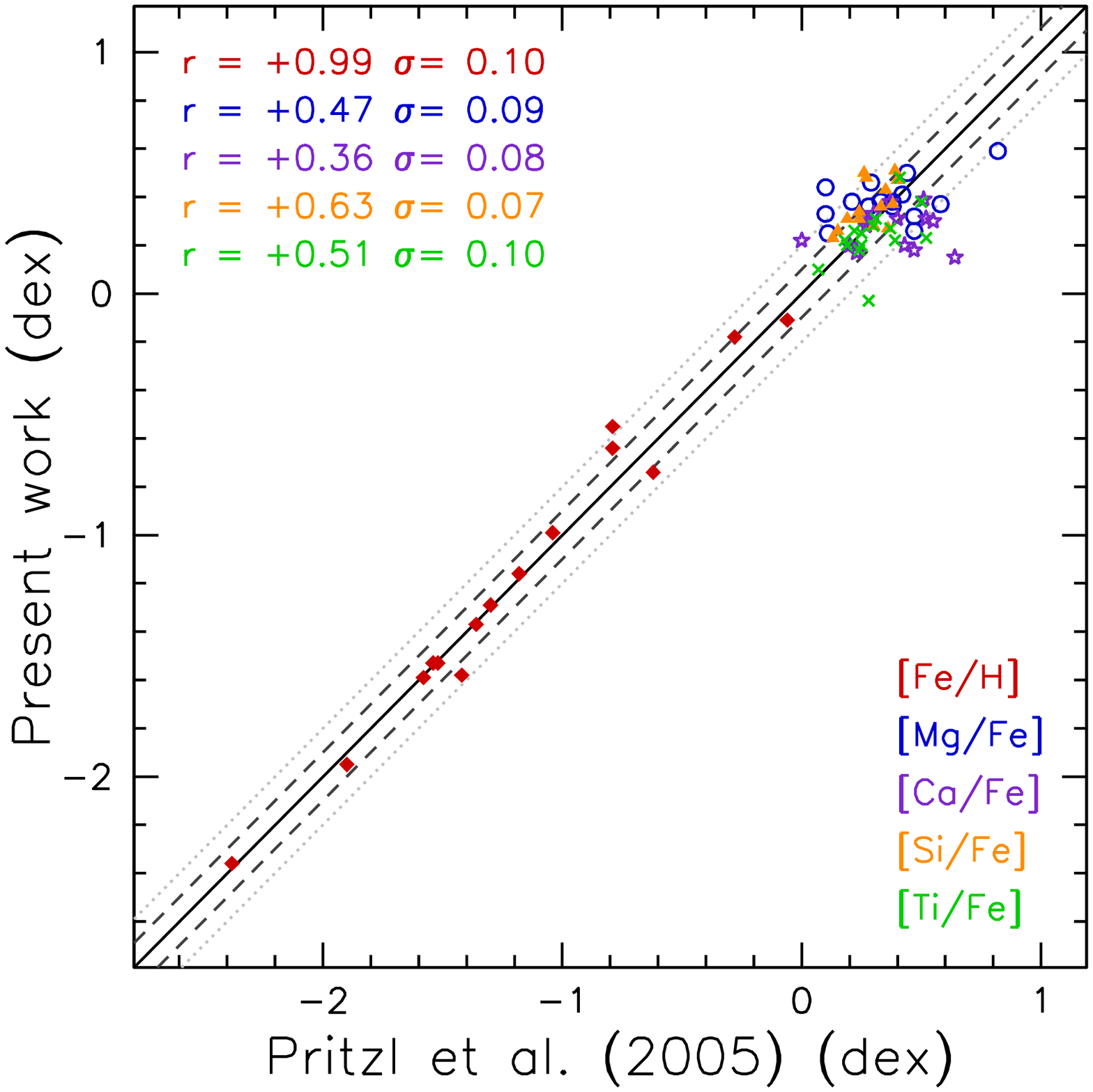}
  \caption{Comparison of our adopted metallicities and $\alpha$-element 
abundance patterns for GGCs from the S05 sample which overlap with that of 
\cite{Pri05}.  The Pearson (linear) correlation coefficient and the 1-$\sigma$ 
dispersion (in logarithmic dex units) of a linear least-squares fit between 
each pair of datasets are indicated at top-left.  The dashed and dotted lines 
correspond to $\pm$0.1 and $\pm$0.2 dex offsets from the locus of equality 
(solid line), respectively.}
  \label{fig:LitvsPri05}
 \end{center}
\end{figure*}

As stated before, the most extensive survey of the literature on GGC chemical 
abundance patterns prior to our compilation was made by Pr05.  We already 
compared their results against ours in terms of the ratios [Mg/Fe] and [Ca/Fe] 
in the right-hand panel of \Fig{CavsMg}.  In \Fig{LitvsPri05} we expand on this 
comparison by plotting our estimates versus theirs of the metallicities and $\alpha$-element abundances (Mg, Ca, Si, Ti) for the 15 GGCs common to both 
samples.  Despite several outliers, a good correspondence clearly exists 
between the Pr05 metallicities and our own.  Conversely, the large scatter and 
low Pearson coefficients (shown at top-left) exhibited by the other sets of 
points in \Fig{LitvsPri05} means we cannot reach the same conclusion regarding 
the abundances of $\alpha$-elements.  Specifically, our results compare least 
favourably to those of Pr05 in terms of [Ca/Fe].  By inspection of Pr05's 
methodology, we find that the most egregious discrepancies between the two sets 
of abundance ratios can be explained by two effects.  These are: (i) our 
inclusion of references that post-dated their work, and (ii) corrections that 
Pr05 implemented to standardize literature data to a common log {\it gf} 
system and solar abundance pattern.  The latter corrections can often be 
significant in size ($\sim$0.2 dex), a point which was already hinted at in 
\Fig{CavsMg}.  As Pr05 point out though, abundance analyses are often published 
without specifying the assumed log {\it gf} values and solar abundances, making 
it difficult (if not impossible) to gauge what those corrections should be.  We 
therefore abstain from attempting such corrections ourselves and instead 
embrace the fact that our adopted abundance patterns likely suffer from the 
full effect of the systematics which underpin spectral syntheses.


\section{Conclusions}\label{sec:Concs}

Drawing on a wealth of literature data up to mid-2012, we have assembled a new 
compilation of the known ages, metallicities, and chemical abundance patterns 
for the 41 GGCs studied by S05.  This extensive compilation represents a 
singular expansion upon similar but more limited previous work on the stellar 
populations of these systems \citep{Ha96,Pri05,MF09}.  We anticipate that it 
will prove to be a key ingredient for stringent evaluations of the absolute 
reality and robustness of predictions from the latest suite of SPS models 
designed to recover the above information for unresolved systems (Roediger et 
al., in prep.).

Given the wide range of parameter space spanned by the S05 sample, our 
compilation should also benefit a wide range of other astrophysical interests.  
The age distribution for these clusters suggests that they arose from a star 
formation history that consisted of a strong peak (12.5-13.0 Gyr ago) 
superimposed upon a relatively smooth background.  Combining this information 
with their metallicities and $\alpha$-element abundances, it appears that each 
GGC was formed rapidly either in situ or in a satellite galaxy and subsequently 
accreted onto the Milky Way.  Furthermore, with our compiled abundance patterns 
we confirm previous claims that (i) the surface abundances of C and N in 
evolved stars are altered by mixing episodes as they ascend the SGB/RGB, (ii) 
many GGCs host at least two distinct stellar populations, and (iii) the 
enrichment of $\alpha$-elements in these systems, specifically Mg and Ca, 
likely occurred through multiple channels.  The fact that the mean chemical 
abundance patterns of GGCs are sensitive to the first two phenomena are 
important caveats that must be considered during SPS model evaluations.  
Similarly, we estimate that absolute age determinations for GGCs are subject to 
systematic uncertainties on the order of $\sim$2-3 Gyr.

While the above results are certainly of some value, it is our opinion that 
many other applications of our compilation have yet to be explored.  To enable 
the community to further pursue such ancillary science or modify the results of 
our compilation as they see fit, we provide electronic tables of the input data 
online\footnote{\url{http://www.astro.queensu.ca/people/Stephane\_Courteau/roediger2013/index.html}} for each one of our clusters.

\bigskip
We thank Bill Harris and Charlie Conroy for insightful discussions and detailed 
comments on an earlier version of this paper which led to valuable 
improvements.  J.~R. and S.~C. acknowledge financial support from the National 
Science and Engineering Council of Canada in the form of an Alexander Graham 
Bell PGS D Fellowship and a Discovery Grant, respectively.


\begin{deluxetable}{cccccccc}
 \tabletypesize{\footnotesize}
 \tablewidth{0pc}  
 \tablecaption{Sample clusters\tablenotemark{1}}
 \tablehead{
  \colhead{NGC} &
  \colhead{Other} &
  \colhead{$l$} &
  \colhead{$b$} &
  \colhead{R$_{GC}$} &
  \colhead{} &
  \colhead{} &
  \colhead{} \\
  \colhead{ID} &
  \colhead{ID} &
  \colhead{(deg)} &
  \colhead{(deg)} &
  \colhead{(kpc)} &
  \colhead{$E$($B-V$)} &
  \colhead{$c$} &
  \colhead{($B-R$)/($B+V+R$)} \\
  \colhead{(1)} &
  \colhead{(2)} &
  \colhead{(3)} &
  \colhead{(4)} &
  \colhead{(5)} &
  \colhead{(6)} &
  \colhead{(7)} &
  \colhead{(8)}
 }
 \startdata
 0104 & 47 Tuc & 305.89 & -44.89 &  7.4 & 0.04 & 2.07 & -0.99 \cr
 1851 &        & 244.51 & -35.03 & 16.6 & 0.02 & 1.86 & -0.36 \cr
 1904 & M79    & 227.23 & -29.35 & 18.8 & 0.01 & 1.70 &  0.89 \cr
 2298 &        & 245.63 & -16.00 & 15.8 & 0.14 & 1.38 &  0.93 \cr
 2808 &        & 282.19 & -11.25 & 11.1 & 0.22 & 1.56 & -0.49 \cr
 3201 &        & 277.23 &   8.64 &  8.8 & 0.24 & 1.29 &  0.08 \cr
 5286 &        & 311.61 &  10.57 &  8.9 & 0.24 & 1.41 &  0.80 \cr
 5904 & M5     &   3.86 &  46.80 &  6.2 & 0.03 & 1.73 &  0.31 \cr
 5927 &        & 326.60 &   4.86 &  4.6 & 0.45 & 1.60 & -1.00 \cr
 5946 &        & 327.58 &   4.19 &  5.8 & 0.54 & 2.50 &     - \cr
 5986 &        & 337.02 &  13.27 &  4.8 & 0.28 & 1.23 &  0.97 \cr
 6121 & M4     & 350.97 &  15.97 &  5.9 & 0.35 & 1.65 & -0.06 \cr
 6171 & M107   &   3.37 &  23.01 &  3.3 & 0.33 & 1.53 & -0.73 \cr
 6218 & M12    &  15.72 &  26.31 &  4.5 & 0.19 & 1.34 &  0.97 \cr
 6235 &        & 358.92 &  13.52 &  4.2 & 0.31 & 1.53 &  0.89 \cr
 6254 & M10    &  15.14 &  23.08 &  4.6 & 0.28 & 1.38 &  0.98 \cr
 6266 & M62    & 353.57 &   7.32 &  1.7 & 0.47 & 1.71 &  0.32 \cr
 6284 &        & 358.35 &   9.94 &  7.5 & 0.28 & 2.50 &     - \cr
 6304 &        & 355.83 &   5.38 &  2.3 & 0.54 & 1.80 & -1.00 \cr
 6316 &        & 357.18 &   5.76 &  2.6 & 0.54 & 1.65 & -1.00 \cr
 6333 & M9     &   5.54 &  10.71 &  1.7 & 0.38 & 1.25 &  0.87 \cr
 6342 &        &   4.90 &   9.72 &  1.7 & 0.46 & 2.50 & -1.00 \cr
 6352 &        & 341.42 &  -7.17 &  3.3 & 0.22 & 1.10 & -1.00 \cr
 6356 &        &   6.72 &  10.22 &  7.5 & 0.28 & 1.59 & -1.00 \cr
 6362 &        & 325.55 & -17.57 &  5.1 & 0.09 & 1.09 & -0.58 \cr
 6388 &        & 345.56 &  -6.74 &  3.1 & 0.37 & 1.75 &     - \cr
 6441 &        & 353.53 &  -5.01 &  3.9 & 0.47 & 1.74 &     - \cr
 6522 &        &   1.02 &  -3.93 &  0.6 & 0.48 & 2.50 &  0.71 \cr
 6528 &        &   1.14 &  -4.17 &  0.6 & 0.54 & 1.50 & -1.00 \cr
 6544 &        &   5.84 &  -2.20 &  5.1 & 0.76 & 1.63 &  1.00 \cr
 6553 &        &   5.26 &  -3.03 &  2.2 & 0.63 & 1.16 & -1.00 \cr
 6569 &        &   0.48 &  -6.68 &  3.1 & 0.53 & 1.31 &     - \cr
 6624 &        &   2.79 &  -7.91 &  1.2 & 0.28 & 2.50 & -1.00 \cr
 6626 & M28    &   7.80 &  -5.58 &  2.7 & 0.40 & 1.67 &  0.90 \cr
 6637 & M69    &   1.72 & -10.27 &  1.7 & 0.18 & 1.38 & -1.00 \cr
 6638 &        &   7.90 &  -7.15 &  2.2 & 0.41 & 1.33 & -0.30 \cr
 6652 &        &   1.53 & -11.38 &  2.7 & 0.09 & 1.80 & -1.00 \cr
 6723 &        &   0.07 & -17.30 &  2.6 & 0.05 & 1.11 & -0.08 \cr
 6752 &        & 336.49 & -25.63 &  5.2 & 0.04 & 2.50 &  1.00 \cr
 7078 & M15    &  65.01 & -27.31 & 10.4 & 0.10 & 2.29 &  0.67 \cr
 7089 & M2     &  53.37 & -35.77 & 10.4 & 0.06 & 1.59 &  0.96 \cr
 \enddata
 \label{tbl:Sample}
 \tablenotetext{1}{All data adopted from the 2010 edition of \cite{Ha96}.}
\end{deluxetable}

\bigskip

\begin{deluxetable}{cccc|cccccc}
 \tabletypesize{\tiny}
 \tablewidth{0pc}  
 \tablecaption{Reference key}
 \rotate
 \tablehead{
  \multicolumn{4}{c|}{Ages} &
  \multicolumn{6}{c}{Metallicities \& abundance ratios} \\
  \colhead{Number} &
  \colhead{Reference} &
  \multicolumn{1}{c}{Number} &
  \multicolumn{1}{c|}{Reference} &
  \colhead{Number} &
  \colhead{Reference} &
  \colhead{Number} &
  \colhead{Reference} &
  \colhead{Number} &
  \colhead{Reference} \\
  \colhead{(1)} &
  \colhead{(2)} &
  \multicolumn{1}{c}{(3)} &
  \multicolumn{1}{c|}{(4)} &
  \colhead{(5)} &
  \colhead{(6)} &
  \colhead{(7)} &
  \colhead{(8)} &
  \colhead{(9)} &
  \colhead{(10)}
 }
 \startdata
  1 & \cite{MF09}   & 47 & \cite{Fu96}   &  1 & \cite{Ca09c} & 47 & \cite{Ba99}   &  92 & \cite{Ca07c}  \cr
  2 & \cite{DAng05} & 48 & \cite{Ji96}   &  2 & \cite{AZ88}  & 48 & \cite{Co99}   &  93 & \cite{Grat07} \cr
  3 & \cite{MW06}   & 49 & \cite{Re96}   &  3 & \cite{Ca09a} & 49 & \cite{Iv99}   &  94 & \cite{Le07}   \cr
  4 & \cite{Do10}   & 50 & \cite{Ri96}   &  4 & \cite{Ca09b} & 50 & \cite{Sn00a}  &  95 & \cite{Va07}   \cr
  5 & \cite{SW02}   & 51 & \cite{Sam96a} &  5 & \cite{SZ78}  & 51 & \cite{Sn00b}  &  96 & \cite{Wa07}   \cr
  6 & \cite{Ro99}   & 52 & \cite{Sam96b} &  6 & \cite{Co79}  & 52 & \cite{Ca01}   &  97 & \cite{Ha08}   \cr
  7 & \cite{AL81}   & 53 & \cite{San96}  &  7 & \cite{BD80}  & 53 & \cite{Co01}   &  98 & \cite{Ki08}   \cr
  8 & \cite{Ha83}   & 54 & \cite{Br97}   &  8 & \cite{DCC80} & 54 & \cite{Gr01}   &  99 & \cite{KMW08}  \cr
  9 & \cite{Bu84}   & 55 & \cite{Kr97}   &  9 & \cite{Pi80}  & 55 & \cite{Iv01}   & 100 & \cite{Mari08} \cr
 10 & \cite{Ca84}   & 56 & \cite{Bu98}   & 10 & \cite{Gr82}  & 56 & \cite{KS01}   & 101 & \cite{Mart08} \cr
 11 & \cite{DC84}   & 57 & \cite{Gu98}   & 11 & \cite{Pi83}  & 57 & \cite{Co02}   & 102 & \cite{MWB08}  \cr
 12 & \cite{SR84}   & 58 & \cite{JP98}   & 12 & \cite{SM83}  & 58 & \cite{Be03}   & 103 & \cite{Or08}   \cr
 13 & \cite{Ca85}   & 59 & \cite{Ri98}   & 13 & \cite{Ge84}  & 59 & \cite{Ca03}   & 104 & \cite{Pa08}   \cr
 14 & \cite{AL86}   & 60 & \cite{SW98}   & 14 & \cite{La85}  & 60 & \cite{Ho03}   & 105 & \cite{YG08}   \cr
 15 & \cite{Bu86}   & 61 & \cite{Al99}   & 15 & \cite{Gr86}  & 61 & \cite{Me03}   & 106 & \cite{Yon08a} \cr
 16 & \cite{GO86}   & 62 & \cite{Br99}   & 16 & \cite{SS86}  & 62 & \cite{Mi03}   & 107 & \cite{Yon08b} \cr
 17 & \cite{He86}   & 63 & \cite{Gi99}   & 17 & \cite{Gr87b} & 63 & \cite{RC03}   & 108 & \cite{Yon08c} \cr
 18 & \cite{He87}   & 64 & \cite{HR99}   & 18 & \cite{CS88}  & 64 & \cite{Yo03}   & 109 & \cite{Ba09}   \cr
 19 & \cite{RF87}   & 65 & \cite{Pi99}   & 19 & \cite{GO89}  & 65 & \cite{Br04}   & 110 & \cite{Fe09}   \cr
 20 & \cite{JH88}   & 66 & \cite{Ch00}   & 20 & \cite{Be90}  & 66 & \cite{Car04a} & 111 & \cite{Ta09}   \cr
 21 & \cite{Sa88}   & 67 & \cite{Da00}   & 21 & \cite{Br90}  & 67 & \cite{Car04b} & 112 & \cite{Vi09}   \cr
 22 & \cite{Al89}   & 68 & \cite{FG00}   & 22 & \cite{Fr91}  & 68 & \cite{Cav04}  & 113 & \cite{Wo09}   \cr
 23 & \cite{Bu89}   & 69 & \cite{He00}   & 23 & \cite{Sn91}  & 69 & \cite{Jas04}  & 114 & \cite{Yo09}   \cr
 24 & \cite{Do89}   & 70 & \cite{Va00}   & 24 & \cite{SS91}  & 70 & \cite{Zo04}   & 115 & \cite{Br10b}  \cr
 25 & \cite{Hu89}   & 71 & \cite{Be01}   & 25 & \cite{BW92}  & 71 & \cite{AB05}   & 116 & \cite{DOM10}  \cr
 26 & \cite{Sa89}   & 72 & \cite{Or01}   & 26 & \cite{Ba92}  & 72 & \cite{Ca05}   & 117 & \cite{DO10}   \cr
 27 & \cite{Al90a}  & 73 & \cite{Te01}   & 27 & \cite{Br92}  & 73 & \cite{Cle05}  & 118 & \cite{KMW10}  \cr
 28 & \cite{Al90b}  & 74 & \cite{vBM01}  & 28 & \cite{Dr92}  & 74 & \cite{Co05}   & 119 & \cite{Val10}  \cr
 29 & \cite{Al90c}  & 75 & \cite{Zo01}   & 29 & \cite{MW92}  & 75 & \cite{Gr05}   & 120 & \cite{Vi10}   \cr
 30 & \cite{Fe91}   & 76 & \cite{Gr02}   & 30 & \cite{Sn92}  & 76 & \cite{Or05}   & 121 & \cite{WC10}   \cr
 31 & \cite{SC91}   & 77 & \cite{FJ02}   & 31 & \cite{Mi93}  & 77 & \cite{SmG05}  & 122 & \cite{Ca11a}  \cr
 32 & \cite{Ch92}   & 78 & \cite{vB02}   & 32 & \cite{Ar94}  & 78 & \cite{SmV05}  & 123 & \cite{Gr11}   \cr
 33 & \cite{DL92}   & 79 & \cite{Gr03}   & 33 & \cite{Dr94}  & 79 & \cite{Va05}   & 124 & \cite{Lai11}  \cr
 34 & \cite{FP92}   & 80 & \cite{LS03}   & 34 & \cite{DCA95} & 80 & \cite{Yo05}   & 125 & \cite{Ma11}   \cr
 35 & \cite{MW92}   & 81 & \cite{Mo03}   & 35 & \cite{Kr95}  & 81 & \cite{AB06}   & 126 & \cite{OC11}   \cr
 36 & \cite{Or92}   & 82 & \cite{Pu03}   & 36 & \cite{Mi95a} & 82 & \cite{Ca06a}  & 127 & \cite{So11}   \cr
 37 & \cite{Wa92}   & 83 & \cite{Ha04}   & 37 & \cite{Mi95b} & 83 & \cite{Ca06b}  & 128 & \cite{Va11}   \cr
 38 & \cite{Al94}   & 84 & \cite{Br05}   & 38 & \cite{NDC95} & 84 & \cite{Gr06}   & 129 & \cite{VG11}   \cr
 39 & \cite{Ri94}   & 85 & \cite{Hu07}   & 39 & \cite{Fu96}  & 85 & \cite{JP06}   & 130 & \cite{Ca12a}  \cr
 40 & \cite{CK95}   & 86 & \cite{Sa07}   & 40 & \cite{Mi96}  & 86 & \cite{Pr06}   & 131 & \cite{Gr12b}  \cr
 41 & \cite{Fe95}   & 87 & \cite{Ba09}   & 41 & \cite{Sh96}  & 87 & \cite{So06}   & 132 & \cite{Gr12c}  \cr
 42 & \cite{Fu95}   & 88 & \cite{BS09}   & 42 & \cite{Ge97}  & 88 & \cite{Yo06}   & 133 & \cite{La12}   \cr
 43 & \cite{Ma95}   & 89 & \cite{DA09}   & 43 & \cite{GL97}  & 89 & \cite{Wy06}   & 134 & \cite{Mo12}   \cr
 44 & \cite{Or95}   & 90 & \cite{Mo09}   & 44 & \cite{Sm97}  & 90 & \cite{Ca07a}  & 135 & \cite{Vi12}   \cr
 45 & \cite{Sa95}   & 91 & \cite{Zo09}   & 45 & \cite{Sn97}  & 91 & \cite{Ca07b}  & 136 & \cite{WC12}   \cr
 46 & \cite{Da96}   & 92 & \cite{Bo10}   & 46 & \cite{GW98}  &    &               &     &               \cr
 \enddata
 \label{tbl:Refs-1}
\end{deluxetable}

\bigskip

\begin{deluxetable}{clll}
 \tabletypesize{\tiny}
 \tablewidth{0pc}  
 \tablecaption{Available age and recommended chemical abundance references per S05 GGC.  Boldface numbers under the second column denote our the sources of our adopted ages.}
 \tablehead{
  \colhead{NGC} &
  \colhead{Age} &
  \colhead{Metallicity} &
  \colhead{Abundance} \\
  \colhead{ID} &
  \colhead{Reference(s)} &
  \colhead{Reference(s)} &
  \colhead{Reference(s)} \\
  \colhead{(1)} &
  \colhead{(2)} &
  \colhead{(3)} &
  \colhead{(4)}
 }
 \startdata
0104 & \textbf{1}-6, 8, 18, 24, 31, 40, 43, 50, 56, 60, 63, 70, 75-76, 79, 84, 86-87, 89 & 1-2, 18, 25, 71, 89, 99, 102         & 3-4, 11, 15, 21, 25, 38, 65-66, 71-72, 89, 99, 113, 117, 136                   \cr
1851 & \textbf{1}-3, 5-6, 21, 27, 37, 40, 50, 53, 56, 60, 70                             & 1-2, 42, 105, 114, 120, 122          & 105, 114, 120, 122, 131-133                                                    \cr
1904 & 2-3, \textbf{5}-6, 16-17, 38, 40, 50, 55-56, 60                                   & 1, 22, 42                            & 3-4, 22                                                                        \cr
2298 & \textbf{1}, 4-5, 14, 16, 20, 29, 32, 35, 40, 50, 56, 60                           & 1, 29                                & 29                                                                             \cr
2808 & \textbf{1}-3, 5-6, 9, 16, 28, 40, 50, 53, 56, 70                                  & 1, 42                                & 4, 10, 59, 67, 82-83, 115, 123                                                 \cr
3201 & \textbf{1}-7, 10, 22, 40, 50-51, 56, 60, 74, 80, 92                               & 1, 20, 42                            & 3-4, 10-11, 19, 46                                                             \cr
5286 & \textbf{1}, 4, 45, 91                                                             & 1                                    & -                                                                              \cr
5904 & \textbf{1}-6, 19, 31, 40, 50, 53, 56, 58, 60, 70                                  & 1, 30, 41, 55, 63, 108               & 3-4, 9, 11, 14-15, 27, 30, 32, 41, 44, 55, 57, 63, 101, 106, 108, 118, 124     \cr
5927 & \textbf{1}-4, 47, 52, 68, 84                                                      & 1-2, 22                              & 22                                                                             \cr
5946 & \textbf{2}                                                                        & 1-2                                  & -                                                                              \cr
5986 & \textbf{1}-4                                                                      & 1, 42                                & 69                                                                             \cr
6121 & \textbf{1}-2, 4-6, 13, 40, 53, 56, 89                                             & 1, 20, 25, 33, 37, 49, 100, 108, 111 & 3-4, 15, 21, 24-25, 28, 49, 78, 96, 100, 106, 108, 116, 125, 129, 134-135      \cr
6171 & \textbf{1}-6, 11-12, 23, 30, 34, 40-41, 48, 60                                    & 1, 5                                 & 3-4, 12, 126                                                                   \cr
6218 & \textbf{1}-2, 4-6, 26, 40, 50, 56, 78, 83                                         & 1, 34, 85                            & 4, 56, 62, 69, 85, 91                                                          \cr
6235 & \textbf{2}                                                                        & 1, 60                                & -                                                                              \cr
6254 & \textbf{1}-2, 4-6, 25, 31, 40, 50, 56, 60, 78                                     & 1, 35, 97                            & 3-4, 11, 19, 35, 62, 77, 97, 101                                               \cr
6266 & 2-\textbf{3}                                                                      & 1                                    & -                                                                              \cr
6284 & 2-\textbf{3}                                                                      & 1                                    & -                                                                              \cr
6304 & \textbf{1}, 3-4                                                                   & 1, 79                                & -                                                                              \cr
6316 & -                                                                                 & 1-2, 95                              & -                                                                              \cr
6333 & -                                                                                 & 1                                    & -                                                                              \cr
6342 & \textbf{2}, 64                                                                    & 1-2, 76                              & 76                                                                             \cr
6352 & \textbf{1}, 4-6, 42, 50, 56, 60, 69, 82                                           & 1, 22, 110                           & 17, 22, 110                                                                    \cr
6356 & \textbf{3}                                                                        & 1-2, 37                              & -                                                                              \cr
6362 & \textbf{1}-6, 14, 56, 62, 65                                                      & 1, 42                                & 17                                                                             \cr
6388 & \textbf{1}, 85, 90                                                                & 1-2, 96, 121                         & 3, 92, 96                                                                      \cr
6441 & \textbf{1}                                                                        & 1-2, 73, 103                         & 84, 93, 103                                                                    \cr
6522 & \textbf{3}, 87                                                                    & 1, 109                               & 109                                                                            \cr
6528 & 36, \textbf{54}, 59, 72, 77, 81, 84                                               & 1-2, 70, 76, 87                      & 52-53, 70, 76                                                                  \cr
6544 & \textbf{2}                                                                        & 1, 119                               & -                                                                              \cr
6553 & 33, 44, \textbf{54}, 57, 71, 75                                                   & 1, 26, 61, 81                        & 26, 47-48, 53, 61, 81                                                          \cr
6569 & -                                                                                 & 1, 79                                & 128                                                                            \cr
6624 & \textbf{1}, 3-5, 39, 69                                                           & 1-2                                  & 128                                                                            \cr
6626 & 46, \textbf{73}                                                                   & 34, 36                               & 43                                                                             \cr
6637 & \textbf{1}-5, 39, 69                                                              & 1, 36                                & 94                                                                             \cr
6638 & \textbf{3}                                                                        & 1, 16                                & -                                                                              \cr
6652 & \textbf{1}-5, 40, 60, 66                                                          & 1-2                                  & -                                                                              \cr
6723 & \textbf{1}-6, 61                                                                  & 1                                    & 13, 39                                                                         \cr
6752 & \textbf{1}-2, 4-6, 15, 31, 40, 49-50, 56, 60, 70, 79, 84                          & 1, 20, 31, 42, 68, 107               & 4, 7-8, 11, 15, 24, 38, 40, 54, 64, 68, 72, 75, 80, 88, 90, 104, 107, 112, 130 \cr
7078 & \textbf{1}-6, 40, 50, 56, 60, 70                                                  & 1-2, 23, 31-32, 45, 86, 98, 111      & 3-4, 6, 23, 32, 40, 45, 50-51, 58, 74, 86, 101, 127                            \cr
7089 & \textbf{1}-4, 67                                                                  & 1-2                                  & 101                                                                            \cr
 \enddata
 \label{tbl:Refs-2}
\end{deluxetable}

\bigskip

\begin{deluxetable}{ccccccc}
 \tabletypesize{\footnotesize}
 \tablewidth{0pc}  
 \tablecaption{Ages, [Fe/H], [Mg/Fe], [C/Fe], [N/Fe], and [Ca/Fe] for S05 GGCs.  Boldface numbers denote less certain entries (see Appendix for details).}
 \tablehead{
  \colhead{NGC} &
  \colhead{Age} &
  \colhead{[Fe/H]} &
  \colhead{[Mg/Fe]} &
  \colhead{[C/Fe]} &
  \colhead{[N/Fe]} &
  \colhead{[Ca/Fe]} \\
  \colhead{ID} &
  \colhead{(Gyr)} &
  \colhead{(dex)} &
  \colhead{(dex)} &
  \colhead{(dex)} &
  \colhead{(dex)} &
  \colhead{(dex)} \\
  \colhead{(1)} &
  \colhead{(2)} &
  \colhead{(3)} &
  \colhead{(4)} &
  \colhead{(5)} &
  \colhead{(6)} &
  \colhead{(7)}
 }
 \startdata
 0104 &      13.1 $\pm$ 0.9  & -0.72 $\pm$ 0.08 &      +0.41 $\pm$ 0.14  &      -0.13 $\pm$ 0.20  & +0.87 $\pm$ 0.55 &      +0.17 $\pm$ 0.15  \cr
 1851 &      10.0 $\pm$ 0.5  & -1.17 $\pm$ 0.08 &      +0.40 $\pm$ 0.10  &      -0.36 $\pm$ 0.51  & +0.84 $\pm$ 0.46 &      +0.36 $\pm$ 0.09  \cr
 1904 & {\bf 11.7 $\pm$ 1.3} & -1.58 $\pm$ 0.12 &      +0.26 $\pm$ 0.07  &              -         &         -        &      +0.22 $\pm$ 0.04  \cr
 2298 &      12.7 $\pm$ 0.7  & -1.95 $\pm$ 0.04 &      +0.59 $\pm$ 0.20  &              -         &         -        &      +0.39 $\pm$ 0.05  \cr
 2808 &      10.9 $\pm$ 0.4  & -1.14 $\pm$ 0.13 &      +0.25 $\pm$ 0.18  &      -0.50 $\pm$ 0.28  & +1.25 $\pm$ 1.06 &      +0.39 $\pm$ 0.14  \cr
 3201 &      10.2 $\pm$ 0.4  & -1.59 $\pm$ 0.20 &      +0.36 $\pm$ 0.11  &      -0.55 $\pm$ 0.23  & +0.56 $\pm$ 0.27 &      +0.20 $\pm$ 0.11  \cr
 5286 &      12.5 $\pm$ 0.5  & -1.70 $\pm$ 0.07 &              -         &              -         &         -        &              -         \cr
 5904 &      10.6 $\pm$ 0.4  & -1.29 $\pm$ 0.08 &      +0.33 $\pm$ 0.10  &      -0.48 $\pm$ 0.26  & +0.68 $\pm$ 0.59 &      +0.28 $\pm$ 0.11  \cr
 5927 &      12.7 $\pm$ 0.9  & -0.49 $\pm$ 0.44 & {\bf -0.12           } &              -         &         -        & {\bf +0.34           } \cr
 5946 & {\bf  9.7 $\pm$ 1.6} & -1.29 $\pm$ 0.14 &              -         &              -         &         -        &              -         \cr
 5986 &      12.2 $\pm$ 0.6  & -1.59 $\pm$ 0.12 & {\bf +0.25           } &              -         &         -        &              -         \cr
 6121 &      12.5 $\pm$ 0.7  & -1.16 $\pm$ 0.09 &      +0.50 $\pm$ 0.08  &      -0.46 $\pm$ 0.28  & +0.65 $\pm$ 0.41 &      +0.30 $\pm$ 0.09  \cr
 6171 &      14.0 $\pm$ 0.8  & -1.02 $\pm$ 0.02 &      +0.51 $\pm$ 0.04  &              -         &         -        & {\bf +0.06 $\pm$ 0.32} \cr
 6218 &      12.7 $\pm$ 0.4  & -1.37 $\pm$ 0.12 &      +0.46 $\pm$ 0.14  & {\bf +1.18 $\pm$ 0.44} &         -        &      +0.31 $\pm$ 0.14  \cr
 6235 & {\bf  9.7 $\pm$ 1.6} & -1.28 $\pm$ 0.31 &              -         &              -         &         -        &              -         \cr
 6254 &      11.4 $\pm$ 0.5  & -1.53 $\pm$ 0.06 &      +0.44 $\pm$ 0.13  &      -0.77 $\pm$ 0.37  & +1.01 $\pm$ 0.45 &      +0.33 $\pm$ 0.11  \cr
 6266 & {\bf 11.6 $\pm$ 0.6} & -1.18 $\pm$ 0.07 &              -         &              -         &         -        &              -         \cr
 6284 & {\bf 11.0          } & -1.26            &              -         &              -         &         -        &              -         \cr
 6304 &      13.6 $\pm$ 1.1  & -0.45 $\pm$ 0.26 &              -         &              -         &         -        &              -         \cr
 6316 &             -        & -0.46 $\pm$ 0.16 &              -         &              -         &         -        &              -         \cr
 6333 &             -        & -1.77            &              -         &              -         &         -        &              -         \cr
 6342 & {\bf 10.2 $\pm$ 0.8} & -0.55 $\pm$ 0.10 &      +0.38 $\pm$ 0.01  &      -0.34 $\pm$ 0.03  &         -        &      +0.38 $\pm$ 0.01  \cr
 6352 &      12.7 $\pm$ 0.9  & -0.64 $\pm$ 0.13 &      +0.38 $\pm$ 0.17  &              -         &         -        &      +0.18 $\pm$ 0.11  \cr
 6356 & {\bf 15.0 $\pm$ 3.0} & -0.40 $\pm$ 0.12 &              -         &              -         &         -        &              -         \cr
 6362 &      13.6 $\pm$ 0.6  & -0.99 $\pm$ 0.26 &      +0.37 $\pm$ 0.16  &              -         &         -        &      +0.15 $\pm$ 0.03  \cr
 6388 &      12.0 $\pm$ 1.0  & -0.55 $\pm$ 0.15 &      +0.24 $\pm$ 0.12  &      -0.66 $\pm$ 0.15  &         -        &      +0.00 $\pm$ 0.13  \cr
 6441 &      11.3 $\pm$ 0.9  & -0.46 $\pm$ 0.06 &      +0.35 $\pm$ 0.14  &      -0.45 $\pm$ 0.17  &         -        &      +0.20 $\pm$ 0.17  \cr
 6522 & {\bf 15.0 $\pm$ 1.1} & -1.34 $\pm$ 0.36 &      +0.27 $\pm$ 0.10  &              -         &         -        &      +0.17 $\pm$ 0.06  \cr
 6528 & {\bf 12.0 $\pm$ 2.0} & -0.12 $\pm$ 0.24 &      +0.25 $\pm$ 0.11  &      -0.35 $\pm$ 0.12  &         -        &      +0.30 $\pm$ 0.09  \cr
 6544 & {\bf  8.8 $\pm$ 1.0} & -1.40 $\pm$ 0.22 &              -         &              -         &         -        &              -         \cr
 6553 & {\bf 12.0 $\pm$ 2.0} & -0.18 $\pm$ 0.03 &      +0.32 $\pm$ 0.08  &      -0.62 $\pm$ 0.10  & +1.15 $\pm$ 0.38 &      +0.20 $\pm$ 0.15  \cr
 6569 &             -        & -0.76 $\pm$ 0.13 &      +0.50 $\pm$ 0.06  &      -0.27 $\pm$ 0.11  &         -        &      +0.31 $\pm$ 0.04  \cr
 6624 &      12.5 $\pm$ 0.9  & -0.44 $\pm$ 0.07 &      +0.42 $\pm$ 0.05  &      -0.29 $\pm$ 0.20  &         -        &      +0.40 $\pm$ 0.04  \cr
 6626 & {\bf 14.0 $\pm$ 1.1} & -1.32 $\pm$ 0.05 &              -         &              -         &         -        & {\bf +0.11 $\pm$ 0.02} \cr
 6637 &      13.1 $\pm$ 0.9  & -0.64 $\pm$ 0.17 &      +0.28 $\pm$ 0.17  &              -         &         -        &      +0.20 $\pm$ 0.17  \cr
 6638 & {\bf 12.0          } & -0.95 $\pm$ 0.13 &              -         &              -         &         -        &              -         \cr
 6652 &      12.9 $\pm$ 0.8  & -0.81 $\pm$ 0.17 &              -         &              -         &         -        &              -         \cr
 6723 &      13.1 $\pm$ 0.7  & -1.10 $\pm$ 0.07 &      +0.44             &              -         &         -        &      +0.33 $\pm$ 0.13  \cr
 6752 &      11.8 $\pm$ 0.6  & -1.53 $\pm$ 0.16 &      +0.38 $\pm$ 0.15  &      -0.45 $\pm$ 0.37  & +0.93 $\pm$ 0.63 &      +0.31 $\pm$ 0.09  \cr
 7078 &      12.9 $\pm$ 0.5  & -2.39 $\pm$ 0.14 &      +0.36 $\pm$ 0.23  &      -0.30 $\pm$ 0.36  & +0.84 $\pm$ 0.63 &      +0.31 $\pm$ 0.14  \cr
 7089 &      11.8 $\pm$ 0.6  & -1.64 $\pm$ 0.08 &              -         &      -0.62 $\pm$ 0.14  &         -        &              -         \cr
 \enddata
 \label{tbl:GGCSPs-1}
\end{deluxetable}

\bigskip

\begin{deluxetable}{cccccc}
 \tabletypesize{\footnotesize}
 \tablewidth{0pc}  
 \tablecaption{[O/Fe], [Na/Fe], [Si/Fe], [Cr/Fe], and [Ti/Fe] for S05 GGCs.  Boldface numbers denote less certain entries (see Appendix for details).}
 \tablehead{
  \colhead{NGC} &
  \colhead{[O/Fe]} &
  \colhead{[Na/Fe]} &
  \colhead{[Si/Fe]} &
  \colhead{[Cr/Fe]} & 
  \colhead{[Ti/Fe]} \\
  \colhead{ID} &
  \colhead{(dex)} &
  \colhead{(dex)} &
  \colhead{(dex)} &
  \colhead{(dex)} &
  \colhead{(dex)} \\
  \colhead{(1)} &
  \colhead{(2)} &
  \colhead{(3)} &
  \colhead{(4)} &
  \colhead{(5)} &
  \colhead{(6)}
 }
 \startdata
 0104 &      +0.24 $\pm$ 0.20  &      +0.35 $\pm$ 0.22  &      +0.31 $\pm$ 0.12  &      +0.10 $\pm$ 0.08  &      +0.28 $\pm$ 0.12  \cr
 1851 &      +0.09 $\pm$ 0.26  &      +0.18 $\pm$ 0.31  &      +0.32 $\pm$ 0.11  &      -0.00 $\pm$ 0.16  &      +0.17 $\pm$ 0.06  \cr
 1904 &      +0.10 $\pm$ 0.19  &      +0.32 $\pm$ 0.25  &      +0.28 $\pm$ 0.03  &      -0.28 $\pm$ 0.14  &      +0.22 $\pm$ 0.10  \cr
 2298 &              -         &              -         &      +0.51 $\pm$ 0.05  &      -0.03 $\pm$ 0.10  &      -0.03 $\pm$ 0.03  \cr
 2808 &      +0.12 $\pm$ 0.38  &      +0.28 $\pm$ 0.27  &      +0.35 $\pm$ 0.13  &      +0.02 $\pm$ 0.08  &      +0.30 $\pm$ 0.09  \cr
 3201 &      +0.15 $\pm$ 0.27  &      +0.12 $\pm$ 0.25  &      +0.34 $\pm$ 0.11  &      -0.06 $\pm$ 0.24  &      +0.20 $\pm$ 0.12  \cr
 5286 &              -         &              -         &              -         &              -         &              -         \cr
 5904 &      +0.15 $\pm$ 0.27  &      +0.19 $\pm$ 0.26  &      +0.31 $\pm$ 0.10  &      -0.08 $\pm$ 0.19  &      +0.22 $\pm$ 0.10  \cr
 5927 & {\bf +0.33           } & {\bf +1.23           } & {\bf +0.74           } & {\bf -0.11           } & {\bf +0.37           } \cr
 5946 &              -         &              -         &              -         &              -         &              -         \cr
 5986 & {\bf +0.65           } & {\bf +0.70           } & {\bf +0.30           } & {\bf +0.15           } & {\bf +0.22           } \cr
 6121 &      +0.31 $\pm$ 0.14  &      +0.29 $\pm$ 0.19  &      +0.50 $\pm$ 0.07  &      -0.03 $\pm$ 0.07  &      +0.31 $\pm$ 0.07  \cr
 6171 &      +0.17 $\pm$ 0.18  &      +0.37 $\pm$ 0.20  &      +0.54 $\pm$ 0.08  &              -         &      +0.40 $\pm$ 0.10  \cr
 6218 &      +0.31 $\pm$ 0.34  &      +0.32 $\pm$ 0.27  &      +0.36 $\pm$ 0.07  &      -0.04 $\pm$ 0.19  &      +0.27 $\pm$ 0.13  \cr
 6235 &              -         &              -         &              -         &              -         &              -         \cr
 6254 &      +0.23 $\pm$ 0.24  &      +0.17 $\pm$ 0.27  &      +0.28 $\pm$ 0.07  &      +0.01 $\pm$ 0.15  &      +0.26 $\pm$ 0.12  \cr
 6266 &              -         &              -         &              -         &              -         &              -         \cr
 6284 &              -         &              -         &              -         &              -         &              -         \cr
 6304 &              -         &              -         &              -         &              -         &              -         \cr
 6316 &              -         &              -         &              -         &              -         &              -         \cr
 6333 &              -         &              -         &              -         &              -         &              -         \cr
 6342 &      +0.31 $\pm$ 0.02  &              -         &      +0.37 $\pm$ 0.04  &              -         &      +0.25 $\pm$ 0.03  \cr
 6352 & {\bf +0.04           } &      +0.19 $\pm$ 0.16  &      +0.27 $\pm$ 0.21  &      -0.05 $\pm$ 0.07  &      +0.23 $\pm$ 0.13  \cr
 6356 &              -         &              -         &              -         &              -         &              -         \cr
 6362 &              -         &      +0.25 $\pm$ 0.62  &      +0.43 $\pm$ 0.04  &      -0.14 $\pm$ 0.28  &      +0.38 $\pm$ 0.06  \cr
 6388 &      -0.06 $\pm$ 0.25  &      +0.42 $\pm$ 0.23  &      +0.30 $\pm$ 0.12  &      -0.06 $\pm$ 0.14  &      +0.20 $\pm$ 0.19  \cr
 6441 &      +0.09 $\pm$ 0.18  &      +0.41 $\pm$ 0.28  &      +0.36 $\pm$ 0.18  &      -0.06 $\pm$ 0.20  &      +0.32 $\pm$ 0.17  \cr
 6522 &      +0.49 $\pm$ 0.17  &      +0.04 $\pm$ 0.21  &      +0.25 $\pm$ 0.08  &              -         &      +0.16 $\pm$ 0.04  \cr
 6528 &      +0.20 $\pm$ 0.14  &      +0.41 $\pm$ 0.13  &      +0.26 $\pm$ 0.13  &      +0.00 $\pm$ 0.06  &      +0.10 $\pm$ 0.22  \cr
 6544 &              -         &              -         &              -         &              -         &              -         \cr
 6553 &      +0.36 $\pm$ 0.18  &      +0.30 $\pm$ 0.30  &      +0.23 $\pm$ 0.15  &      +0.04 $\pm$ 0.09  &      +0.18 $\pm$ 0.21  \cr
 6569 &      +0.48 $\pm$ 0.09  &              -         &      +0.49 $\pm$ 0.08  &              -         &      +0.40 $\pm$ 0.04  \cr
 6624 &      +0.41 $\pm$ 0.14  &              -         &      +0.38 $\pm$ 0.09  &              -         &      +0.37 $\pm$ 0.04  \cr
 6626 &              -         &              -         & {\bf +0.60 $\pm$ 0.07} &              -         & {\bf +0.02 $\pm$ 0.21} \cr
 6637 &      +0.20 $\pm$ 0.31  &      +0.35 $\pm$ 0.29  &      +0.45 $\pm$ 0.12  &              -         &      +0.24 $\pm$ 0.13  \cr
 6638 &              -         &              -         &              -         &              -         &              -         \cr
 6652 &              -         &              -         &              -         &              -         &              -         \cr
 6723 &              -         &              -         &      +0.68 $\pm$ 0.13  &              -         &      +0.24 $\pm$ 0.15  \cr
 6752 &      +0.26 $\pm$ 0.25  &      +0.32 $\pm$ 0.26  &      +0.47 $\pm$ 0.19  &      -0.13 $\pm$ 0.12  &      +0.20 $\pm$ 0.11  \cr
 7078 &      +0.27 $\pm$ 0.23  &      +0.36 $\pm$ 0.34  &      +0.48 $\pm$ 0.20  &      -0.23 $\pm$ 0.08  &      +0.48 $\pm$ 0.20  \cr
 7089 &              -         &              -         &              -         &              -         &              -         \cr
 \enddata
 \label{tbl:GGCSPs-2}
\end{deluxetable}


\clearpage

\clearpage
\begin{figure*}
 \begin{center}
  \includegraphics[width=0.9\textwidth]{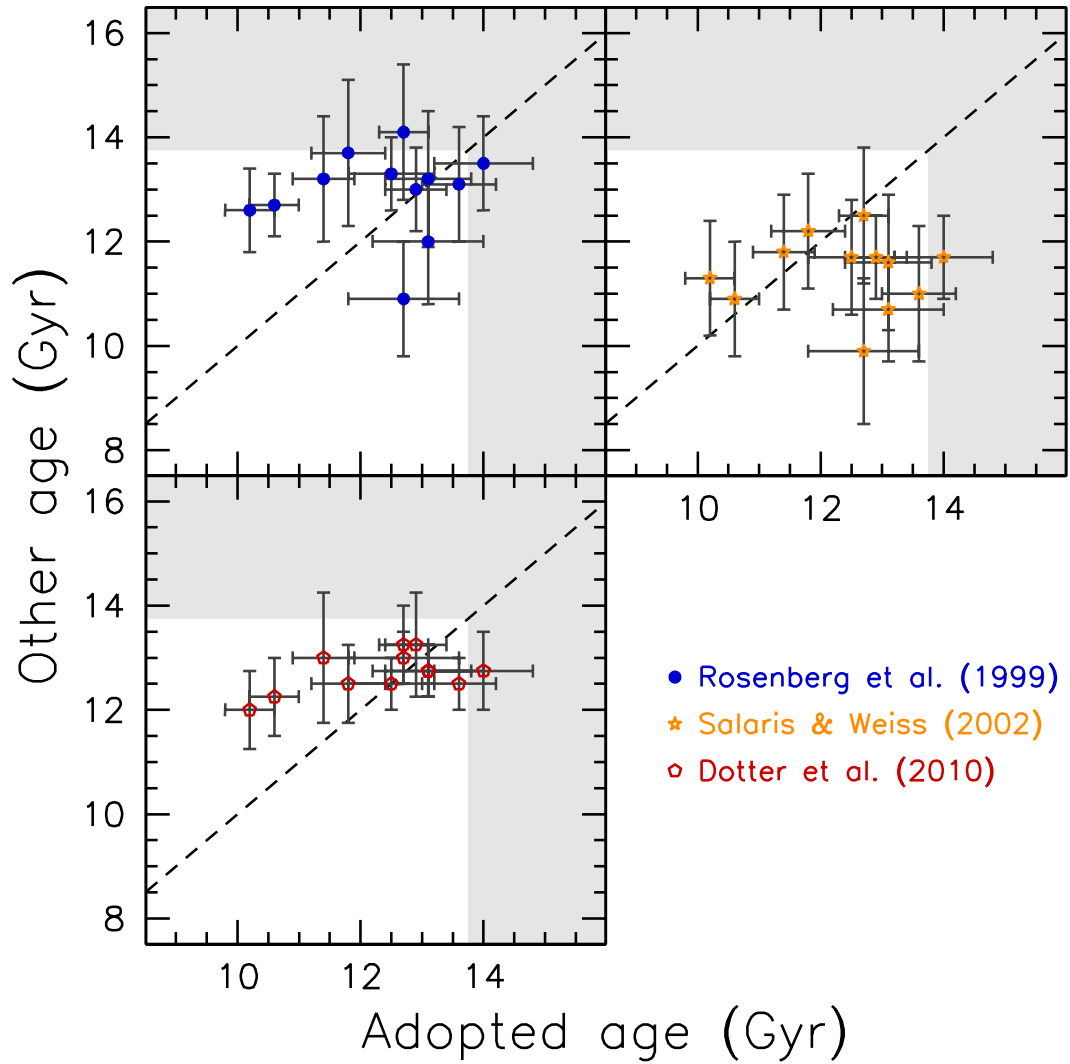}
  \caption{Comparison of our adopted ages (x-axis) against similar independent 
estimates from other homogeneous age compilations (y-axis) for a sub-sample of 
12 GGCs from S05 common to \cite{Ro99}, \cite{SW02}, \cite{Do10} and 
\citet[MF09]{MF09}.  The dashed line in each panel represents equality between 
our adopted and other age estimates and the shaded regions demarcate ages in 
excess of that of the Universe \citep{Ko11}.}
  \label{fig:CompAge}
 \end{center}
\end{figure*}

\clearpage

\clearpage

\clearpage

\clearpage

\clearpage

\clearpage

\clearpage

\clearpage

\clearpage


\clearpage


\appendix
\section{Notes on Individual Clusters}\label{sec:Notes}
In this Appendix, we provide on a cluster-by-cluster basis all known sources to 
us of ages and chemical abundance ratios for each of the S05 GGCs.  We also 
list sources of metallicity information, but restricted to those we have 
adopted since an exhaustive list for each cluster would be prohibitively large 
while adding little benefit.  Regarding the chemical abundance references, we 
explicitly identify those included in our compilation and their respective 
statistical richnesses (\ie number of measurements, $N$), outline our 
justification(s) for any exclusions thereof, and point out instances where we 
think systematic discrepancies might be an issue.  For those clusters in our 
sample known to harbour multiple stellar populations, we also comment on 
possible signatures of this phenomenon within our adopted abundance patterns 
and relate them to other evidence.  On the other hand, for the age references 
we concentrate on comparing the independent measurements against our adopted 
values or the estimated age of the Universe \cite[13.76 $\pm$ 0.11 Gyr;][]{Ko11} in order to highlight the (sometimes severe) degree of systematic error 
inherent to this quantity.  Pinpointing the various sources of this error would 
be best accomplished on a per cluster basis though and so is beyond the scope 
of the present work.

Unless stated otherwise, our adopted ages come from MF09.  We omit the ages 
from \cite{Ch96} on the grounds that nearly all of them for the S05 clusters 
exceed the age of the Universe.  Also note that the relative ages from \cite{DAng05} refer to the \cite{CG97} metallicity scale and were transformed assuming a 
normalization of 10.9 Gyr.  The \cite{Bu98} and \cite{Ro99} ages were re-cast 
into absolute terms by adding to them a zeropoint of 15.0 and 13.2 Gyr, 
respectively.  The zeropoint for the ages from \cite{Ri96} varies from 14.0 Gyr 
for the most metal-rich clusters to 16.0 Gyr for the most metal-poor ones.

\subsection{NGC 0104}
Following Ha10, we use the measured metallicities from \cite{AZ88}, \cite{CS88}, \cite{BW92}, \cite{AB05}, \cite{Wy06}, \cite{KMW08}, \cite{MWB08} and \cite{Ca09c} to calculate the weighted [Fe/H] for NGC 0104.  The large body of work on 
the chemical composition of this cluster covers all major evolutionary stages, 
including the MS \citep{Br04,Car04a,Ca05,KMW08,DO10} [$N$ = 173], RGB \citep{Pi83,Gr86,Br90,BW92,NDC95,Car04a,AB05,Ca05,Wy06,KMW08,Ca09a,Ca09b} [$N$ = 190], HB 
\citep{AB05} [$N$ = 1] and AGB \citep{Br90,BW92,Wy06,Wo09,WC12} [$N$ = 21], all 
of which we fold into our compiled abundance pattern.  Conversely, we excluded 
the results of \cite{Br91,Br96} and \cite{Ca98} from the present work on the 
grounds that they did not provide abundances for the individual stars 
comprising their respective samples.  A plethora of evidence on this cluster 
strongly suggests that it hosts multiple stellar populations, which includes 
variations in either the spectroscopic band strengths (CN, CH) or light-element 
abundances (C, N, O, Na, Mg, Al) of individual stars over a wide range of 
evolutionary stages \citep{Ma78,HB80,CDC81,Br94,Br96,Car04a,Cam06,Ca09a,Ca09b,DO10,Pa10}, distinct photometric main and red giant branch sequences \citep{An09,dC10,Pi12} and radial gradients in both colour and He abundance \citep{HB85,Na11}.  From our compilation, we find some support for the notion that NGC 0104 
hosts multiple populations via the slightly enhanced rms envelopes on our mean 
[C/Fe], [N/Fe], [Na/Fe] and [O/Fe] ratios ($\ge$0.2 dex, compared to $\le$0.15 
dex for other species).  Given this cluster's well-established abundance anti-correlations \citep{Ca09a}, we suggest that systematics between the available 
abundance measurements of Ca, Si, Cr and Ti may be responsible for weakening 
the multiple-population signal in our data.  Finally, the age of NGC 0104 has 
been determined in several other studies to date \citep{He87,Do89,SC91,CK95,Ma95,Ri96,Bu98,Ro99,Zo01,Br05,MW06,MWB08,Do10}, the values from which span a range 
of 12.0-14.7 Gyr for this parameter and are formally consistent with the one we 
have adopted (13.1 $\pm$ 0.9 Gyr).  Exceptional cases concern those 
measurements which either fall below our adopted value by 1.1-2.4 Gyr \citep[1.2-1.8$\sigma$;][]{Va00,Gr02,Gr03,SW02,DAng05,Sa07,BS09} [3.9 Gyr (2.9$\sigma$) 
in the case of \citealt{SW98}] or, worse yet, exceed the estimated age of the 
Universe by 2.7-6.2 Gyr \citep{Ha83,Gi99}.

\subsection{NGC 1851}
Following Ha10, we use the measured metallicities from \cite{AZ88}, \cite{Ge97}, \cite{Vi10}, \cite{YG08}, \cite{Yo09} and \cite{Ca09c,Ca11a} to calculate the 
weighted [Fe/H] for NGC 1851.  A large number of studies have also addressed 
the chemical composition of this cluster, covering nearly all evolutionary 
stages, including the MS \citep{La12} [$N$ = 64], SGB \citep{Gr12c} [$N$ = 77], 
RGB \citep{YG08,Yo09,Vi10,Ca11a,Gr12b} [$N$ = 8, 4, 15, 124 and 12 stars, 
respectively] and HB \citep{Gr12b} [$N$ = 91]; our recommended abundance 
pattern incorporates all of these results.  We omitted \cite{RH87} and \cite{Ca12b} from our compilation however on the grounds that neither work provided 
abundances for the individual stars comprising their respective samples.  Our 
recommended abundance pattern exhibits large rms envelopes about the mean 
abundance ratios for C (0.51 dex), N (0.46 dex), O (0.26 dex) and Na (0.31 
dex), while those for other species are typically $\leq$0.10 dex.  The large 
abundance scatters in this cluster (partly) reflect its known C$-$N and Na$-$O 
anti-correlations \citep{YG08,Yo09,Vi10,Ca10,Ca11a} and thus that it harbours 
multiple stellar populations.  Additional evidence of the multiple population 
phenomenon in NGC 1851 includes its radial colour gradient \citep{Ba88} and 
split SGB and RGB \citep{Ca08,Mi08,DA09,Ha09,Le09,Ve09,Ca11b,Pi12}.  Finally, 
the age of NGC 1851 has been determined in several other studies to date, the 
values from a minority of which \citep{Ro99,SW02,MW06} span a range of 9.2-10.6 
Gyr for this parameter and are consistent with the one we have adopted (10.0 
$\pm$ 0.5 Gyr).  Exceptional cases include measurements which either exceed our 
adopted value by 1.0-5.0 Gyr \citep[1.1-3.6$\sigma$;][]{Sa88,Wa92,CK95,Ri96,San96,Bu98,Va00} or fall below it by 1.2-2.1 Gyr \citep[1.9-2.2$\sigma$;][]{SW98,DAng05}.  Worse yet, the age from \cite{Al90a} lies in excess of that estimated 
for the Universe by 2.2 Gyr (1.1$\sigma$).

\subsection{NGC 1904}
Following Ha10, we use the measured metallicities from \cite{Fr91}, \cite{Ge97} 
and \cite{Ca09c} to calculate the weighted [Fe/H] for NGC 1904.  For its 
recommended abundance pattern, we combine results from the few studies of the 
chemical abundances of its RGB population \citep{Fr91,Ca09a,Ca09b} [$N$ = 72, 
total].  Other studies of this cluster's chemical composition include \cite{GO89}, \cite{Fa05} and \cite{SL09} but those are omitted from our compilation on 
the grounds that their results are at strong odds with those from our adopted 
references (Gratton \& Ortolani), or correspond to an advanced (post-AGB; {\c 
S}ahin \& Lambert) or some exotic (extreme HB; Fabbian) stage of stellar 
evolution.  From our compiled abundance pattern, we find that NGC 1904 has 
significant rms envelopes about its mean [Mg/Fe], [O/Fe] and [Na/Fe] ratios 
(0.2-0.3 dex) compared to those of other species ($\leq$0.13 dex), which 
reflects the known Mg-Al and Na-O anti-correlations exhibited by its RGB 
members \citep{Ca09a,Ca09b}.  These anti-correlations, along with this 
cluster's extreme HB morphology \citep{Gr10} and radial colour gradient \citep{CA83,Hi96}, provide strong evidence for the existence of multiple stellar 
population(s) within it.  Finally, since NGC 1904 was not part of the MF09 
sample, we adopt the age estimated for it by \citet[11.7 $\pm$ 1.3 Gyr]{SW02}, 
but only in a {\it tentative} sense on the grounds that a significant fraction 
of the latter's estimates are systematically younger than those of the former 
(\Fig{CompAge}).  Several other age determinations appear in the literature for 
this cluster \citep{CK95,Ri96,SW98,Ro99,MW06}, which span a range of 10.1-13.3 
Gyr and are consistent with the one we have adopted.  Exceptional cases include 
measurements which either exceed our adopted value by 2.7 Gyr \citep[1.1$\sigma$;][]{Bu98} or fall below it by 2.2 Gyr \citep[1.4$\sigma$;][]{DAng05}.  Worse 
yet, the ages from \cite{GO86}, \cite{He86}, \cite{Al94} and \cite{Kr97} lie in 
excess of that estimated for the Universe by 2.2-4.2 Gyr.

\subsection{NGC 2298}
Following Ha10, we use the measured metallicities from \cite{MW92} and \cite{Ca09c} to calculate the weighted [Fe/H] for NGC 2298.  Our only knowledge of the 
chemical abundances in this cluster comes from \cite{MW92}, who analysed three 
of its RGB members.  We therefore adopt their results for our compilation.  The 
age of NGC 2298 has been determined in several other studies to date \citep{Al90c,Ch92,CK95,SW98,SW02,Do10}, the values from which span a range of 11.7-15.0 
Gyr for this parameter and are formally consistent with the one we have adopted 
(12.7 $\pm$ 0.7 Gyr).  Exceptional cases concern those measurements which 
exceed either our adopted value by 2.8 Gyr \citep[1.5$\sigma$;][]{MW92} or, 
worse yet, the estimated age of the Universe by 1.3-4.3 Gyr \citep{AL86,GO86,JH88,Ri96,Bu98}.

\subsection{NGC 2808}
Following Ha10, we use the measured metallicities from \cite{Ge97} and \cite{Ca09c} to calculate the weighted [Fe/H] for NGC 2808.  For our recommended 
abundance pattern, we combine results from the many studies which have 
addressed the chemical composition of this cluster's MS \citep{Br10b} [$N$ = 
2], RGB \citep{Gr82,Ca03,Car04b,Ca06a,Ca06b,Ca09b} [$N$ = 1, 81, 20, 20, 120 
and 12, respectively] and HB \citep{Gr11} [$N$ = 26].  We exclude the work of 
\cite{Pa06} from our compilation on the grounds that the majority of stars in 
their sample are drawn from an exotic (extreme HB) stage of stellar evolution.  
NGC 2808 is a special case given that its multiple stellar populations are 
reflected by many signatures, such as a wide range of CN band strengths amongst 
its red giants, three distinct MSs, an extreme HB and peculiar colour 
distribution amongst its blue stragglers \citep{NS83,DC04,DAnt05,Le05,Pi07,Br10a,DE10,Gl10,Da11,DE12}.  Our compiled abundance pattern also supports the idea 
of multiple populations in this cluster via the large rms envelopes on our mean 
[Mg/Fe], [C/Fe], [N/Fe], [O/Fe] and [Na/Fe] ratios ($\gtrsim$0.2 dex), which we 
interpret as reflecting the underlying Mg$-$Al, C$-$N and Na$-$O anti-correlations that have been found amongst its member stars of {\it all} evolutionary 
phases \citep{Ca06b,Ca09b,Br10b,Gr11}.  Finally, the age of NGC 2808 has been 
determined in several other studies to date \citep{CK95,San96,Ro99,Va00}, the 
values from which span a range of 10.7-12.4 Gyr for this parameter and are 
consistent with the one we have adopted (10.9 $\pm$ 0.4 Gyr).  Exceptional 
cases include measurements which either exceed our adopted value by 2.6-3.1 Gyr 
\citep[1.5-2.1$\sigma$;][]{Ri96,Bu98} or fall below it by 1.2-2.6 Gyr \citep[1.4-2.6$\sigma$;][]{SW02,DAng05,MW06}.  Worse yet, the ages from \cite{Bu84}, 
\cite{GO86} and \cite{Al90b} lie in excess of that estimated for the Universe 
by 2.2-4.2 Gyr.

\subsection{NGC 3201}
Following Ha10, we use the measured metallicities from \cite{Be90}, \cite{Ge97} 
and \cite{Ca09c} to calculate the weighted [Fe/H] for NGC 3201.  For our 
recommended abundance pattern, we combine the results from the many studies 
that have analyzed the chemical composition of its RGB members: \cite{Gr82} 
[$N$ = 2], \cite{Pi83} [$N$ = 4], \cite{GO89} [$N$ = 3], \cite{GW98} [$N$ = 18] 
and \cite{Ca09a,Ca09b} [$N$ = 149 and 10, respectively].  \cite{RH89} have 
analysed the Ca abundances of HB stars belonging to this cluster but we omit 
their result from our compilation given that these authors only provide the 
mean [Ca/Fe] ratio for their sample.  Based on several lines of evidence, 
including a split RGB \citep{Kr10}, radial gradients in spectroscopic index 
strengths \citep{Ch88}, a bimodal CN distribution amongst its red giants \citep{SN82} and chemical abundance anti-correlations \citep{Ca09a,Ca09b}, it has been 
suggested that NGC 3201 hosts multiple stellar populations.  Finding signatures 
of this phenomenon in our compiled abundance pattern is complicated by the fact 
that the rms envelopes for the anti-correlated elements (C, N, O, Na; 0.23-0.27 
dex) are similar to that of Cr (0.24 dex); the latter's size could be an 
artifact of systematic bias though between the metallicities determined by the 
sources of our [Cr/Fe] ratios \citep{Gr82,Pi83,GO89}.  Finally, the age of NGC 
3201 has been determined in several other studies to date \citep{AL81,Ca84,CK95,SW98,SW02}, the values from which span a range of 9.9-12.0 Gyr for this 
parameter and are consistent with the one we have adopted (10.2 $\pm$ 0.4 
Gyr).  Exceptional cases include measurements which either exceed our adopted 
value by 1.0-5.8 Gyr \citep[1.6-3.3$\sigma$;][]{Al89,Ri96,Bu98,Ro99,LS03,Bo10,Do10} [3.2 Gyr (5.0$\sigma$) in the case of \citealt{LS03}] or fall below it by 
1.2-2.2 Gyr \citep[3.0-3.9$\sigma$;][]{DAng05,MW06}.  Worse yet, the ages from 
\cite{Sam96a} \cite{vBM01} lie in excess of that estimated for the Universe by 
0.24 and 4.2 Gyr, respectively.

\subsection{NGC 5286}
Little is known about the stellar population(s) of NGC 5286, except for its age 
and metallicity.  Following Ha10, we simply adopt the measured value of this 
parameter from \cite{Ca09c}.  Alternative determinations of the age of this 
cluster have been made by \cite{Sa95}, \cite{Zo09} and \cite{Do10}, only the 
latter of which (13.0 $\pm$ 1.0 Gyr) is consistent with the one we have adopted 
(12.5 $\pm$ 0.5 Gyr).  In both of the other cases, the determinations lie in 
excess of the estimated age of the Universe by 2.5-3.5 Gyr.

\subsection{NGC 5904}
Following Ha10, we use the measured metallicities from \cite{Sn92}, \cite{Sh96}, \cite{Iv01}, \cite{RC03}, \cite{Yon08c} and \cite{Ca09c} to calculate the 
weighted [Fe/H] for NGC 5904.  The chemical composition of this cluster has 
been extremely well-studied across nearly all relevant evolutionary stages, 
including the MS \citep{RC03} [$N$ = 6], SGB \citep{Br92,Co02} [$N$ = 51], RGB 
\citep{Ar94,Br92,Ca09a,Ca09b,Gr86,Iv01,KMW10,Lai11,La85,Mart08,Pi80,Pi83,RC03,Sh96,Sm97,Sn92,Yon08a,Yon08c} [$N$ = 276], HB \citep{Lai11} [$N$ = 2] and AGB 
\citep{Iv01,KMW10,Lai11,Sn92} [$N$ = 21]; we fold all of the results from these 
studies into our compilation.  However, we omit the [O/Fe] ratios from \cite{Gr87a} since the authors determined them for two different assumptions of their 
stars' carbon abundances and it is not clear which they prefer.  From our 
recommended abundance pattern, we find that this cluster's carbon, oxygen and 
sodium abundances are each spread over a range of $\sim$0.3 dex, and $\sim$0.6 
dex for its nitrogen abundance.  These comparitively large spreads (most other 
elements have rms envelopes of 0.1 dex) reflect the abundance anti-correlations 
that have been found amongst this cluster's MS and RGB populations \citep{Os71,RC03,Ca09a,Ca09b,Lai11}, a hallmark of the multiple population phenomenon in 
globular clusters.  Additional evidence suggesting that NGC 5904 hosts more 
than one stellar population is the radial colour gradient found by \cite{Bu81}.  Finally, the age of NGC 5904 has been determined in several other studies to 
date \citep{San96,JP98,SW98,SW02,MW06}, the values from which span a range of 
9.9-10.9 Gyr for this parameter and are consistent with the one we have adopted 
(10.6 $\pm$ 0.4 Gyr).  Exceptional cases include [concern those] measurements 
which either exceed our adopted value by 0.9-4.3 Gyr \citep[1.0-2.9$\sigma$;][]{CK95,Ri96,Bu98,Ro99,Va00,Do10} or fall below it by 1.8 Gyr \citep[2.0$\sigma$;][]{DAng05}.  Worse yet, the ages from \cite{RF87} and \cite{SC91} lie in excess 
of that estimated for the Universe by 2.2-3.2 Gyr.

\subsection{NGC 5927}
Following Ha10, we use the measured metallicities from \cite{AZ88}, \cite{Fr91} 
and \cite{Ca09c} to calculate the weighted [Fe/H] for NGC 5927.  The generous 
error on our adopted metallicity (-0.49 $\pm$ 0.44 dex) stems from the wide 
spread ($\sim$0.8 dex) amongst the input measurements.  Our only knowledge of 
the chemical abundances in this cluster comes from \cite{Fr91}, who analysed 
one of its RGB members.  As such, we adopt their results in our compilation, 
but with some reservations since their measured metallicity (-1.08 dex) falls 
well outside the error on our adopted value for this parameter, while their 
[Mg/Fe], [Na/Fe] and [Si/Fe] ratios (-0.12, +1.23 and +0.74 dex, respectively) 
seem suspiciously low.  The age of NGC 5927 has been determined in several 
other studies to date \citep{Fu96,FG00,Br05,MW06,Do10}, the values from which 
span a range of 10.9-13.0 Gyr for this parameter and are consistent with the 
one we have adopted (12.7 $\pm$ 0.9 Gyr).  Exceptional cases concern those 
measurements which either fall below our adopted value by 2.7 Gyr \citep[2.5$\sigma$;][]{DAng05} or, worse yet, exceed the estimated age of the Universe by 
1.24 Gyr \citep{Sam96b}.

\subsection{NGC 5946}
Little is known about the stellar population(s) of NGC 5946, except for its age 
and metallicity.  Following Ha10, we use the measured values of this parameter 
from \cite{AZ88} and \cite{Ca09c} to calculate the weighted [Fe/H] for this 
cluster.  These input values are identical though, so we adopt the uncertainty 
quoted by Carretta to provide some idea of that associated with our 
metallicity.  Since NGC 5946 was not part of the MF09 sample, we adopt the age 
estimated for it by \citet[9.7 $\pm$ 1.6 Gyr]{DAng05}, but only in a {\it 
tentative} sense on the grounds that the latter's estimates are biased to 
younger values relative to those of the former.

\subsection{NGC 5986}
Following Ha10, we use the measured metallicities from \cite{Ge97} and \cite{Ca09c} to calculate the weighted [Fe/H] for NGC 5986.  Our only knowledge of the 
chemical abundances in this cluster comes from \cite{Jas04}, who analysed two 
highly-evolved members, one which they hypothesize as being well into its 
post-AGB phase and experienced a third dredge-up, and the other as just 
beginning to leave the AGB sequence.  As such, we only adopt the latter's 
abundances into our compilation.  Despite the good agreement between the 
measured metallicity of this star (-1.65 dex) and our adopted value for this 
cluster (-1.59 $\pm$ 0.12 dex), we notice that its [O/Fe], [Na/Fe] and [Cr/Fe] 
ratios all seem rather large in comparison to those for the rest of the S05 
sample.  We therefore advise caution when interpreting results based on this 
single star's abundance pattern.  Alternative determinations of the age of NGC 
5986 have been made by \cite{DAng05}, \cite{MW06} and \cite{Do10}, the latter 
two of which span a range of 12.0-13.2 Gyr for this parameter and are 
consistent with the one we have adopted (12.2 $\pm$ 0.6 Gyr).  The age 
determined by De Angeli \etal falls below our adopted value by 2.6 Gyr 
(3.1$\sigma$).

\subsection{NGC 6121}
Following Ha10, we use the measured metallicities from \cite{Be90}, \cite{BW92}, \cite{Dr94}, \cite{Mi95b}, \cite{Iv99}, \cite{Mari08}, \cite{Yon08c}, \cite{Ta09} and \cite{Ca09c} to calculate the weighted [Fe/H] for NGC 6121.  For our 
recommended abundance pattern, we combine the results from the many studies of 
the chemical composition of this cluster's MS/SGB \citep{Mo12} [$N$ = 91], RGB 
\citep{Br90,BW92,Ca09a,Ca09b,DOM10,Dr92,Gr86,Iv99,Mari08,SmV05,SS91,VG11,Wa07,Yon08a,Yon08c} [$N$ = 425], HB \citep{Ma11,Vi12} [$N$ = 28] and AGB \citep{Iv99} 
[$N$ = 10] populations.  Many of the above studies suggest that NGC 6121 plays 
host to multiple stellar populations on the basis of anti-correlations observed 
amongst the light-element abundances of its MS, RGB and HB members \citep{SmV05,Mari08,Ma11,Ca09a,Ca09b,DOM10,VG11,Vi12,Mo12}.  In addition, \cite{Mari08} 
found that this cluster's RGB is split in colour-magnitude diagrams based, in 
part, on U$-$band information.  Our compilation also supports the existence of 
multiple populations in this cluster through the comparatively large rms 
envelopes on our mean [C/Fe], [N/Fe], [O/Fe] and [Na/Fe] ratios (0.14-0.41 
dex), whereas those of other light elements remain below 0.1 dex.  Finally, the 
age of NGC 6121 has been determined in several other studies to date \citep{Ca85,CK95,San96,Ro99,SW02,DA09,Do10}, the values from which span a range of 11.7-13.3 Gyr for this parameter and are consistent with the one we have adopted (12.5 
$\pm$ 0.7 Gyr).  Exceptional cases concern those measurements which either fall 
below our adopted value by 2.6 Gyr \citep[3.2$\sigma$;][]{DAng05} or, worse 
yet, exceed the estimated age of the Universe by 2.1 Gyr \citep[1.3$\sigma$;][]{Bu98}.

\subsection{NGC 6171}
Following Ha10, we use the measured metallicities from \cite{SZ78} and \cite{Ca09c} to calculate the weighted [Fe/H] for NGC 6171.  Ha10's value for this 
parameter can be exactly reproduced if Searle \& Zinn's measurement is 
transformed onto Carretta's scale (-1.00 dex; Harris, {\it priv. comm.}); we 
have therefore adopted this transformation here.  Nearly all of our recommended 
abundance pattern for this cluster is based on the results of the few studies 
that have focussed on the elemental abundances amongst its RGB population: 
\cite{Ca09a,Ca09b} and \cite{OC11} [$N$ = 33, 5 and 13, respectively].  For its 
Ca abundance, we draw on the mean [Ca/Fe] ratio from \cite{SM83} with the 
caveat that it is based on an exotic phase of stellar evolution (RR Lyrae) and 
that their quoted [Fe/H] (-0.84 $\pm$ 0.25 dex) is systematically lower than 
(but nevertheless consistent with) that of \cite{Ca09c}.  \cite{SP82} also 
measured the mean Ca abundance of NGC 6171, but given that their result is in 
excellent agreement with that of Smith \& Manduca, we opted to omit it from our 
compilation.  Owing to the CN/CH band strength variations \citep{Sm88} and a 
Na-O anti-correlation observed amongst its RGB members \citep{Ca09a,Ca09b}, NGC 
6171 is suspected of hosting multiple stellar populations, a suggestion which 
is supported by the comparitively large rms envelopes we find on its mean Na 
and O abundances ($\sim$0.2 dex, compared to $\lesssim$0.1 dex for other 
species).  Finally, the age of NGC 6171 has been determined in several other 
studies to date \citep{DC84,Fe91,CK95,Ji96,Ro99}, the values from which span a 
range of 13.5-16.0 Gyr for this parameter and are formally consistent with the 
one we have adopted (14.0 $\pm$ 0.8 Gyr).  Exceptional cases concern those 
measurements which either fall below our adopted value by 1.2-3.6 Gyr \citep[1.1-2.8$\sigma$;][]{SW98,SW02,DAng05,MW06,Do10} [5.0 Gyr (6.2$\sigma$) in the case 
of \citealt{MW06}] or, worse yet, exceed the estimated age of the Universe by 
3.2-6.2 Gyr \citep{SR84,Bu89,FP92,Fe95}.  Note that the age we have adopted for 
this cluster is statistically consistent with being younger than that of the 
Universe.

\subsection{NGC 6218}
Following Ha10, we use the measured metallicities from \cite{DCA95}, \cite{JP06} and \cite{Ca09c} to calculate the weighted [Fe/H] for NGC 6218.  Our 
recommended abundance pattern for this cluster combines results from several 
studies that have targetted members of its RGB and AGB phases: \cite{JP06} and 
\cite{Ca07b,Ca09b} [$N$ = 21, 79 and 11, respectively; RGB], and \cite{KS01}, 
\cite{Mi03} and \cite{Jas04} [$N$ = 3 total; AGB].  We omit the results of 
\cite{Kl03} from our compilation on the grounds that they correspond to an 
advanced stage of stellar evolution (post-AGB); indeed, their abundances for 
several species, although formally consistent, exhibit (sometimes large) 
differences from our adopted values.  The mean [O/Fe] and [Na/Fe] ratios in our 
abundance pattern are distinguished by larger rms envelopes (0.34 and 0.27 dex, 
respectively) compared to those of other species (0.14 dex, in the median) on 
account of the fact that NGC 6218 is known to exhibit a Na-O anti-correlation 
\citep{Ca07b}, an established hallmark of the multiple populations phenomenon 
in GGCs.  Finally, the age of NGC 6218 has been determined in several other 
studies to date \citep{CK95,Ri96,Bu98,SW02,Ha04,Do10}, the values from which 
span a range of 11.8-14.5 Gyr for this parameter and are formally consistent 
with the one we have adopted (12.7 $\pm$ 0.4 Gyr).  Exceptional cases include 
measurements which either exceed our adopted value by 1.4 Gyr \citep[1.0$\sigma$;][]{Ro99} or fall below it by 2.7 Gyr \citep[2.7$\sigma$;][]{DAng05}.  Worse 
yet, the ages from \cite{Sa89} and \cite{vB02} lie in excess of that estimated 
for the Universe by 2.2-3.2 Gyr ($>$3.2$\sigma$).

\subsection{NGC 6235}
Little is known about the stellar population(s) of NGC 6235, except for its age 
and metallicity.  Following Ha10, we use the measured values of this parameter 
from \cite{Ho03} and \cite{Ca09c} to calculate the weighted [Fe/H] for this 
cluster.  The generous error on our adopted metallicity (-1.28 $\pm$ 0.31 dex) 
stems from the wide spread ($\sim$0.4 dex) amongst the input measurements.  
Since NGC 6235 was not part of the MF09 sample, we adopt the age estimated for 
it by \citet[9.7 $\pm$ 1.6 Gyr]{DAng05}, but only in a {\it tentative} sense on 
the grounds that the latter's estimates are biased to younger values relative 
to those of the former.

\subsection{NGC 6254}
Following Ha10, we use the measured metallicities from \cite{Kr95}, \cite{Ha08} 
and \cite{Ca09c} to calculate the weighted [Fe/H] for NGC 6254.  A large number 
of studies addressing the abundance pattern of this cluster (predominantly 
through its RGB population) are found within the literature, the majority of 
which we incorporate in our compilation.  These studies include: \cite{Pi83} 
[$N$ = 3], \cite{GO89} [$N$ = 2], \cite{Kr95} [$N$ = 14], \cite{Mi03} [$N$ = 
2], \cite{SmG05} [$N$ = 15], \cite{Ha08} [$N$ = 5], \cite{Mart08} [$N$ = 8] and 
\cite{Ca09a,Ca09b} [$N$ = 14 and 147, respectively].  Omitted works include 
\cite{Gr80}, \cite{GL97}, \cite{Mo01,Mo04} and \cite{Mi09} either because they 
focussed on an advanced/exotic stage of stellar evolution or expressed their 
abundances relative to another cluster not included in the present work.  NGC 
6254 is suspected of hosting multiple stellar populations on the basis of CN/CH 
band strength variations \citep{SF97}, large scatters in both [C/Fe] and [N/Fe] 
ratios at fixed luminosity \citep{Os71,SmG05} and a Na-O anti-correlation 
observed amongst its RGB members \citep{Ca09a,Ca09b}.  Our adopted abundance 
pattern for this cluster supports such a notion via the large rms envelopes 
attached to our mean [C/Fe], [N/Fe], [O/Fe] and [Na/Fe] ratios (0.37, 0.45, 
0.24 and 0.27 dex, respectively), whereas other ratios have envelopes 
$\leq$0.15 dex in size.  Finally, while the age of NGC 6254 has been determined 
in several other studies to date, only that from \citet[11.8 $\pm$ 1.1 Gyr]{SW02} is consistent with the one we have adopted (11.4 $\pm$ 0.5 Gyr).  The many 
exceptional cases include  measurements which either exceed our adopted value 
by 1.6-3.8 Gyr \citep[1.2-1.8$\sigma$;][]{Bu98,Ro99,Do10} or fall below it by 
1.3-2.0 Gyr \citep[1.1-2.1$\sigma$;][]{SW98,DAng05}.  Worse yet, the ages from 
several sources \citep{Hu89,SC91,CK95,Ri96,vB02} lie in excess of that 
estimated for the Universe by 2.2-6.2 Gyr ($>$1.1$\sigma$).

\subsection{NGC 6266}
Little is known about the stellar population(s) of NGC 6266, except for its age 
and metallicity.  Following Ha10, we simply adopt the measured value of this 
parameter from \cite{Ca09c}.  To our knowledge, the age of this cluster has 
only been estimated by \citet[10.0 $\pm$ 0.6 Gyr]{DAng05} and \citet[11.0 $\pm$ 0.6 Gyr]{MW06}.  Since NGC 6266 was not part of the MF09 sample, we adopt the 
latter estimate for our compilation on the grounds that the De Angeli ages are 
biased to younger values relative to those of MF09.  We also note that, in this 
instance, the age estimate from De Angeli \etal is younger than that of 
Meissner \& Weiss by 1.6 Gyr, a difference of 1.9$\sigma$.

\subsection{NGC 6284}
Little is known about the stellar population(s) of NGC 6284, except for its age 
and metallicity.  Since \cite{Ca09c} derive the metallicity of this cluster 
(-1.31 $\pm$ 0.09 dex) from the 2003 edition of \cite{Ha96}, we simply adopt 
the corresponding value listed in Ha10 (-1.26 dex), at the expense of not 
having an error estimate.  \cite{SP82} measured the abundance of calcium in two 
RR Lyrae stars from this cluster ([Ca/Fe] = +0.48 $\pm$ 0.32 dex), but we omit 
their result from our compilation since the mean metallicity they quoted for 
those same stars (-0.91 $\pm$ 0.25 dex) seems anomalously high compared to our 
adopted value.  To our knowledge, the age of NGC 6284 has only been estimated 
by \citet[9.5 $\pm$ 0.5 Gyr]{DAng05} and \citet[11.0 Gyr]{MW06}.  Since this 
cluster was not part of the MF09 sample, we adopt the latter estimate for our 
compilation on the grounds that the De Angeli ages are biased to younger values 
relative to those of MF09.  We also note that, in this instance, the age 
estimate from De Angeli \etal is younger than that of Meissner \& Weiss by 1.5 
Gyr, a difference of $\lesssim$3.0$\sigma$.

\subsection{NGC 6304}
Little is known about the stellar population(s) of NGC 6304, except for its age 
and metallicity.  Following Ha10, we use the measured values of this parameter 
from \cite{Va05} and \cite{Ca09c} to calculate the weighted [Fe/H] for this 
cluster.  The generous error on our adopted metallicity (-0.45 $\pm$ 0.26 dex) 
stems from the wide spread ($\sim$0.3 dex) amongst the input measurements.  
Alternative determinations of the age of NGC 6304 have been made by \cite{MW06} 
and \cite{Do10}, which together span a range of 12.8-13.6 Gyr for this 
parameter and are consistent with the one we have adopted (13.6 $\pm$ 1.1 Gyr).

\subsection{NGC 6316}
Little is known about the stellar population(s) of NGC 6316, except for its 
metallicity.  Following Ha10, we use the measured values of this parameter from 
\cite{AZ88}, \cite{Va07} and \cite{Ca09c} to calculate the weighted [Fe/H] for 
this cluster.

\subsection{NGC 6333}
Little is known about the stellar population(s) of NGC 6333, except for its 
metallicity.  Since \cite{Ca09c} derive their value for this parameter (-1.79 
$\pm$ 0.09 dex) from the 2003 edition of \cite{Ha96}, we simply adopt the 
corresponding [Fe/H] listed for this cluster in Ha10 (-1.77 dex), at the 
expense of not having an error estimate.

\subsection{NGC 6342}
Following Ha10, we use the measured metallicities from \cite{AZ88}, \cite{Or05} 
and \cite{Ca09c} to calculate the weighted [Fe/H] for NGC 6342.  Our only 
knowledge of the chemical abundances in this cluster comes from \cite{Or05} as 
well, who analysed four of its RGB members.  We therefore adopt their results 
for our compilation.  To our knowledge, the age of NGC 6342 has only been 
estimated by \citet[14.5 $\pm$ 0.4 Gyr]{HR99} and \citet[10.2 $\pm$ 0.8 Gyr]{DAng05}.  Since this cluster was not part of the MF09 sample, we adopt the latter 
estimate for our compilation on the grounds that that from Heitsch \& Richtler 
exceeds the age of Universe by 0.7 Gyr, or 1.8$\sigma$.  However, we 
nevertheless regard our adopted value with skepticism as the De Angeli ages are 
biased to younger values relative to those of MF09.

\subsection{NGC 6352}
Following Ha10, we use the measured metallicities from \cite{Fr91}, \cite{Ca09c} and \cite{Fe09} to calculate the weighted [Fe/H] for NGC 6352.  Our 
recommended abundance pattern for this cluster also combines results from \cite{Fr91} and \cite{Fe09}, as well as \cite{Gr87b}, whom have collectively 
targetted its RGB [$N$ = 4] and HB [$N$ = 9] stars.  We omit the [O/Fe] data 
obtained by \cite{Gr87a} on the grounds that they were computed for two 
different [C/Fe] ratios, which itself has yet to be constrained for this 
cluster.  Based on variations observed in the CN/CH band strengths of its RGB 
members, \cite{Pa10} have claimed that NGC 6352 hosts multiple stellar 
populations.  Signatures of this phenomenon are difficult to come by in our 
adopted abundance pattern however: our [O/Fe] ratio comes from \cite{Fr91}, 
whose results correspond to a single star (and are thus statistically ill-defined), while the rms envelopes for our [Mg/Fe] and [Na/Fe] ratios (0.17 and 0.16 
dex, respectively) are not remarkably different than those for unaffected 
species (\eg 0.21 and 0.13 dex for Si and Ti, respectively).  The comparitively 
large rms envelope on our mean [Ti/Fe] ratio may be explained by Gratton's use 
of overestimated equivalent widths, small numbers of lines and a different 
treatment for collisional broadening \citep{Fe09}.  Finally, the age of NGC 
6352 has been determined in several other studies to date, the values from a 
minority of which \citep{Fu95,Ri96,Do10} span a range of 13.0-15.2 Gyr for this 
parameter and are formally consistent with the one we have adopted (12.7 $\pm$ 
0.9 Gyr).  Exceptional cases include measurements which either exceed our 
adopted value by 2.6 Gyr \citep[1.4$\sigma$;][]{Bu98} or fall below it by 
1.8-3.3 Gyr \citep[1.3-3.1$\sigma$;][]{SW98,SW02,Ro99,Pu03}.  Worse yet, the 
age from \cite{He00} lie in excess of that estimated for the Universe by 0.24 
Gyr (2.2$\sigma$).

\subsection{NGC 6356}
Little is known about the stellar population(s) of NGC 6356, except for its age 
and metallicity.  Following Ha10, we use the measured values of this parameter 
from \cite{AZ88}, \cite{Mi95b} and \cite{Ca09c} to calculate the weighted 
[Fe/H] for this cluster.  To our knowledge, the age of NGC 6356 has only been 
estimated by \citet[15.0 $\pm$ 3.0 Gyr]{MW06} and thus we adopt it for our 
compilation.  Note that this result is statistically consistent with being less 
than the age of the Universe.

\subsection{NGC 6362}
Following Ha10, we use the measured metallicities from \cite{Ge97} and \cite{Ca09c} to calculate the weighted [Fe/H] for NGC 6362.  In terms of its abundance 
pattern, we only adopt the results of \cite{Gr87b}, whom analysed the chemical 
abundances of two red giant branch stars belonging to this cluster.  The Ca 
abundance of NGC 6362 was also measured by \cite{SP82}, but we omit this result 
from our compilation on the grounds that they correspond to an exotic phase of 
stellar evolution (RR Lyrae), despite the fact that their [Fe/H] and [Ca/Fe] 
ratios compare well with our adopted values.  The age of NGC 6362 has been 
determined in several other studies to date, the values from a minority of 
which \citep{Bu98,Pi99,Ro99} span a range of 13.1-15.1 Gyr for this parameter 
and are formally consistent with the one we have adopted (13.6 $\pm$ 0.6 Gyr).  
Exceptional cases concern those measurements which either fall below our 
adopted value by 1.1-3.6 Gyr \citep[$>$1.4$\sigma$;][]{Br99,SW02,DAng05,MW06,Do10} or, worse yet, exceed the estimated age of the Universe by 2.4 Gyr 
\citep[1.5$\sigma$;][]{AL86}.

\subsection{NGC 6388}
Following Ha10, we use the measured metallicities from \cite{AZ88}, \cite{Wa07}, \cite{WC10} and \cite{Ca09c} to calculate the weighted [Fe/H] for NGC 6388.  
Our recommended abundance pattern for this cluster combines the results from 
the few studies which have focussed on the elemental abundances amongst its RGB 
population: \cite{Ca07c,Ca09a} and \cite{Wa07} [$N$ = 7, 36 and 8, 
respectively].  The mean [O/Fe] and [Na/Fe] ratios in our abundance pattern are 
distinguished by larger rms envelopes (0.25 and 0.23 dex, respectively) 
compared to those of other species (0.14 dex, in the median), likely on account 
of the fact that NGC 6388 is known to exhibit a Na-O anti-correlation.  The 
small spread in [Mg/Fe] ratios (0.12 dex) however seems at odds with this 
cluster's purported Mg-Al anti-correlation \citep{Ca07c,Ca09a}.  The light 
abundance variations observed within NGC 6388, along with its split SGB \citep{Mo09,Pi12} and bimodal HB \citep{SC98,Bu07,Yoo08}, provide strong evidence that 
it is home to multiple stellar populations.  Alternative determinations of the 
age of this cluster have been made by \cite{Hu07} and \cite{Mo09}, which 
together span a range of 11.5-12.0 Gyr for this parameter and are consistent 
with the one we have adopted (12.0 $\pm$ 1.0 Gyr).

\subsection{NGC 6441}
Following Ha10, we use the measured metallicities from \cite{AZ88}, \cite{Cle05}, \cite{Or08} and \cite{Ca09c} to calculate the weighted [Fe/H] for NGC 6441.  
Our recommended abundance pattern for this cluster combines the results from 
the few studies that have focussed on the elemental abundances amongst its RGB 
population: \cite{Gr06}, \cite{Grat07} and \cite{Or08} [$N$ = 5, 25 and 8, 
respectively].  Based on the Na-O and Mg-Al anti-correlations exhibited within 
their respective datasets, these authors argue that NGC 6441 harbours multiple 
stellar populations, a suggestion which coincides with the popular 
interpretation of a helium enhancement as the source of its highly extended 
(bimodal) HB \citep{SC98,La99,Bu07,CD07,Yoo08}.  The notion of multiple 
populations within this cluster is supported by the rather large rms spread in 
our compiled Na abundance (+0.41 $\pm$ 0.28 dex), while systematics may be to 
blame for the lack of any other abundance signature to this effect (the 
remaining rms values range from 0.15-0.20 dex).  To our knowledge, the only age 
estimate available in the literature for NGC 6441 is the one we have adopted 
from MF09.

\subsection{NGC 6522}
Following Ha10, we use the measured metallicities from \cite{Ba09} and \cite{Ca09c} to calculate the weighted [Fe/H] for NGC 6522.  Our only knowledge of the 
chemical abundances in this cluster comes from \cite{Ba09} as well, who 
analysed eight of its RGB members.  We therefore adopt their results for our 
compilation.  To our knowledge, the age of NGC 6522 has only been estimated by 
\citet[15.0 $\pm$ 1.1 Gyr]{MW06} and \citet[14.7 $\pm$ 0.4 Gyr]{Ba09}, both of 
which exceed the known age of the Universe to significant degrees (1.1 and 
2.3$\sigma$, respectively).  Since this cluster was not part of the MF09 
sample, we adopt the former estimate for our compilation, but only in a {\it 
tentative} sense on the grounds that it could suffer from unknown systematic 
errors relative to those of MF09.

\subsection{NGC 6528}
Following Ha10, we use the measured metallicities from \cite{AZ88}, \cite{Zo04}, \cite{Or05}, \cite{So06} and \cite{Ca09c} to calculate the weighted [Fe/H] for 
NGC 6528.  The generous error on our adopted metallicity (-0.12 $\pm$ 0.24 dex) 
stems from the wide spread ($\sim$0.6 dex) amongst the individual 
measurements.  The few studies that have analysed the chemical composition of 
this cluster so far have focussed on either its RGB \citep{Co01,Or05,Zo04} [$N$ 
= 14, 4 and 2, respectively] or HB \citep{Ca01,Zo04} [$N$ = 6 and 1, 
respectively] populations.  For our recommended abundance pattern, we combine 
the results from these studies, except for the C, N and Ca abundances from 
Zoccali, which for each species are identical amongst all three of their stars 
(and thus suspect).  Moreover, Zoccali \etal find sub-solar abundances of Ti 
for two stars in their sample, which helps inflate the rather large rms 
envelope on our mean [Ti/Fe] ratio.  Finally, since NGC 6528 was not part of 
the MF09 sample, we adopt the age estimated for it by \citet[12.0 $\pm$ 2.0 Gyr]{Br97} but only in a {\it tentative} sense on the grounds that the latter 
estimate could suffer from unknown systematic errors relative to those of the 
former.  Several other age determinations appear in the literature for this 
cluster \citep{Ri98,Or01,FJ02,Mo03,Br05}, which span a range of 11.0-13.0 Gyr 
and are consistent with the one we have adopted.  On the other hand, the age 
estimated by \cite{Or92} lies in excess of that estimated for the Universe by 
0.24 Gyr ($\lesssim$2.2$\sigma$).

\subsection{NGC 6544}
Little is known about the stellar population(s) of NGC 6544, except for its age 
and metallicity.  Following Ha10, we use the measured values of this parameter 
from \cite{Ca09c} and \cite{Val10} to calculate the weighted [Fe/H] for this 
cluster.  The generous error on our adopted metallicity (-1.40 $\pm$ 0.22 dex) 
stems from the wide spread ($\sim$0.3 dex) amongst the input measurements.  
Since NGC 6544 was not part of the MF09 sample, we adopt the age estimated for 
it by \citet[8.8 $\pm$ 1.0 Gyr]{DAng05}, but only in a {\it tentative} sense on 
the grounds that the latter's estimates are biased to younger values relative 
to those of the former.

\subsection{NGC 6553}
Following Ha10, we use the measured metallicities from \cite{Ba92}, \cite{Me03}, \cite{AB06} and \cite{Ca09c} to calculate the weighted [Fe/H] for NGC 6553.  
For our recommended abundance pattern, we also draw on the above works 
(excluding Carretta et al.) as well as \cite{Ba99}, \cite{Co99} and \cite{Co01}, whom have each determined the chemistry of this cluster, largely based on its 
RGB population (except Coelho et al., who focussed on HB stars).  These studies 
used samples of 1, 5, 4, 2, 5 and 8 stars, respectively.  We do not include in 
our compilation the C or O abundances from \cite{Or02} on the grounds that 
those results lack a rigorous definition for each of the two stars in their 
sample.  To our knowledge, no evidence exists in the literature to suggest that 
NGC 6553 harbours multiple stellar populations.  The large rms envelopes that 
we find on this cluster's mean [N/Fe] and [Na/Fe] ratios ($\geq$0.3 dex), 
however, might be first evidence to this effect.  Finally, since NGC 6553 was 
not part of the MF09 sample, we adopt the age estimated for it by \citet[12.0 $\pm$ 2.0 Gyr]{Br97} but only in a {\it tentative} sense on the grounds that the 
latter estimate could suffer from unknown systematic errors relative to those 
of the former.  Several other age determinations appear in the literature for 
this cluster \citep{DL92,Or95,Be01,Zo01}, which span a range of 10.5-13.2 Gyr 
and are consistent with the one we have adopted.  On the other hand, the age 
determined by \cite{Gu98} lies in excess of that estimated for the Universe by 
2.24 Gyr.

\subsection{NGC 6569}
Following Ha10, we use the measured metallicities from \cite{Va05} and \cite{Ca09c} to calculate the weighted [Fe/H] for NGC 6569.  Our only knowledge of the 
chemical abundances in this cluster comes from \cite{Va11} who analysed six of its RGB members.  We therefore adopt their values for our compilation.  We are 
not aware of any age determinations for NGC 6569 in the literature.

\subsection{NGC 6624}
Following Ha10, we use the measured metallicities from \cite{AZ88} and \cite{Ca09c} to calculate the weighted [Fe/H] for NGC 6624.  Our only knowledge of the 
chemical abundances in this cluster comes from \cite{Va11}, who analysed five 
of its RGB members.  We therefore adopt their results for our compilation.  
Alternative determinations of the age of NGC 6624 have been made by \cite{MW06} 
and \cite{Do10}, which together span a range of 12.0-13.0 Gyr for this 
parameter and are consistent with the one we have adopted (12.5 $\pm$ 0.9 
Gyr).  In other cases, the determinations either fall below our adopted value 
by 1.9 Gyr \citep[1.1$\sigma$;][]{SW02} or, worse yet, exceed the estimated age 
of the Universe by 0.2-4.2 Gyr \citep[$\gtrsim$2.2$\sigma$;][]{Ri94,He00}.

\subsection{NGC 6626}
Our only solid knowledge about the stellar population(s) of NGC 6626 concerns 
its metallicity and, to a lesser extent, its age.  Following Ha10, we use the 
measurements of the former parameter from \cite{DCA95} and \cite{Mi95a} to 
calculate the weighted [Fe/H] for this cluster.  \cite{GL97} measured the 
calcium, silicon and titanium abundance for a highly-evolved and variable 
(post-AGB, RV Tau) star belonging to this cluster.  We {\it tentatively} adopt 
their results in our compilation, given the good agreement between their 
measured [Fe/H] (-1.31 $\pm$ 0.10 dex) and the value we have adopted for this 
parameter (-1.32 $\pm$ 0.05 dex).  To our knowledge, the age of NGC 6626 has 
only been estimated by \citet[16.0 Gyr]{Da96} and \citet[14.0 $\pm$ 1.1 Gyr]{Te01}, where the former exceeds the known age of the Universe by 2.2 Gyr.  Since 
this cluster was not part of the MF09 sample, we adopt the latter estimate for 
our compilation, but only in a {\it tentative} sense as well on the grounds 
that it could suffer from unknown systematic errors relative to those of MF09.

\subsection{NGC 6637}
Following Ha10, we use the measured metallicities from \cite{Mi95a} and \cite{Ca09c} to calculate the weighted [Fe/H] for NGC 6637.  In terms of its abundance 
pattern, we only adopt the results of \cite{Le07}, whom analysed the chemical 
abundances of two RGB and three (red) HB stars belonging to this cluster.  We 
omit the Fe, Mg and Si abundances measured by \cite{Ge84} on the grounds that 
they are statistically ill-defined (corresponding to a single star), lack error 
estimates and favour a much more metal-poor designation for the cluster (-1.21 
dex, as opposed to our adopted value of -0.64 dex).  Moreover, the Si abundance 
from Geisler is greater than our adopted value by 0.56 dex (4.7$\sigma$).  
Unfortunately, \cite{Le07} failed to notice this discrepancy, making the 
precise reason(s) for it unclear; increasing Geisler's [Fe/H] value would 
certainly reduce the discrepancy, but at the expense of introducing a new one 
between their [Mg/Fe] measurement (+0.21 dex) and that of Lee (+0.28 dex).  
\cite{Le07} found evidence of a Na-O anti-correlation amongst their sample, 
which is reflected through the rather large uncertainties we quote for the mean 
abundances of these two species ($\sim$0.3 dex each) and agrees with the 
variations/bimodality in the observed CN band strengths of this cluster's RGB 
population \citep{Ge86,Sm89}.  Finally, the age of NGC 6637 has been determined 
in several other studies to date, the values from two of which \citep{MW06,Do10} span a range of 12.5-13.2 Gyr for this parameter and are consistent with the 
one we have adopted (13.1 $\pm$ 0.9 Gyr).  Exceptional cases concern those 
measurements which either fall below our adopted value by 2.5-3.2 Gyr \citep[1.5-2.3$\sigma$;][]{SW02,DAng05} or, worse yet, exceed the estimated age of the 
Universe by 0.2-4.2 Gyr \citep[$\gtrsim$2.2$\sigma$;][]{Ri94,He00}.

\subsection{NGC 6638}
Little is known about the stellar population(s) of NGC 6638, except for its age 
and metallicity.  Following Ha10, we use the measured values of this parameter 
from \cite{SS86} and \cite{Ca09c} to calculate the weighted [Fe/H] for this 
cluster.  Since NGC 6638 was not part of the MF09 sample, we adopt the age 
estimated for it by \citet[12.0 Gyr]{MW06}, but only in a {\it tentative} sense 
on the grounds that it could suffer from unknown systematic errors relative to 
the estimates of MF09.

\subsection{NGC 6652}
Our only knowledge about the stellar population(s) of NGC 6652 concerns its age 
and metallicity.  Following Ha10, we use the measured values of this parameter 
from \cite{AZ88} and \cite{Ca09c} to calculate the weighted [Fe/H] for this 
cluster.  The age of NGC 6652 has been determined in several other studies to 
date, the values from two of which \citep{Ch00,Do10} span a range of 11.7-13.2 
Gyr for this parameter and are consistent with the one we have adopted (12.9 
$\pm$ 0.8 Gyr).  Exceptional cases concern those measurements which fall below 
our adopted value by 1.3-4.9 Gyr \citep[1.2-3.6$\sigma$;][]{CK95,SW98,SW02,DAng05,MW06}.

\subsection{NGC 6723}
Following Ha10, we simply adopt the measured value of this cluster's 
metallicity from \cite{Ca09c}.  For its abundance pattern, we combine results 
from the work of \cite{Ge84} and \cite{Fu96} whom, respectively, measured the 
chemical abundances of one and three of its RGB members.  The only overlap 
between these two works is with regards to the ratio [Si/Fe], where they 
together agree that its value lies in the narrow range of +0.64 $-$ +0.68 
dex\footnote{This agreement may be illegitimate though since Fullton find a 
metallicity that is lower than that from Geisler by 0.12 dex (1.3$\sigma$) and 
Carretta \etal by 0.16 dex (1.4$\sigma$).}.  Since Fullton does not provide 
abundance information on a star-by-star basis, we opt to use their mean [Si/Fe] 
ratio alone in our compilation on the grounds that it carries an error estimate 
and is statistically more representative than Geisler's single-star 
measurement.  The age of NGC 6723 has been determined in several other studies 
to date, the values from two of which \citep{Ro99,Do10} span a range of 
12.8-13.2 Gyr for this parameter and are consistent with the one we have 
adopted (13.1 $\pm$ 0.7 Gyr).  Exceptional cases concern those measurements 
which either fall below our adopted value by 1.5-2.5 Gyr \citep[1.0-1.7$\sigma$;][]{SW02,DAng05} [3.1 Gyr (4.4$\sigma$) in the case of \citealt{MW06}] or, 
worse yet, exceed the estimated age of the Universe by 2.4 Gyr \citep[3.4$\sigma$;][]{Al99}.

\subsection{NGC 6752}
Following Ha10, we use the measured metallicities from \cite{Be90}, \cite{Mi93}, \cite{Ge97}, \cite{Cav04}, \cite{Yon08b} and \cite{Ca09c} to calculate the 
weighted [Fe/H] for NGC 6752.  Our recommended abundance pattern for this 
cluster folds in results from several sources, the majority of which have 
focussed on the chemical composition of its RGB members \citep{BD80,Ca07a,Ca09b,Ca12a,Cav04,DCC80,Gr86,Gr05,Mi96,NDC95,Pi83,SS91,Yo03,Yo05,Yo06,Yon08b} [$N$ = 
421], while some others have done likewise for its MS \citep{Gr01,Ca05,Pa08} 
[$N$ = 11], SGB \citep{Gr01,Ca05} [$N$ = 9] and HB \citep{Vi09} [$N$ = 6] 
populations.  The significant effort invested in analysing the chemistry of NGC 
6752 has yielded many detections of anti-correlations in the abundances of 
several light elements amongst the members of multiple evolutionary stages 
\citep[MS/RGB/HB;][]{CDC81,Cav04,Ca07a,Gr01,Pa05,Sh10,Vi09,Yo05,Yo06,Yon08a,Yon08b,Yon08c}.  This, along with corresponding detections of a photometric split 
in its MS and RGB \citep{Ca09a,Ca11b,Mi10}, strongly suggest that NGC 6752 
harbours more than one stellar population.  Our recommended abundance pattern 
supports the idea of multiple populations within this cluster, based on the 
large spreads in our compiled [C/Fe] (0.37 dex), [N/Fe] (0.63 dex), [O/Fe] 
(0.25 dex) and [Na/Fe] (0.26 dex) ratios, compared to those for the other 
elements we have tabulated (0.12 dex, in the median).  Abundance analyses of 
NGC 6752 which we excluded from our compilation include \cite{Gr87a}, \cite{Gl89}, \cite{Pa05} and \cite{Hu09} either because they did not provide unique 
results (Gratton, Pasquini), their $\alpha$-abundances strongly disagree with 
those from the above works (Glaspey) or their sample corresponds to an exotic 
stage of stellar evolution (Hubrig).  Finally, the age of this cluster has been 
determined in several other studies to date \citep{Ri96,Va00,SW02,DAng05,Do10}, 
the values from which span a range of 11.2-13.3 Gyr for this parameter and are 
consistent with the one we have adopted (11.8 $\pm$ 0.6 Gyr).  Exceptional 
cases include measurements which either exceed our adopted value by 1.9-3.0 Gyr 
\citep[1.2-1.6$\sigma$;][]{CK95,Bu98,Ro99,Gr03} [$\gtrsim$1.2 Gyr (2.0$\sigma$) 
in the case of \citealt{Bu86}] or fall below it by 2.2 Gyr \citep[1.8$\sigma$;][]{SW98}.  Worse yet, the ages from \cite{SC91}, \cite{Re96} and \cite{Br05} lie 
in excess of that estimated for the Universe by 0.7-2.2 Gyr ($>$2.2$\sigma$).

\subsection{NGC 7078}
Following Ha10, we use the measured metallicities from \cite{AZ88}, \cite{Sn91}, \cite{Mi93}, \cite{Ar94}, \cite{Sn97}, \cite{Pr06}, \cite{Ki08}, \cite{Ta09} 
and \cite{Ca09c} to calculate the weighted [Fe/H] for NGC 7078.  The chemical 
composition of this cluster has been the subject of numerous studies in the 
literature, albeit largely focussed on its RGB members \citep{Ar94,Ca09a,Ca09b,Co79,Mart08,Mi96,Sn91,Sn97,Sn00a,Sn00b,So11} [$N$ = 163], while similar studies 
of this cluster's other evolutionary stages include \cite{Co05} [$N$ = 68, 
SGB], and \cite{Be03}, \cite{Pr06} and \cite{So11} [$N$ = 23 total, HB]; these 
results form the basis of our recommended abundance pattern.  A number of 
studies were excluded from our compilation on the grounds that either their 
analysis concerned an advanced stage of stellar evolution \citep[post-AGB;][]{Bi01,Mo04}, their results could not be transformed into ratios of the form 
[$X$/Fe] \citep{Tr83} or they found highly sub-solar [Mg/Fe] and [Si/Fe] ratios 
\citep{Jas04}.  Our compiled data for NGC 7078 show relatively enhanced rms 
envelopes about the mean abundances for carbon (0.36 dex), nitrogen (0.63 dex) 
and sodium (0.34 dex), while those for other species are modest (0.08-0.23 dex) 
in size.  We interpret this as a signature of multiple populations within this 
cluster.  A wide variety of other evidence exists to this effect, such as CN/CH 
band strength variations \citep{Pa10}, anti-correlations of light-element 
abundances \citep{Ca09a,Ca09b}, the colour spread amongst this cluster's RGB 
population \citep{Lar11}, and a radial colour gradient \citep{St94,Lar11}.  
Finally, the age of NGC 7078 has been determined in several other studies to 
date \citep{CK95,Ro99,Va00,Do10}, the values from which span a range of 
13.0-14.1 Gyr for this parameter and are formally consistent with the one we 
have adopted (12.9 $\pm$ 0.5 Gyr).  Exceptional cases concern those 
measurements which either fall below our adopted value by 1.2-2.5 Gyr \citep[1.2-2.4$\sigma$;][]{SW98,SW02,DAng05} or, worse yet, exceed the estimated age of 
the Universe by 1.9-3.0 Gyr \citep[1.3-1.9$\sigma$;][]{Ri96,Bu98}.

\subsection{NGC 7089}
Following Ha10, we use the measured metallicities from \cite{AZ88} and \cite{Ca09c} to calculate the weighted [Fe/H] for NGC 7089.  In terms of its abundance 
pattern, we only adopt the results of \cite{Mart08}, whom analysed the carbon 
abundances of six RGB stars belonging to this cluster.  We omit the Fe, Mg, Ca 
and Ti abundances measured by \cite{GL97} on the grounds that they correspond 
to a single, highly-evolved (post-AGB) star, and thus are likely ill-defined.  
Despite several lines of evidence which point towards the existence of multiple 
populations within this cluster, such as its split SGB and RGB, bimodal CN/CH 
band strengths and a positive radial colour gradient \citep{MH81,SM90,So96,Da09,Lar11,Sm11,Pi12}, the rms spread about its mean [C/Fe] ratio does not appear 
especially large (0.14 dex).  Alternative determinations of the age of NGC 7089 
have been made by \cite{MW06} and \cite{Do10}, which together span a narrow 
range of 12.5-12.8 Gyr for this parameter and are consistent with the one we 
have adopted (11.8 $\pm$ 0.6 Gyr).  In other cases, the determinations either 
fall below our adopted value by 1.9 Gyr \citep[1.4$\sigma$;][]{DAng05} or, 
worse yet, exceed the estimated age of the Universe by 2.2 Gyr \citep{Da00}.

\end{document}